\documentclass[a4paper,12pt]{report}

\usepackage[utf8]{inputenc}
\usepackage[T2A]{fontenc}
\usepackage[ukrainian, english]{babel}  

\usepackage{amsmath, amssymb, amsthm} 
\usepackage{graphicx}                 
\graphicspath{{Images/}}               
\usepackage{caption, subcaption}       
\usepackage{geometry}                  
\usepackage{hyperref}                  
\usepackage{cleveref}                  
\usepackage{tikz}
\usepackage{todonotes}
\usepackage{rotating}
\usepackage[braket, qm]{qcircuit}

\begin{document}

\begin{titlepage}
    \centering

    {\Large \textbf{Ivan Franko National University of Lviv}} \\[1cm]
    {\large \textbf{ Department of Theoretical Physics}} \\[3cm]
   
    \textbf{MASTER THESIS} \\[0.6cm]

    \rule{\textwidth}{0.4pt} \\[0.5cm]
    {\Large \textbf{Utilization of SU(2) Symmetry for Efficient Simulation of Quantum Systems}} \\[0.4pt]
    \rule{\textwidth}{0.4pt} \\[1cm]

    \large \textbf{Author: Oleksa Hryniv} \\[2cm]

    \includegraphics[width=0.5\textwidth]{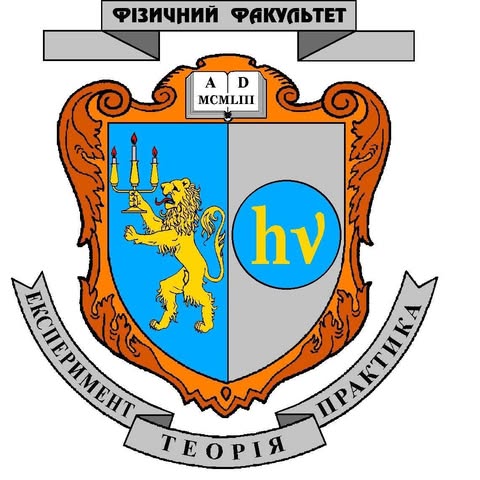}

    A thesis submitted in fulfillment of the requirements for the degree of Master of Science in the Department of Theoretical Physics\\
    Faculty of Physics

    \vfill
    Lviv-2025
\end{titlepage}

\begin{abstract}
\addcontentsline{toc}{chapter}{Abstract}
This work investigates variational compilation methods for simulating quantum systems with internal \textit{SU(2)} symmetry. The central component of the research is the application of the Dynamic Mode Decomposition (DMD) method to extrapolate trained variational circuit parameters beyond the initial optimization range. An approach is proposed for predicting variationally compiled quantum states with a larger number of Trotter steps using extrapolated parameters, eliminating the need for retraining. The efficiency of the method is validated by comparing it with classical Trotterization and the results of variational training.

The proposed method demonstrates an effective integration of symmetry-consistent quantum circuit architecture with spectral prediction techniques. The methodology shows promise for scalable modeling of strongly correlated systems, particularly in condensed matter physics problems, such as the Heisenberg model on Kagome lattices.\vspace{2cm}
\end{abstract}

\tableofcontents
\listoffigures

\chapter{Introduction}
\label{chap:introduction}

Quantum computing offers a transformative advantage in tackling computational tasks that are challenging or infeasible for classical computers. A natural domain for quantum computers is the simulation of quantum physical systems—a concept pioneered by Richard Feynman as a key motivation for developing quantum machines \cite{Feynman1982}. However, achieving practical quantum advantages is hindered by the limitations of current quantum devices, known as Noisy Intermediate-Scale Quantum (NISQ) systems, which are constrained by a limited number of qubits, short coherence times, and modest precision.

Under these constraints, \textbf{variational quantum algorithms (VQAs)} have gained prominence. These algorithms employ a hybrid approach, combining classical optimization with parameterized quantum computations. VQAs exhibit strong scalability in NISQ settings and are applied in fields such as quantum chemistry, optimization, machine learning, and quantum system modeling. Of particular interest is the simulation of the dynamics of strongly correlated many-body systems, such as the Heisenberg or Hubbard models.

A foundational method for simulating the evolution of a quantum system with a Hamiltonian $H$ is the approximation of the unitary evolution operator $U(t) = e^{-iHt}$ through \textbf{Trotterization}—a decomposition into elementary unitary operations. This approach enables the representation of system evolution in terms of operators suitable for implementation on quantum hardware.

The primary objective of this work is to investigate quantum systems possessing fundamental physical symmetries, specifically \textit{SU(2)} symmetry, and to implement their dynamical evolution on quantum computers using Trotterization techniques. Particular emphasis is placed on the development and analysis of variational quantum circuits (ansatzes) that preserve \textit{SU(2)} symmetry during the simulation of system evolution. The theoretical foundations for constructing \textit{SU(2)}-equivariant variational circuits are detailed in Section~\ref{sec:su2_theoretical}, the advantages of this approach are justified, and practical aspects of implementing quantum circuits on modern quantum devices are discussed in Section~\ref{sec:su2_circuit}.

A central focus of this work is addressing the challenge of extrapolating parameters of variational quantum circuits. Accurate simulation of quantum system dynamics often requires a large number of Trotter steps, which significantly increases the depth of quantum circuits and complicates their training. To overcome this issue, a novel approach based on the \textbf{Dynamic Mode Decomposition (DMD)} method is proposed in Section~\ref{sec:extrapolation_parameters}. The method involves initially training variational circuits on a limited (smaller) number of Trotter steps and then using the resulting parameters to extrapolate their behavior for a larger number of Trotter steps without requiring retraining.\\

The proposed algorithm consists of the following steps:
\begin{enumerate}
    \item Training the ansatz on a limited number of Trotter steps, incorporating specialized regularization constraints on the parameters to prevent overfitting and enhance training stability.
    \item Constructing a parameter matrix based on the training results and applying DMD to identify dominant modes of parameter behavior in the quantum circuit.
    \item Extrapolating the obtained circuit parameters to a larger number of Trotter steps using predictions derived from the DMD analysis.
\end{enumerate}

Numerical experiments (Section~\ref{ch:resuts}) validate the effectiveness of the proposed approach. The results demonstrate that the extrapolated parameters achieve high accuracy, closely matching those of direct Trotterization, while requiring significantly lower computational resources.

Thus, the method developed in this work combines the advantages of symmetry-oriented variational circuits with spectral approaches to parameter prediction, paving the way for efficient simulation of complex quantum systems on contemporary quantum devices.            
\chapter{Theoretical Framework}

\section{Variational Quantum Algorithms}

Quantum computers unlock new possibilities for solving problems that are computationally complex or even intractable for classical computers. This is due to the unique properties of quantum systems, such as superposition, entanglement, and quantum parallelism. In particular, quantum algorithms demonstrate advantages in areas such as:

\begin{itemize}
    \item Quantum simulations (studying electronic and quantum phases in superconductors, molecular modeling, quantum chromodynamics);
    \item Classical simulations (computational fluid dynamics, biophysical processes);
    \item Solving optimization problems;
    \item Cryptography and security;
    \item Machine learning and data analysis.
\end{itemize}

Despite the significant potential of quantum algorithms, their practical application is currently limited by the capabilities of modern quantum devices. One of the primary obstacles to widespread adoption of quantum computing is the need to develop fault-tolerant quantum computers capable of correcting errors during computations.

According to estimates from Google, IBM, and Microsoft, achieving practical fault-tolerant quantum advantage may require millions of physical qubits due to the need for quantum error-correcting codes. Current implementations have only hundreds of qubits, making full error correction infeasible. Leading companies estimate that developing fully fault-tolerant quantum computers could take decades. IBM’s quantum roadmap projects the creation of a $100,000$-qubit device by 2033--2040, marking a significant step toward fault-tolerant computing \cite{ibm_quantum_2025, mit_tech_review_2025}.

Modern quantum computers belong to the class of Noisy Intermediate-Scale Quantum (NISQ) devices, characterized by short coherence times, which significantly limit the depth of quantum circuits, a restricted number of qubits, and high noise levels, which complicate the scaling of quantum algorithms.

Variational Quantum Algorithms (VQAs) have emerged as a leading strategy for achieving quantum advantage on current NISQ devices. The core idea of VQAs is to combine quantum computations with classical optimization algorithms. A critical component of variational quantum algorithms is the use of classical resources for optimization; their structure involves the utilization of Parameterized Quantum Circuits (PQCs) executed on a quantum computer, while the optimization of parameters is delegated to a classical optimizer~\cite{cerezo2021variational}.

This approach offers several key advantages that make it suitable for NISQ devices. In particular, the hybrid quantum-classical framework allows the integration of quantum and classical computational capabilities, using the quantum component only for specific operations, thereby minimizing the depth of the quantum circuit and reducing the impact of noise.

Most variational quantum algorithms share a similar structure, consisting of several key components: a \textit{loss function}, an \textit{ansatz} (PQC), and a set of \textit{training data} tailored to the specific problem.

\subsubsection{Loss Function}
The \textbf{loss function} $C(\boldsymbol{\theta})$ is used to encode the solution to a problem. It quantifies how well a parameterized quantum circuit (PQC, or ansatz) approximates the target function or state. Optimizing this function is a central process in variational quantum algorithms, as it guides the updating of the quantum circuit's parameters to achieve the desired outcome.

Formally, the minimization task is defined as
\begin{equation}
\boldsymbol{\theta}^* = \arg\min_{\boldsymbol{\theta}} C(\boldsymbol{\theta}).
\end{equation}
Depending on the problem, the loss function can take various forms. For optimization tasks, it is typically expressed as $C(\boldsymbol{\theta}) = \langle \psi(\boldsymbol{\theta}) | H | \psi(\boldsymbol{\theta}) \rangle$, where $H$ is the system's Hamiltonian. In quantum machine learning, it may be formulated in terms of cross-entropy, KL-divergence, Hellinger distance, or other similarity metrics.

Without loss of generality, the loss function can be represented as
\begin{equation}
C(\boldsymbol{\theta}) = f \left( \{ \rho_k\}, \{ O_k \}, U(\boldsymbol{\theta}) \right),
\end{equation}
where $f$ is a certain function, $U(\boldsymbol{\theta})$ is a unitary operator dependent on the parameters $\boldsymbol{\theta}$, $\{ O_k \}$ is a set of observables used for measurements, and $\{ \rho_k \}$ is a set of input states from the training sample. Often, it is convenient to express the loss function in the form
\begin{equation}
C(\boldsymbol{\theta}) = \sum_k f_k \left( \text{Tr} \left[ O_k U(\boldsymbol{\theta}) \rho_k U^\dagger(\boldsymbol{\theta}) \right] \right),
\label{eq:loss_function}
\end{equation}
where $\{f_k\}$ is a set of weights determining the contribution of each term.

Although there are no explicit requirements for the function $f$ or its individual components $\{f_k\}$, the loss function $C(\boldsymbol{\theta})$ must satisfy several important criteria~\cite{cerezo2021variational}:
\begin{itemize}
    \item \textbf{Correctness}: The minimum must correspond to the correct solution to the problem.
    \item \textbf{Relevance to the problem structure}: Smaller values of the function should correspond to better solutions, not merely satisfy the condition $\boldsymbol{\theta}^* = \arg\min_{\boldsymbol{\theta}} C(\boldsymbol{\theta})$.
    \item \textbf{Trainability}: The loss function must enable efficient optimization of parameters.
\end{itemize}

The chosen loss function, combined with the known structure of the ansatz, defines a hypersurface often referred to as the cost landscape. The optimizer’s task is to find the set of parameters corresponding to the global minimum of the loss function.

The choice of ansatz and loss function is critical for successfully finding the optimal solution. If the cost landscape is overly complex, with numerous local minima or plateaus, optimization may be inefficient, complicating the search for the optimal solution.

One of the most significant challenges in optimizing variational quantum algorithms is the phenomenon of \textit{barren plateaus}~\cite{McClean_2018}—regions in the parameter space where the gradients of the loss function decay exponentially with increasing system size. Since information from a quantum computer can only be obtained through a limited set of measurements, the presence of barren plateaus implies the need for an exponentially large number of measurements to determine the direction of minimization of the loss function~\cite{larocca2024reviewbarrenplateausvariational}. This results in optimization algorithms being unable to effectively update parameters, effectively halting optimization. Consequently, even the use of classical gradient-based methods, such as stochastic gradient descent, fails to ensure convergence due to the exponential vanishing of gradients as the system size increases.

\subsubsection{Ansatz}
The \textbf{ansatz} is the next critical component of a variational quantum algorithm. Its structure, specifically the use and arrangement of single- and multi-qubit quantum operators, determines the set of states the ansatz can span and the degree of entanglement it can generate.

The choice of a specific ansatz typically depends on the problem itself, as in many cases, prior knowledge about the problem can be used to construct it. Such ansatzes are referred to as \textit{problem-inspired ansatzes}. For instance, if it is known that the solution lies within a specific subspace of the Hilbert space, it is advantageous to use an ansatz that ensures the prepared state remains within that subspace. This can significantly reduce the number of parameters required for optimization and enhance the algorithm’s efficiency. For example, in the Variational Quantum Eigensolver (VQE) for molecular simulations, ansatzes are often used that preserve certain quantum numbers, such as total spin or particle number, corresponding to the physical constraints of the system. Similarly, in VQE for computing the ground state of the Kagome lattice~\cite{bosse2021probinggroundstateproperties, Kattem_lle_2022}, it is beneficial to use an \textit{SU(2)}-equivariant ansatz~\cite{east2023needspinsu2equivariant}, as the Kagome Hamiltonian itself is \textit{SU(2)}-equivariant.

Conversely, there are \textit{problem-agnostic ansatzes}—universal architectures applied when no additional information about the problem’s structure is available. These ansatzes are typically constructed as repeated identical blocks (layers) consisting of single- and multi-qubit operators, providing the necessary expressivity and the ability to generate complex entangled states~\cite{sim2019expressibility}. However, due to the lack of physically motivated constraints, such ansatzes may suffer from the barren plateau problem.

\begin{figure}[ht]
\begin{center}
    \begin{tikzpicture}
        \node (top) at (0,0) {\includegraphics[width=0.6\textwidth]{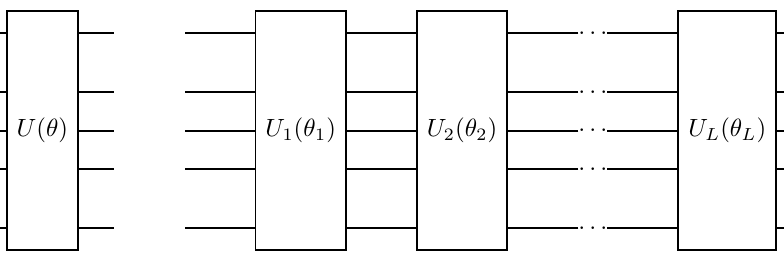}}; 
    \end{tikzpicture}

    \begin{tikzpicture}
        \node (bottom) at (0,-3.5) {\includegraphics[width=0.95\textwidth]{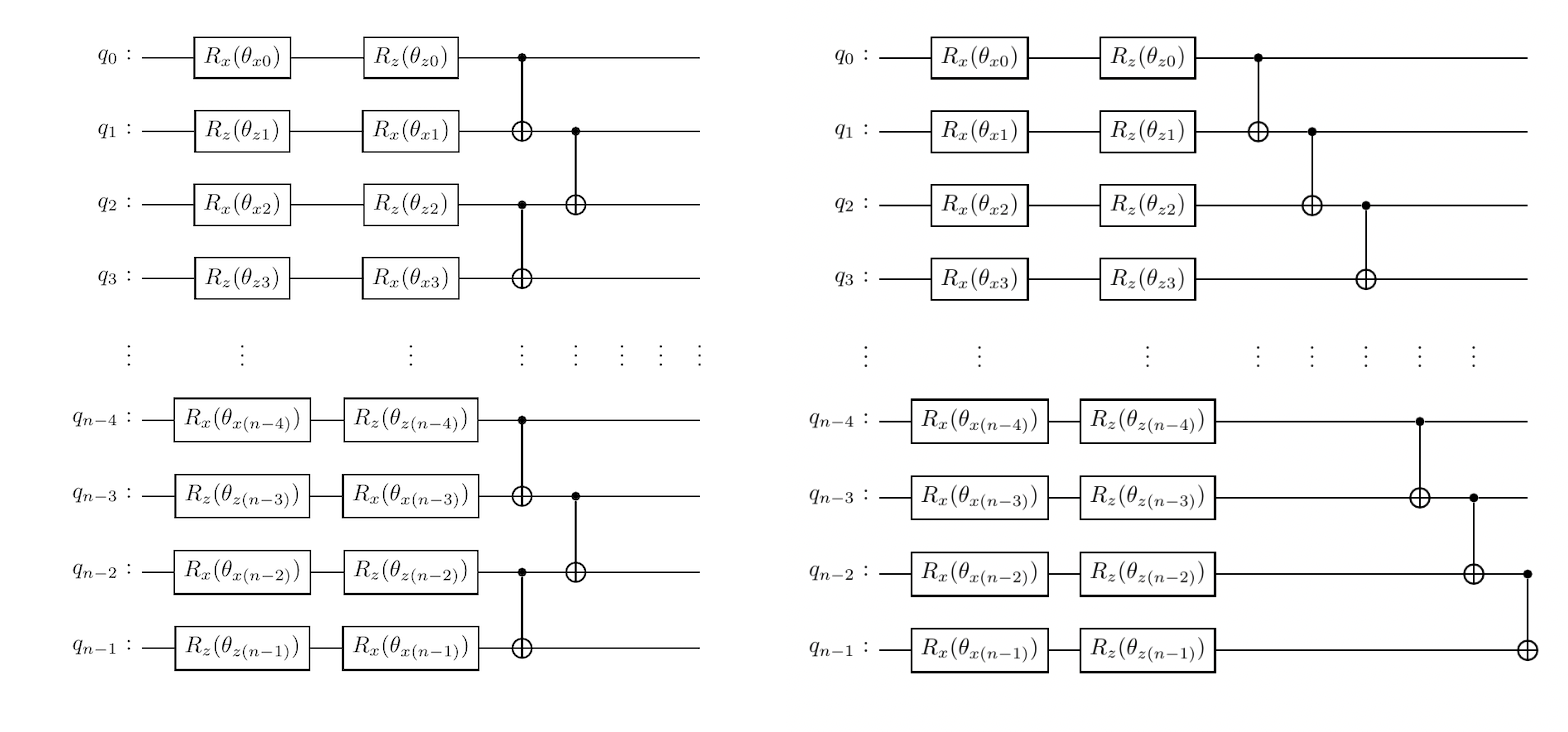}}; 
        
        \draw[thick, ->] (0.5, 1) -- (-1.5,-0.2); 
        \draw[thick, ->] (1, 1) -- (3, -0.2);   
    \end{tikzpicture}
\end{center}
    \caption{Schematic structure of the ansatz}
    \label{fig:ansatz_structure_execution}
\end{figure}

For a loss function defined by Equation~\eqref{eq:loss_function}, the parameters are typically encoded in the unitary operator $U(\boldsymbol{\theta})$, which is applied to the initial state of the quantum circuit. As shown in Figure~\ref{fig:ansatz_structure_execution}, the operator can be expressed as a composition of sequentially applied unitary operators:
\begin{equation}
U(\boldsymbol{\theta}) = U_L(\boldsymbol{\theta}_L) \cdots U_2(\boldsymbol{\theta}_2) U_1(\boldsymbol{\theta}_1).
\end{equation}
Figure~\ref{fig:ansatz_structure_execution} illustrates two popular ansatz structures: linear (left) and brickwall (right). Here, each unitary operator takes the form
\begin{equation}
U_l(\boldsymbol{\theta}_l) = \prod_m e^{-i \theta_m H_m} W_m,
\end{equation}
where $W_m$ is a parameter-free unitary operator, and $H_m$ is a Hermitian operator~\cite{cerezo2021variational}. The types and structures of ansatzes may depend on the problem’s requirements, with commonly used ansatzes including:

\begin{itemize}
    \item[-] \textbf{QAOA Ansatz} --- Used in the Quantum Approximate Optimization Algorithm (QAOA) for solving combinatorial optimization problems~\cite{farhi2014quantumapproximateoptimizationalgorithm}. The QAOA ansatz has an alternating structure, often referred to as the Quantum Alternating Operator Ansatz~\cite{Hadfield_2019}. Its goal is to map an initial state \( |\psi_0\rangle \) to the ground state of a given problem Hamiltonian \( H_P \) by sequentially applying the operator \( e^{-i\gamma_l H_P} \) (called the \textit{problem} operator) and the operator \( e^{-i\beta_l H_M} \) (the \textit{mixer} operator). The ansatz takes the form \( U(\gamma, \beta) = \prod_{l=1}^{p} e^{-i\beta_l H_M} e^{-i\gamma_l H_P} \), where \( \theta = (\gamma, \beta) \), and \( H_M \) is a Hermitian operator known as the mixing Hamiltonian. A key advantage of this ansatz is that the solution space is significantly smaller than the full Hilbert space, enhancing the algorithm’s efficiency~\cite{cerezo2021variational}.

    \item[-] \textbf{Hardware-Efficient Ansatz} --- A class of ansatzes designed to minimize the depth of the quantum circuit when implementing \( U(\theta) \) on a specific quantum device~\cite{Kandala_2017}. These ansatzes are constructed using a set of unitary operators \( W_m \) and \( e^{-i\theta_m H_m} \), selected from the set of operations supported by the device, with multi-qubit operators placed between qubits that are physically connected. The primary advantage of this approach is that it avoids increasing the circuit depth, particularly by eliminating the need for additional qubits (\textit{ancilla}) or extra operations that arise when using ansatzes with structures incompatible with the device’s topology or supported operations.

    \item[-] \textbf{Variable Structure Ansatz} --- While most ansatzes employ a fixed quantum circuit structure and optimize a set of parameters, variable structure ansatzes allow adaptive modification of the circuit structure during optimization. One of the earliest approaches to such adaptability is the ADAPT-VQE algorithm, which adds or removes operations to minimize circuit complexity while maintaining efficiency in quantum chemistry tasks~\cite{cerezo2021variational, Grimsley_2019}.
\end{itemize}

Despite the variety of structures and application goals, an ansatz must possess certain properties to effectively solve a problem. Two critical criteria are \textit{expressibility} and \textit{entangling capability}, as proposed in~\cite{Sim_2019}.

\smallskip

\label{exrp_and_ent}
\textbf{Expressibility} refers to the ability of a quantum circuit to generate (pure) states that effectively span the relevant Hilbert space. For a single qubit, this corresponds to the circuit’s ability to efficiently explore the entire Bloch sphere~\cite{Sim_2019}. To quantitatively assess expressibility, one can compute the deviation from an ideal uniform distribution (Haar distribution), evaluating how the selectivity of the parameterized quantum circuit affects incomplete coverage of the space of possible states.

For a specific class of state ensembles generated by uniformly sampling parameters, we can define the non-uniformity operator \( A \) as
\begin{equation}
    A = \int_{\text{Haar}} (|\psi\rangle\langle\psi|)^{\otimes t} d\psi 
    - \int_{\Theta} (|\psi_{\theta}\rangle\langle\psi_{\theta}|)^{\otimes t} d\theta,
\end{equation}
where the first integral is taken over all pure states according to the Haar distribution, and the second integral is taken over all states \( |\psi_{\theta}\rangle \) obtained from an ensemble formed by uniformly sampling the parameters \( \theta \) of the quantum circuit. A corresponding measure of expressibility can be defined as the squared Hilbert-Schmidt norm of the operator \( A \), \( \| A \|_{\text{HS}}^2 = \text{Tr} \left( A^\dagger A \right) \). Thus, the smaller the norm of the non-uniformity operator \( A \), the more expressive the quantum circuit, meaning it is better able to approximate random states from the Haar distribution.

To evaluate the \textbf{entangling capability}, the Meyer-Wallach (MW) measure \( Q \) is used, which quantifies the multi-partite entanglement of pure quantum states and is easily computed on quantum computers~\cite{Sim_2019}. The Meyer-Wallach measure quantifies the average entanglement of individual qubits with respect to the rest of the system and is often expressed as~\cite{meyer2002global}
\begin{equation}
    Q(|\psi\rangle) = 2 - \frac{2}{n} \sum_{j=1}^{n} \text{Tr}(\rho_j^2),
\end{equation}
where \( n \) is the number of qubits in the system, and \( \rho_j \) is the single-qubit density matrix obtained after computing the partial trace over the complementary subsystem. A value of \( Q = 0 \) is achieved only for separable states, while highly entangled states have values close to \( Q = 1 \).

The entangling capability can then be assessed by averaging over an ensemble of states generated by the quantum circuit:
\begin{equation}
    \text{Ent} = \frac{1}{|S|} \sum_{\theta_i \in S} Q(|\psi_{\theta_i} \rangle),
\end{equation}
where \( S = \{\theta_i\} \) is a set of randomly sampled parameter vectors.

Analyzing these criteria for a chosen ansatz enables an assessment of its potential prior to the training process. This facilitates the selection of an optimal structure that best aligns with the problem’s specifics and the hardware constraints of the quantum device on which it will be implemented~\cite{10821244}.

\subsection{Challenges in the Training Process}

In classical machine learning, the training of models often encounters challenges such as overfitting, the presence of local minima in the loss function, and the vanishing gradient problem in deep neural networks. These limitations have spurred the development of new optimization techniques, such as adaptive gradient methods (e.g., Adam) and regularization mechanisms.

Unlike classical quantum algorithms (e.g., Shor’s or Grover’s algorithms)~\cite{Nielsen2000}, variational quantum algorithms (VQAs) are heuristic in nature, making their convergence less predictable. As the number of qubits increases, variational algorithms become increasingly difficult to optimize, motivating active research in this area~\cite{McClean_2018, larocca2024reviewbarrenplateausvariational, Ragone_2024}. One of the primary reasons for the inability to effectively train variational quantum algorithms is the previously mentioned phenomenon of \textit{barren plateaus}.

To investigate the causes of \textit{barren plateaus}, it is necessary to evaluate how the value of the loss function \( \ell_{\boldsymbol{\theta}}(\rho, O) \) changes with variations in the parameters \( \boldsymbol{\theta} \). In particular, it is useful to characterize the variance of the loss function in the parameter space:
\begin{equation}
    \text{Var}_{\boldsymbol{\theta}}[\ell_{\boldsymbol{\theta}}(\rho, O)] = \mathbb{E}_{\boldsymbol{\theta}}[\ell_{\boldsymbol{\theta}}(\rho, O)^2] - (\mathbb{E}_{\boldsymbol{\theta}}[\ell_{\boldsymbol{\theta}}(\rho, O)])^2.
\end{equation}
The loss function exhibits a \textit{barren plateau} if its variance decays exponentially with increasing system size~\cite{Ragone_2024}:
\begin{equation}
    \text{Var}_{\boldsymbol{\theta}}[\ell_{\boldsymbol{\theta}}(\rho, O)] \in O(1/b^n), \quad b > 1.
\end{equation}

In the literature, a common approach to detecting \textit{barren plateaus} involves analyzing the concentration of partial derivatives of the loss function, i.e., evaluating \( \text{Var}_{\boldsymbol{\theta}} \left[ \frac{\partial \ell_{\boldsymbol{\theta}}(\rho, O)}{\partial \theta_\mu} \right] \). However, since the concentration of the loss function itself implies the concentration of its derivatives~\cite{Arrasmith_2022}, it is sufficient to consider only the variance of the loss function.

The presence of \textit{barren plateaus} can be caused by various aspects of variational problems, including the expressibility of the parameterized quantum circuit, the locality of the measurement operator \( O \), and the entanglement of the initial state \( \rho \)~\cite{Cerezo_2021, McClean_2018, Holmes_2022}.

Despite the apparent independence of the sources of \textit{barren plateaus}, a unified expression for the variance of the loss function can be derived for sufficiently deep parameterized quantum circuits. Specifically, the work in~\cite{Ragone_2024} presents a general theory based on Lie algebras, which links the scaling of the loss variance to the dimension of the Lie algebra \( \mathfrak{g} \) of the generators of unitary operators, given by \( \text{Var}_{\boldsymbol{\theta}}[\ell_{\boldsymbol{\theta}}(\rho, O)] \sim \frac{1}{\dim(\mathfrak{g})} \).

To compute the variance of the loss function, we introduce the concept of the Dynamic Lie Algebra (DLA) for a parameterized circuit. The DLA is a closed subspace of the algebra \( \mathfrak{u}(2^n) \), generated by the circuit’s operators under the action of commutators:
\begin{equation}
    \mathfrak{g} = \langle iG \rangle_{\text{Lie}},
\end{equation}
where \( \mathfrak{g} \) is a subspace of \( \mathfrak{u}(2^n) \), closed under commutators, generated by nested commutators of the generators \( iG \).

Since \( \mathfrak{g} \) is a subalgebra of \( \mathfrak{u}(2^n) \) (the algebra of all skew-Hermitian operators on the Hilbert space \( H \)), it is a reductive Lie algebra and can thus be decomposed into a direct sum of commuting ideals:
\begin{equation}
    \mathfrak{g} = \mathfrak{g}_1 \oplus \dots \oplus \mathfrak{g}_{k-1} \oplus \mathfrak{g}_k;
    \label{eq:lie_decompose}
\end{equation}
here, \( \mathfrak{g}_1, \dots, \mathfrak{g}_{k-1} \) are simple Lie algebras, and \( \mathfrak{g}_k \) is an abelian algebra (the center \( Z(\mathfrak{g}) \) of the algebra), whose commutator with any other element is zero. This implies that the dynamic Lie algebra consists of two parts: a semisimple part \( [\mathfrak{g}, \mathfrak{g}] = \mathfrak{g}_1 \oplus \dots \oplus \mathfrak{g}_{k-1} \), which governs the non-trivial dynamics of the quantum circuit, and a central part \( \mathfrak{g}_k \), which contains mutually commuting operators.

We introduce the concept of \( \mathfrak{g} \)-purity for a Hermitian operator \( H \in i \mathfrak{u}(2^n) \):
\begin{equation}
    P_{\mathfrak{g}}(H) = \text{Tr} \left( H_{\mathfrak{g}}^2 \right) 
    = \sum_{j=1}^{\dim(\mathfrak{g})} \left| \text{Tr} \left( B_j^\dagger H \right) \right|^2,
\end{equation}
where \( H_{\mathfrak{g}} \) denotes the orthogonal projection of the operator \( H \) onto the complex extension \( \mathfrak{g}_\mathbb{C} = \operatorname{span}_{\mathbb{C}}(\mathfrak{g}) \) of the algebra \( \mathfrak{g} \), and \( \{B_j\}_{j=1}^{\dim(\mathfrak{g})} \) is an orthonormal basis (over \( \mathbb{C} \)) of the algebra \( \mathfrak{g}_\mathbb{C} \) with respect to the Hilbert-Schmidt inner product \( \langle A, B \rangle = \operatorname{Tr}(A^\dagger B) \).

Let \( O \in i\mathfrak{g} \) or \( \rho \in i\mathfrak{g} \), where our dynamic Lie algebra \( \mathfrak{g} \) has the structure defined in Equation~\eqref{eq:lie_decompose}. Then, the expected value of the loss function is zero for the semisimple components \( \mathfrak{g}_1 \oplus \dots \oplus \mathfrak{g}_{k-1} \), with a non-zero contribution only from the abelian component:
\begin{equation}
    \mathbb{E}_{\boldsymbol{\theta}}[\ell_{\boldsymbol{\theta}}(\rho, O)] = \text{Tr} \left[ \rho_{\mathfrak{g}_k} O_{\mathfrak{g}_k} \right].
\end{equation}
In contrast, the variance of the loss function vanishes for the central component \( \mathfrak{g}_k \), with contributions only from the simple components:
\begin{equation}
    \text{Var}_{\boldsymbol{\theta}}[\ell_{\boldsymbol{\theta}}(\rho, O)] = \sum_{j=1}^{k-1} \frac{P_{\mathfrak{g}_j}(\rho) P_{\mathfrak{g}_j}(O)}{\dim(\mathfrak{g}_j)},
\end{equation}
where \( P_{\mathfrak{g}_j}(\cdot) \) is the \( \mathfrak{g}_j \)-purity of the corresponding operator.

As seen, if \( \mathfrak{g} \) is abelian and if \( \rho \) or \( O \) commutes with \( \mathfrak{g} \) (a slightly more general condition than belonging to \( \mathfrak{g} \)), the loss function landscape is completely flat. This result elucidates the causes of \textit{barren plateaus} and the contributions of the components \( \mathfrak{g}_j \) from the decomposition~\eqref{eq:lie_decompose}.

Since the variance of the loss function is inversely proportional to the dimension \( \dim(\mathfrak{g}) \), the latter directly determines the expressibility of the quantum circuit, as a larger \( \dim(\mathfrak{g}) \) leads to greater concentration of the loss function. Specifically, as shown in~\cite{Ragone_2024}, if the dimension of the DLA grows as \( \Omega(b^n) \) for some \( b > 2 \), the loss function achieves exponential concentration regardless of the initial state or measurement operator. This implies that highly expressive quantum circuits with exponentially large DLAs are untrainable.

Consider the \( \mathfrak{g} \)-purity of the density operator of the initial state \( \rho \). A smaller \( \mathfrak{g} \)-purity corresponds to greater generalized entanglement, which, in turn, leads to a smaller variance of the loss function. Thus, if \( P_{\mathfrak{g}}(\rho) = O(1/b^n) \), meaning the state has high generalized entanglement, the loss function concentrates exponentially, regardless of the quantum circuit’s expressibility.

Similarly, consider the \( \mathfrak{g} \)-purity of the measurement operator \( O \). We call an operator \( O \) \textit{generalized-local} if it belongs to the subspace of observable operators defined by the algebra \( \mathfrak{g} \). In this case, the condition \( O_{\mathfrak{g}} = O \) holds. If the measurement operator is \textit{generalized-nonlocal}, i.e., \( O_{\mathfrak{g}} = O(1/b^n) \), the loss function exhibits a \textit{barren plateau} effect, independent of the dimension of the dynamic Lie algebra.

The presence of these criteria imposes certain constraints on the design of the training process for variational quantum algorithms. In particular, the use of deep and highly expressive ansatzes, which may initially seem appealing for spanning the space of possible solutions, can conversely lead to significant optimization challenges. Similarly, employing highly entangled initial states, which are useful for encoding information, or global measurement operators, which provide access to the entire quantum system, may degrade the algorithm’s trainability.

Thus, constructing an effective training process requires a balanced choice of ansatz, initial state, and measurement operator to avoid both insufficient expressibility and excessive concentration of the loss function, which hinders optimization.

\subsection{Optimization Process}

After defining the loss function and selecting the ansatz, the next step is to find the solution to the problem through the optimization of the parameters \( \boldsymbol{\theta} \). For many variational quantum algorithms, it is possible to analytically compute the gradients of the loss function, leading to the parameter-shift rule~\cite{Mitarai_2018}.

For simplicity, consider a loss function of the form given in Equation~\eqref{eq:loss_function}. Let the parameter \( \theta_l \) be the \( l \)-th element of the parameter vector \( \boldsymbol{\theta} \), defining a unitary operator of the form
\begin{equation}
    U_l(\theta_l) = e^{-i\theta_l G_l},
\end{equation}
where \( G_l \) is a Hermitian operator, such as a Pauli matrix \( X, Y, Z \) or their combination.

Suppose the loss function is defined as the expectation value of some observable operator \( O \):
\begin{equation}
    C(\boldsymbol{\theta}) = \langle \psi | U^\dagger(\boldsymbol{\theta}) O U(\boldsymbol{\theta}) | \psi \rangle.
\end{equation}
Then, the partial derivative of the loss function with respect to the parameter \( \theta_l \) can be computed using the rule:
\begin{equation}
    \frac{\partial C(\boldsymbol{\theta})}{\partial \theta_l} = c \left[ C(\boldsymbol{\theta} + s \mathbf{e}_l) - C(\boldsymbol{\theta} - s \mathbf{e}_l) \right],
\end{equation}
where \( s \) is a fixed shift, and \( \mathbf{e}_l \) is the unit vector in the direction of the parameter \( \theta_l \).

The values of the coefficient \( c \) and the shift \( s \) depend on the form of the generator \( G_l \). For Pauli matrices (e.g., \( G_l = \sigma_i /2 \)), the shift is \( s = \frac{\pi}{4} \), the coefficient \( c \) is \( \frac{1}{2} \), and the gradient can be expressed as:
\begin{equation}
    \frac{\partial C(\boldsymbol{\theta})}{\partial \theta_l} = \frac{1}{2} \left[ C(\boldsymbol{\theta} + \frac{\pi}{4} \mathbf{e}_l) - C(\boldsymbol{\theta} - \frac{\pi}{4} \mathbf{e}_l) \right].
\end{equation}
Thus, computing the gradient requires two additional quantum measurements for the shifted values of the parameter \( \theta_l \).

The derivation of this rule is based on applying quantum perturbation to the parameterized unitary operator. Since
\begin{equation}
    \frac{d}{d\theta_l} e^{-i \theta_l G_l} = -i G_l e^{-i \theta_l G_l},
\end{equation}
we have:
\begin{equation}
    \frac{\partial C(\boldsymbol{\theta})}{\partial \theta_l} = \langle \psi | U^\dagger(\boldsymbol{\theta}) (-i G_l) O U(\boldsymbol{\theta}) | \psi \rangle.
\end{equation}
Assuming \( G_l \) has only two eigenvalues \( \pm \lambda \), using the spectral decomposition, we obtain:
\begin{equation}
    \frac{\partial C(\boldsymbol{\theta})}{\partial \theta_l} = \lambda \left[ C(\boldsymbol{\theta} + \frac{\pi}{4} \mathbf{e}_l) - C(\boldsymbol{\theta} - \frac{\pi}{4} \mathbf{e}_l) \right].
\end{equation}
This result generalizes to any generators with a discrete spectrum~\cite{Schuld_2019}. While the parameter-shift rule is an effective method for evaluating gradients, it has several significant limitations:
\begin{itemize}
    \item \textbf{Restrictions on generators} \( G_l \). The method works only for operators \( G_l \) with a spectrum containing a few discrete values (e.g., Pauli matrices).
    \item \textbf{Exponential number of measurements for large circuits}. Since each parameter requires at least two additional evaluations of the loss function, circuits with many parameters may demand a significant number of computations.
    \item \textbf{Noise and errors in quantum measurements}. In real NISQ devices, measurements are subject to noise, which can lead to significant deviations in gradient calculations.
\end{itemize}

Although the parameter-shift rule is a valid method for computing gradients in variational quantum algorithms, its practical application is often impractical due to high computational costs, sensitivity to noise, and scalability issues on modern quantum computers. Classical optimization methods offer faster convergence and are better suited for navigating the multidimensional and complex loss function landscapes that frequently arise in practice.

Optimization methods for variational quantum algorithms can be broadly divided into two main categories:
\begin{enumerate}
    \item \textbf{Gradient-based methods}, which utilize information about the derivatives of the loss function. The most common include:
    \begin{itemize}
        \item Stochastic Gradient Descent (SGD), well-suited for quantum problems where gradients can only be obtained as statistical estimates~\cite{ruder2017overviewgradientdescentoptimization};
        \item Adaptive optimization methods (e.g., Adam), which adjust the step size during optimization, enabling more efficient and accurate solutions compared to basic SGD~\cite{kingma2014adam}.
    \end{itemize}
    
    \item \textbf{Gradient-free methods}, which do not rely on computing derivatives and may be more robust to noise. These include:
    \begin{itemize}
        \item SPSA (Simultaneous Perturbation Stochastic Approximation)~\cite{spall1998overview};
        \item Bayesian optimization methods~\cite{snoek2012practical};
        \item Evolutionary algorithms (e.g., CMA-ES)~\cite{hansen2001completely}.
    \end{itemize}
\end{enumerate}

Clearly, gradient-based methods such as SGD or Adam directly depend on gradient information from the loss function, making them particularly vulnerable to the \textit{barren plateau} phenomenon. As gradients decay exponentially, optimization becomes infeasible due to the lack of clear direction for parameter updates, halting the optimization process. There is a hypothesis that gradient-free methods may be less susceptible to \textit{barren plateaus} since they do not rely on gradient information, instead using alternative strategies for parameter updates, such as stochastic search or evolutionary algorithms. However, research indicates that gradient-free optimizers do not fully resolve the \textit{barren plateau} problem~\cite{Arrasmith_2021}. The primary reason is that even gradient-free algorithms rely on differences in loss function values, which also become exponentially small. This means that decision-making for parameter updates, whether using gradient-based or gradient-free methods, still faces the challenge of insufficient signal for effective optimization.

Nevertheless, in practice, gradient-free methods are often more effective in variational quantum algorithms, as they perform better under high noise conditions in NISQ devices and can partially mitigate the \textit{barren plateau} effect through flexible, adaptive approaches to parameter selection.

We now examine two commonly used gradient-free optimization methods employed in this work in greater detail.

\subsubsection{SPSA}

The Simultaneous Perturbation Stochastic Approximation (SPSA) algorithm~\cite{spall1992multivariate, spall1998overview} is an effective gradient-free optimization method specifically designed for high-dimensional problems where computing gradients is costly or infeasible. SPSA is used in quantum machine learning, neural network optimization, control systems, and statistical parameter estimation. Its key feature is the ability to estimate an approximate gradient using only two evaluations of the loss function, regardless of the number of parameters. This significantly distinguishes it from traditional methods, such as the finite difference method, where the number of gradient evaluations grows linearly with the number of parameters.

Consider the problem of minimizing a loss function \( L(\boldsymbol{\theta}) \):
\begin{equation}
    \boldsymbol{\theta}^{*} = \arg\min_{\boldsymbol{\theta}} L(\boldsymbol{\theta}),
\end{equation}
where \( \boldsymbol{\theta} \) is the parameter vector. The classical finite difference method estimates the gradient as follows:
\begin{equation}
    \frac{\partial L}{\partial \theta_i} \approx \frac{L(\boldsymbol{\theta} + c \mathbf{e}_i) - L(\boldsymbol{\theta} - c \mathbf{e}_i)}{2c},
\end{equation}
where \( \mathbf{e}_i \) is a unit vector that perturbs only one parameter. This approach requires \( 2d \) evaluations of the loss function per iteration, which is inefficient for large problems.

SPSA employs simultaneous perturbation of all parameters:
\begin{equation}
    \hat{g}_i(\boldsymbol{\theta}) = \frac{L(\boldsymbol{\theta} + c \boldsymbol{\Delta}) - L(\boldsymbol{\theta} - c \boldsymbol{\Delta})}{2c \Delta_i},
\end{equation}
where:
\begin{itemize}
    \item[-] \( c \) is a small positive constant,
    \item[-] \( \boldsymbol{\Delta} \) is a random vector whose elements \( \Delta_i \) are typically drawn from a symmetric Bernoulli distribution \( \pm 1 \),
    \item[-] \( \hat{g}_i(\boldsymbol{\theta}) \) is the gradient estimate for the parameter \( \theta_i \).
\end{itemize}
Parameter updates are performed using the standard stochastic rule:
\begin{equation}
    \boldsymbol{\theta}_{k+1} = \boldsymbol{\theta}_k - a_k \hat{\nabla} L(\boldsymbol{\theta}_k),
\end{equation}
where \( \hat{\nabla} L(\boldsymbol{\theta}_k) \) is the gradient estimate, and \( a_k \) is an adaptive step size.

Despite its advantages, including efficient handling of high-dimensional problems, robustness to noise, and ease of implementation, SPSA has certain limitations. Compared to gradient-based methods, SPSA exhibits slower convergence, with a convergence rate of \( O(k^{-1/3}) \), whereas gradient-based methods achieve \( O(k^{-1/2}) \). Additionally, the effectiveness of SPSA heavily depends on the appropriate choice of \( a_k \) and \( c_k \).

Despite these drawbacks, due to its robustness to noise, fixed number of function evaluations, and partial mitigation of the \textit{barren plateau} phenomenon, SPSA is a popular optimization method for variational optimization tasks.

\subsubsection{NFT}

The sequential minimal optimization algorithm, proposed in~\cite{Nakanishi_2020} and known as NFT after the authors (Nakanishi, Fujii, Todo), is a gradient-free optimization method specifically designed for quantum-classical hybrid algorithms. It improves upon traditional gradient-based and gradient-free methods by leveraging the periodic structure of the loss function in parameterized quantum circuits. This optimization method is particularly well-suited for NISQ devices, where limited coherence times, statistical noise, and \textit{barren plateaus} can reduce the effectiveness of standard gradient-based approaches.

The NFT method requires certain conditions to be effective in variational quantum algorithms:
\begin{enumerate}
    \item \textbf{Parameter independence}: The parameters \( \theta_j \) in the parameterized quantum circuit must be \textbf{independent} of one another.
    \item \textbf{Structure of the parameterized circuit}: The circuit consists of two types of operators:
    \begin{itemize}
        \item \textbf{Fixed unitary operators} (e.g., Hadamard, CNOT, controlled-$Z$).
        \item \textbf{Rotation operators} of the form:
    \end{itemize}
    \begin{equation}
        R_j(\theta_j) = \exp \Bigl( - i\frac{\theta_j}{2} A_j \Bigr),
    \end{equation}
    where \( A_j^2 = I \) (e.g., \( R_x, R_y, R_z \)).
    \item \textbf{Form of the loss function}: The loss function must be expressed as a weighted sum of expectation values:
    \begin{equation}
        C(\boldsymbol{\theta}) = \sum_{k=1}^{K} w_k \langle \psi_k | U^\dagger (\boldsymbol{\theta}) H_k U (\boldsymbol{\theta}) | \psi_k \rangle,
    \end{equation}
    where \( H_k \) are Hermitian operators, \( \{ |\psi_k\rangle \} \) are input states, and \( w_k \) are scalar weight coefficients.
\end{enumerate}

Most quantum-classical hybrid algorithms, such as the Variational Quantum Eigensolver (VQE) and the Quantum Approximate Optimization Algorithm (QAOA), satisfy these requirements, making them suitable for NFT optimization.

NFT iteratively updates parameters by selecting one or a few at a time and finding their optimal values based on the structure of the loss function. For a parameter \( \theta_j \), the loss function takes a trigonometric form:
\begin{equation}
    L_j^{(n)}(\theta_j) = a_1^{(n)} \cos(\theta_j - a_2^{(n)}) + a_3^{(n)},
\end{equation}
where \( a_1^{(n)}, a_2^{(n)}, a_3^{(n)} \) are constants determined by evaluating the loss function at three different values of the parameter \( \theta_j \). Since \( L_j(\theta_j) \) has a sinusoidal structure, the value of \( \theta_j \) that minimizes \( L_j(\theta_j) \) can be determined analytically:
\begin{equation}
    \theta_j^{(n+1)} = \arg\min_{\theta_j} L_j^{(n)}(\theta_j),
\end{equation}
enabling precise parameter updates with a minimal number of computations.

Since trigonometric loss functions arise naturally in parameterized quantum circuits, NFT effectively exploits this structure and, unlike stochastic methods (e.g., SPSA, Adam), can precisely find the optimal value without additional hyperparameters, demonstrating significantly faster convergence than traditional gradient-based or gradient-free approaches.

However, NFT is not a universal solution—its effectiveness depends on the structure of the quantum circuit, the form of the loss function, and the impact of noise. Despite these limitations, the NFT algorithm holds significant potential for accelerating the optimization of quantum machine learning models and variational algorithms, making it a valuable tool for quantum computing.
\newpage

\section{Trotterization}
\subsection{Core Concept and Approach}

One of the most promising applications of quantum computers is the simulation of physical systems, particularly those governed by the laws of quantum mechanics. Many fundamental problems in physics, chemistry, and materials science require understanding the behavior of quantum systems, a task that poses significant challenges for classical computers.

Physical systems with many interacting particles exhibit exponential growth in computational complexity as the number of particles \( N \) increases. This is because the complete description of the quantum state of such a system requires storing the wave function in a Hilbert space of dimension \( 2^N \). Consequently, memory and computational resources grow exponentially with \( N \), limiting the applicability of classical simulation methods for large systems.

For example:
\begin{itemize}
    \item In quantum chemistry, predicting molecular structures and reaction mechanisms requires solving the Schrödinger equation for many-particle systems, a computationally intensive task for large molecules.
    \item In condensed matter physics, understanding high-temperature superconductivity or exotic phases of matter necessitates modeling strongly correlated electronic systems, which is extremely challenging for classical methods.
    \item In high-energy physics, simulating lattice gauge theories (e.g., quantum chromodynamics) demands enormous computational resources and often relies on approximations.
\end{itemize}

Unlike classical methods, which face exponential complexity, quantum devices can directly encode and simulate the evolution of the wave function using unitary operations. The dynamics of a closed quantum system are described by the Schrödinger equation:
\begin{equation}
    i \hbar \frac{d}{dt} |\psi(t)\rangle = H |\psi(t)\rangle,
\end{equation}
where \( H \) is the system’s Hamiltonian, which determines its energy structure and evolution.

The formal solution to this equation is the unitary time-evolution operator \( U(t) = e^{-i H t} \), which transforms the initial state according to \( |\psi(t)\rangle = U(t) |\psi(0)\rangle \).

Accurately implementing the operator \( U(t) \) on a quantum computer is crucial for simulating quantum systems. Typically, the Hamiltonian \( H \) is expressed as a sum of several terms: \( H = \sum_j H_j \). Due to hardware limitations, a quantum computer has access only to operations corresponding to individual terms \( e^{-i H_j t} \), but not to the full evolution operator \( e^{-i H t} \) directly. The main challenge arises from the non-commutativity of the Hamiltonian components: \( [H_i, H_j] \neq 0, \ i \neq j \).

Thus, the exact unitary evolution operator \( U(t) = e^{-i H t} \) cannot be simply represented as a product of exponentials of the individual Hamiltonian components. This necessitates the use of \textbf{Trotterization}—a method that approximates the full evolution by sequentially applying the exponentials of the individual Hamiltonian components in a controlled manner.

\subsection{Mathematical Formulation}

We assume that the system’s Hamiltonian can be written as \( H = \sum_{j=1}^{m} H_j \). Since the components \( H_j \) do not commute with each other, the evolution operator \( e^{-i H t} \) does not have a simple decomposition into a product of exponentials, and the Trotter-Suzuki decomposition is used to approximate it~\cite{suzuki1992}.

\subsubsection{First-Order Trotterization}
The simplest approximation, known as the first-order Trotter formula, is based on the Lie-Trotter formula:
\begin{equation}
    e^{-i H t} = \lim_{n \to \infty} \left( \prod_{j=1}^{m} e^{-i H_j t/n} \right)^n.
\end{equation}
For a finite number of steps \( n \), the following approximation is used:
\begin{equation}
    U_{\text{Trotter},1}(t, n) = \left( \prod_{j=1}^{m} e^{-i H_j t/n} \right)^n.
\end{equation}
This approximation introduces an error due to the non-commutativity of the terms, which can be evaluated using the Baker-Campbell-Hausdorff (BCH) formula:
\begin{equation}
    e^{X} e^{Y} = e^{X + Y + \frac{1}{2} [X,Y] + \mathcal{O}(\|X\|\|Y\|^2)}.
\end{equation}
Accordingly, the error of first-order Trotterization is expressed as:
\begin{equation}
    U_{\text{Trotter},1}(t, n) = e^{-i H t} + \mathcal{O} \left(\frac{t^2}{n} \sum_{i < j} \| [H_i, H_j] \| \right),
\end{equation}
scaling as \( \mathcal{O}(t^2/n) \). Increasing \( n \) reduces the error but requires a greater number of quantum operations~\cite{childs2018}.

\subsubsection{Higher-Order Trotterization}
To reduce the error, a symmetrized decomposition can be used. The second-order (Suzuki) formula is given by:
\begin{equation}
    U_{\text{Trotter},2}(t, n) = \left( \prod_{j=1}^{m} e^{-i H_j t/(2n)} \prod_{j=m}^{1} e^{-i H_j t/(2n)} \right)^n.
\end{equation}
The error of this approximation can be estimated as:
\begin{equation}
    \mathcal{O} \left(\frac{t^3}{n^2} \sum_{i < j} \| [H_i, [H_i, H_j]] \| \right).
\end{equation}

More generally, the recursive Suzuki formula allows the construction of approximations of order \( 2k \):
\begin{equation}
    S_{2k}(t) = \left[ S_{2k-2}(p_k t) \right]^2 S_{2k-2}((1-4p_k) t) \left[ S_{2k-2}(p_k t) \right]^2,
\end{equation}
where \( p_k = (4 - 4^{1/(2k-1)})^{-1} \). For order \( 2k \), the error scales as:
\begin{equation}
    \mathcal{O} \left(\frac{t^{2k+1}}{n^{2k}} \right).
\end{equation}
Thus, increasing \( k \) significantly reduces the error but requires a substantially larger number of operations.

A key challenge in simulating quantum systems with low error is the need for a large number of steps \( n \), which significantly increases the depth of the quantum circuit. Since each Trotter step involves applying a sequence of unitary operators, the number of elementary operations (e.g., two-qubit CNOT gates) grows with increasing approximation accuracy. This poses serious limitations for implementation on modern quantum devices. Another significant issue is the accuracy of long-term quantum system simulations. Since the Trotterization error scales as \( \mathcal{O}(t^{2k+1}/n^{2k}) \), increasing the simulation time \( t \) leads to a growing error. To keep the error at an acceptable level, \( n \) must be increased, which further increases the circuit depth. In long-term simulations, the accumulation of these errors can result in outcomes that lose physical meaning.

In addition to the classical Trotterization approach, other methods exist that can reduce the quantum circuit depth and more efficiently implement the evolution of quantum systems.

\subsection{Adaptive Trotterization Methods}
In the work~\cite{K_kc__2022}, an algorithm for simulating a Hamiltonian is presented, utilizing the Cartan decomposition to generate quantum circuits with constant depth, independent of the simulation time \( t \). This contrasts with traditional Trotterization approaches, where the required circuit depth increases with \( t \) to maintain accuracy.

Consider a general time-independent Hamiltonian:
\begin{equation}
    H = \sum_{j} H_j \sigma_j,
\end{equation}
where \( H_j \) are real coefficients, and \( \sigma_j \) are Pauli operators, elements of the \( n \)-qubit Pauli group: \( P_n = \{ I, X, Y, Z \}^{\otimes n} \). The corresponding evolution operator can be expressed as:
\begin{equation}
    U(t) = e^{-i H t} = \prod_{\sigma_i \in \mathfrak{su}(2^n)} e^{i \kappa_i \sigma_i},
    \label{lie_unit}
\end{equation}
where, in general, \( \mathcal{O}(4^n) \) parameters \( \kappa_i \) are required for the Pauli strings \( \sigma_i \), which form a basis for the Lie algebra \( \mathfrak{su}(2^n) \).

However, it is often sufficient to restrict the decomposition of the evolution operator in Equation~\eqref{lie_unit} to a subset of the algebra \( \mathfrak{su}(2^n) \). According to the Baker-Campbell-Hausdorff (BCH) theorem, only nested commutators of the individual terms in \( H \) contribute to the exact expression of the evolution. Thus, it is sufficient to consider only the elements of the subset \textbf{\( \mathfrak{g}(H) \)}, which is a closed set under the commutation operation for the Pauli strings in Equation~\eqref{lie_unit}. We can now limit the decomposition in Equation~\eqref{lie_unit} to elements of the set \textbf{\( \mathfrak{g}(H) \)}, which we call the \textit{Hamiltonian algebra}.

To determine the parameters \( \kappa_i \), we consider the decomposition in Equation~\eqref{lie_unit} consistent with the Hamiltonian algebra \( \mathfrak{g}(\mathcal{H}) \), and for this, we compute its Cartan decomposition (see Appendix~\ref{sec:appA}). Recall that the Cartan decomposition of a Lie algebra \( \mathfrak{g} \) is its orthogonal decomposition with respect to the Killing form:
\begin{equation}\label{eq:Cartan}
    \mathfrak{g} = \mathfrak{k} \oplus \mathfrak{m},
\end{equation}
into a compact subalgebra \( \mathfrak{k} \) and a subspace \( \mathfrak{m} \), satisfying the following conditions:
\begin{equation}
    [\mathfrak{k}, \mathfrak{k}] \subset \mathfrak{k}, \quad 
    [\mathfrak{m}, \mathfrak{m}] \subset \mathfrak{k}, \quad
    [\mathfrak{k}, \mathfrak{m}] = \mathfrak{m}.
\end{equation}
The \textit{Cartan subalgebra} \( \mathfrak{h} \) is the maximal abelian subalgebra of the space \( \mathfrak{m} \). If \( \mathfrak{g} \) has a Cartan decomposition as in Equation~\eqref{eq:Cartan}, then, according to the KHK decomposition (detailed in Appendix~\ref{appendix:A3}), for any \( m \in \mathfrak{m} \), there exist \( K \in e^{i\mathfrak{k}} \) and \( h \in \mathfrak{h} \) such that \( m = K h K^\dagger \).

By directly applying the Cartan decomposition and the KHK decomposition to our Hamiltonian algebra \( \mathfrak{g}(\mathcal{H}) \), we obtain the form of the unitary evolution operator:
\begin{equation}
    U(t) = e^{-i\mathcal{H} t} = K e^{-i h t} K^\dagger.
\end{equation}
Since \( \mathfrak{h} \) is an abelian algebra, each Pauli string in \( h \in \mathfrak{h} \) commutes, allowing straightforward construction of a quantum circuit for \( e^{-i h t} \).

The final task is to find \( K \), which requires solving the equation:
\begin{equation}
    K = \prod_{i} e^{i a_i k_i},
\end{equation}
where \( k_i \) are elements of the subalgebra \( \mathfrak{k} \), and \( a_i \) are parameters to be optimized. The optimization function is defined as:
\begin{equation}
    f(\theta) = \langle K(\theta) v K^\dagger(\theta), H \rangle,
\end{equation}
where \( v \in \mathfrak{h} \) is an arbitrary element whose exponential generates \( e^{ih} \).

The optimization is local, meaning the number of parameters is significantly smaller than the size of the full Lie algebra. The proposed method offers substantial advantages over classical Trotterization, including:
\begin{itemize}
    \item \textbf{Fixed quantum circuit depth}: This method factorizes the evolution operator such that the circuit structure remains fixed, with dependence on time \( t \) only as a parameter.
    \item \textbf{Elimination of error accumulation}: Classical Trotterization has an error of order \( \mathcal{O}(t^2/n) \), whereas the Cartan method provides an exact assessment of the algebraic structure of the evolution operator, avoiding errors that accumulate over time.
    \item \textbf{Natural representation of Hamiltonians}: Certain Hamiltonians (e.g., free fermionic models and specific spin systems) have a natural Lie algebraic structure, making the Cartan decomposition straightforward and enabling more efficient representation of the system’s evolution.
\end{itemize}

Despite its advantages, the method has certain limitations:
\begin{itemize}
    \item \textbf{Growth of the Hamiltonian algebra}: The method’s efficiency depends on the size of the algebra \( \mathfrak{g}(H) \). If \( |\mathfrak{g}(H)| \) is an exponentially large function of the system size \( n \), constructing the quantum circuit requires significant resources. For instance, in the Heisenberg model, \( |\mathfrak{g}(H)| \sim 4^n \), rendering this approach impractical for large systems.
    \item \textbf{Challenges in finding the Cartan decomposition}: This method requires computing involutions and identifying Cartan subalgebras, which is a complex task for systems lacking a natural Cartan decomposition.
    \item \textbf{Absence of an abelian subalgebra}: If the Cartan decomposition lacks an abelian component \( \mathfrak{h} \), the evolution operator cannot be decomposed as \( U(t) = K e^{-i h t} K^\dagger \). In such cases, the time-dependent component is distributed across multiple exponentials of non-commuting operators, approximating the method to classical Trotterization, where the complexity of the evolution scales with \( t \).
\end{itemize}

Another approach to reducing the quantum circuit depth in simulating the evolution of quantum systems is adaptive Trotterization (ADA-Trotter), proposed in~\cite{zhao2023makingtrotterizationadaptiveenergyselfcorrecting}.

Adaptive Trotterization allows dynamic adjustment of the discretization step \( \Delta t \) based on the rate of change of the wave function. This enables the use of small \( \Delta t \) when the system evolves rapidly and larger \( \Delta t \) when the system changes slowly, balancing accuracy and quantum circuit depth.

A key component of adaptive Trotterization is the measurement of energy errors to control the accuracy of the evolution. In the first stage, an initial Trotter step \( \Delta t_0 \) is set, along with acceptable error bounds for the energy \( d_E \) and its variance \( d_{E^2} \). After performing one step of time evolution using standard Trotterization, \( |\psi_{m+1}\rangle = U(\Delta t_m) |\psi_m\rangle \), the mean energy \( E_{m+1} = \langle \psi_{m+1} | H | \psi_{m+1} \rangle \) and energy variance \( \delta E_{m+1}^2 = \langle H^2 \rangle - \langle H \rangle^2 \) are measured. These values are compared to the acceptable bounds. If the conditions \( |E_{m+1} - E| < d_E \) and \( |\delta E_{m+1}^2 - \delta E^2| < d_{E^2} \) are met, the current step \( \Delta t \) is considered accurate, and it can be increased to minimize computational costs. If the energy error exceeds the acceptable bounds, \( \Delta t \) must be reduced to improve accuracy. The Trotter step is adjusted using the bisection method or other optimization techniques.

The adaptive approach also allows the inclusion of additional constraints, such as preserving system symmetries or invariant quantities (e.g., particle number conservation).

One of the primary advantages of ADA-Trotter is its control over error accumulation in long-term time evolution. The error is managed through dynamic adjustment of the step size \( \Delta t \), preventing uncontrolled accumulation of deviations in measured physical quantities. The error estimation in this case is based on deviations from the diagonal representation of physical observables in the long-term regime.

To quantitatively assess the effectiveness of ADA-Trotter, consider an observable \( O \), whose mean value, according to the Eigenstate Thermalization Hypothesis (ETH), is given by:
\begin{equation}
    O_{\text{diag}} = \sum_k |c_k|^2 O_{kk},
\end{equation}
where \( |c_k|^2 \) is the probability of the system being in the eigenstate \( |k\rangle \) of the Hamiltonian, and \( O_{kk} = \langle k | O | k \rangle \) are the diagonal elements of the operator \( O \) in the eigenbasis of \( H \).

This value corresponds to the expectation value of the observable for long-term evolution (assuming system ergodicity). The error in long-term evolution can then be computed as the difference in expectation values between the true observable and that calculated using the adaptive algorithm. This error is:
\begin{equation}
    O_{\text{ADA}} - O_{\text{diag}} = dE \cdot O'(E) + \mathcal{O}(dE^2),
\end{equation}
where \( dE = E - E_{\text{diag}} \) is the deviation of the energy from its exact value, and \( O'(E) \) is the derivative of \( O(E) \) with respect to energy, describing the change in the observable for small energy variations. We see that long-term errors scale linearly with the energy error \( dE \), in contrast to classical Trotterization, where the error scales as \( \mathcal{E}_{\text{Trotter}} \sim \mathcal{O} (t^2 / n) \).

Similar to the approach in~\cite{K_kc__2022}, the adaptive Trotterization method optimizes quantum circuit depth through dynamic adjustment of the Trotter step \( \Delta t \). It also provides better control over long-term errors and offers high flexibility, allowing the incorporation of additional physical constraints during system evolution simulation.

However, the method has certain drawbacks, one of the main ones being the limited control over energy, as adaptation relies solely on the first two moments (mean and variance), which is insufficient for complex quantum systems where higher moments play a significant role. Additionally, there is the so-called freezing problem, where overly strict constraints on acceptable energy deviations force the algorithm to select very small steps \( \Delta t \), rendering the method inefficient.

\subsection{Trotterization of the Heisenberg Hamiltonian}

The Heisenberg Hamiltonian is a fundamental model in quantum mechanics that describes the interaction of spins in quantum magnets. Its general form for a system with \( n \) spins can be written as:
\begin{equation}
    H = \sum_{i}\left( J_x \sigma_i^x \sigma_{i+1}^x + J_y \sigma_i^y \sigma_{i+1}^y + J_z \sigma_i^z \sigma_{i+1}^z \right)
    + \sum_{i} \left(h_x \sigma_i^x + h_z \sigma_i^z \right),
\end{equation}
where:
\begin{itemize}
    \item[-] \( \sigma_i^x, \sigma_i^y, \sigma_i^z \) are Pauli operators acting on the \( i \)-th qubit,
    \item[-] \( J_x, J_y, J_z \) are the exchange interaction coefficients between neighboring spins along the \( x \), \( y \), and \( z \) axes,
    \item[-] \( h_x, h_z \) represent the external magnetic field along the \( x \) and \( z \) axes.
\end{itemize}

This Hamiltonian encompasses three main variants of the Heisenberg model: the \textbf{XXX model} (\( J_x = J_y = J_z \)), which corresponds to isotropic interactions; the \textbf{XXZ model} (\( J_x = J_y \neq J_z \)), which describes anisotropic systems; and the \textbf{XX model} (\( J_z = 0 \)), used to model superconducting qubit systems. In quantum simulations, this Hamiltonian is crucial for studying magnetism, quantum phase transitions, and non-equilibrium dynamics of many-particle quantum systems.

Applying the first-order Trotter decomposition \( e^{-i H t} \approx \left( \prod_{j} e^{-i H_j \delta t} \right)^{t / \delta t} \) to the Heisenberg Hamiltonian, we obtain:
\begin{equation}
    e^{-i H t} \approx \left( e^{-i H_{XX} \delta t} e^{-i H_{YY} \delta t} e^{-i H_{ZZ} \delta t} e^{-i H_{Z} \delta t} e^{-i H_{X} \delta t} \right)^{t / \delta t},
\end{equation}
where:
\begin{align}
    H_{XX} &= \sum_{i} J_x \sigma_i^x \sigma_{i+1}^x, \\
    H_{YY} &= \sum_{i} J_y \sigma_i^y \sigma_{i+1}^y, \\
    H_{ZZ} &= \sum_{i} J_z \sigma_i^z \sigma_{i+1}^z, \\
    H_X &= \sum_{i} h_x \sigma_i^x, \\
    H_Z &= \sum_{i} h_z \sigma_i^z.
\end{align}
Thus, the time evolution can be implemented as the sequential application of the unitary operators \( e^{-i H_{XX} \delta t} \), \( e^{-i H_{YY} \delta t} \), \( e^{-i H_{ZZ} \delta t} \), \( e^{-i H_{X} \delta t} \), and \( e^{-i H_{Z} \delta t} \) in a quantum circuit.

In the following discussion, we focus on the \textbf{isotropic Heisenberg model} without external fields due to its \textit{SU(2)} symmetry properties:
\begin{equation}
    H = J \sum_{i} (\sigma_i^x \sigma_{i+1}^x + \sigma_i^y \sigma_{i+1}^y + \sigma_i^z \sigma_{i+1}^z).
\end{equation}
We consider two approaches to implementing Trotterized time evolution on a quantum computer: the \textit{standard approach}, which uses separate two-qubit operators for \( XX \), \( YY \), and \( ZZ \), and the \textit{optimized approach}, which employs a single two-qubit operator \( XXYYZZ \) to reduce the number of operations.

For the isotropic Heisenberg model, the Trotter decomposition takes the form:
\begin{equation}
    e^{-i H t} \approx \left( e^{-i H_{XX} \delta t} e^{-i H_{YY} \delta t} e^{-i H_{ZZ} \delta t} \right)^{t / \delta t},
\end{equation}
where each of the operators \( e^{-i H_{XX} \delta t} \), \( e^{-i H_{YY} \delta t} \), and \( e^{-i H_{ZZ} \delta t} \) can be decomposed into two-qubit operators. For example, for the \( XX \)-interaction between two neighboring qubits, the evolution operator is:
\begin{equation}
    e^{-i J \sigma^x_i \sigma^x_{i+1} \delta t} = \cos(J \delta t) I - i \sin(J \delta t) \sigma^x_i \sigma^x_{i+1}.
\end{equation}

\begin{figure}[h]
    \centering
    \includegraphics[width=0.4\textwidth]{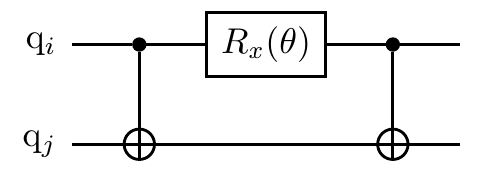}
    \caption{Decomposition of the $R_{xx}(\theta)$ operator using two CNOT gates and a single-qubit rotation}
    \label{fig:rxx_decomposition}
\end{figure}

This can be implemented using a standard quantum circuit that includes CNOT gates and single-qubit rotations (see Figure~\ref{fig:rxx_decomposition}).Similar circuits are constructed for the \( YY \) and \( ZZ \) interactions.

While this circuit is generally efficient, it involves a large number of CNOT gates, motivating the use of the \textit{optimized approach}, which implements all three interactions \( XX \), \( YY \), and \( ZZ \) with a single operator. Instead of applying separate operators \( e^{-i H_{XX} \delta t} \), \( e^{-i H_{YY} \delta t} \), and \( e^{-i H_{ZZ} \delta t} \), a single effective operator \( e^{-i H_{\text{XXYYZZ}} \delta t} \) can be used to realize the interaction across all three axes simultaneously. The optimized operator is constructed based on principles discussed in~\cite{Vatan_2004}, where it was shown that any two-qubit unitary operator can be implemented using at most 4 CNOT gates and 12 single-qubit gates. In this case, the operator \( e^{-i H_{\text{XXYYZZ}} \delta t} \) can be written as:
\begin{equation}
    e^{-i H_{\text{XXYYZZ}} \delta t} = e^{-i \tau (\sigma^x_i \sigma^x_{i+1} + \sigma^y_i \sigma^y_{i+1} + \sigma^z_i \sigma^z_{i+1})},
\end{equation}
where the parameter \( \tau \) determines the interaction duration.

\begin{figure}[h!]
    \centering
    \includegraphics[width=0.9\textwidth]{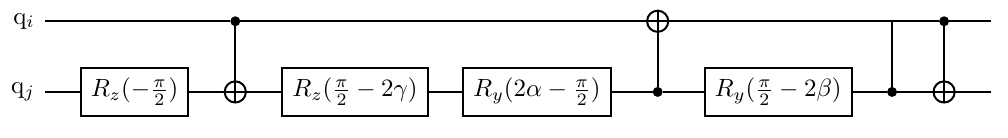}
    \caption{Optimized $XXYYZZ$ operator implemented using CNOT gates and single-qubit rotations}
    \label{fig:xx_yy_zz_decomposition}
\end{figure}

According to the results in~\cite{Vatan_2004}, this operator can be implemented using the quantum circuit shown in Figure~\ref{fig:xx_yy_zz_decomposition}.

To evaluate the effectiveness of the two Trotterization approaches, we compare how quickly the approximated state converges to the exact quantum state. Since the Heisenberg Hamiltonian is a well-studied model, for small systems (e.g., \( N = 6, 8, 10, 12 \) qubits), the exact state can be obtained through direct diagonalization of the evolution operator.

Let \( |\psi_{\text{exact}}\rangle \) be the exact evolved state, obtained by diagonalizing the Hamiltonian:
\[
H | \psi_k \rangle = E_k | \psi_k \rangle,
\]
and
\[
|\psi_{\text{exact}}(t) \rangle = \sum_k e^{-i E_k t} c_k |\psi_k \rangle,
\]
where \( c_k = \langle \psi_k | \psi(0) \rangle \). In the adaptive or standard Trotterization approach, the state is approximated as:
\[
|\psi_{\text{Trotter}}(t) \rangle = U_{\text{Trotter}}(t) |\psi(0)\rangle,
\]
where \( U_{\text{Trotter}}(t) \) is the approximated evolution operator. To assess the accuracy of the approximation, we use the fidelity metric, defined as:
\[
F = \left|\langle \psi_{\text{exact}}(t) | \psi_{\text{Trotter}}(t) \rangle\right|^2.
\]

\begin{figure}[h]
    \centering
    \includegraphics[width=1\textwidth]{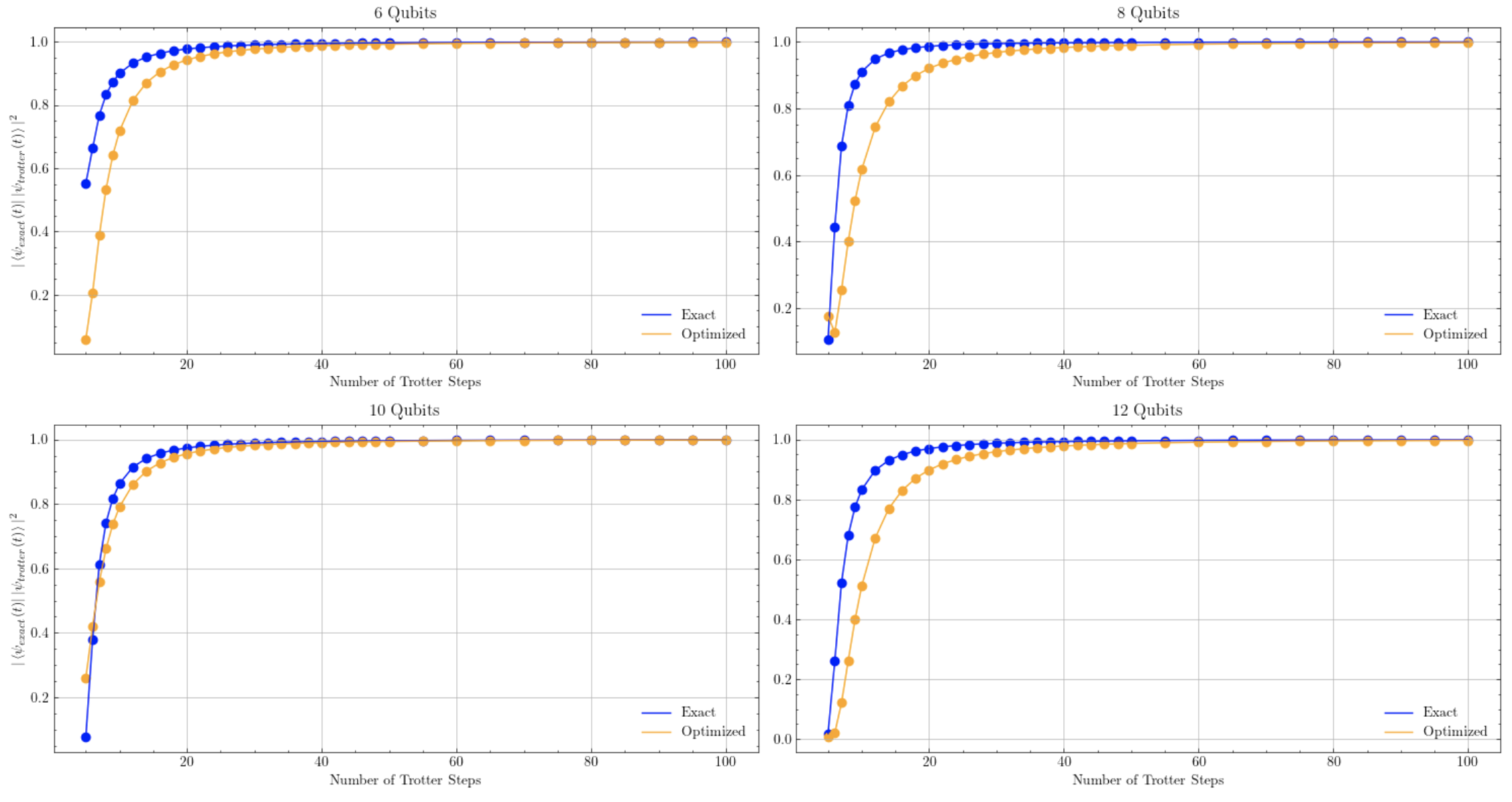}
    \caption[Dependence of state fidelity $F$ on the number of Trotter steps]{Dependence of state fidelity $F$ on the number of Trotter steps for \( N = 6, 8, 10, 12 \) qubits}
    \label{fig:trott_comp}
\end{figure}

\newpage
As shown in the plots in Figure~\ref{fig:trott_comp}, both methods exhibit monotonic growth in fidelity toward \( F \approx 1 \) as the number of Trotter steps increases. This trend holds for all system sizes, but larger \( N \) requires more steps to achieve high accuracy. For small \( t / \delta t \), the standard approach demonstrates higher fidelity compared to the optimized approach, but beyond approximately 50 Trotter steps, the convergence becomes equivalent for both methods.

The choice of method may depend on hardware capabilities. If minimizing quantum circuit depth is critical, the optimized approach is preferable. However, if high accuracy with fewer Trotter steps is required, the standard approach should be used.
\newpage

\section{\textit{SU(2)} Equivariance}
\label{sec:su2_theoretical}

Symmetry is a key element in quantum computing, particularly in the design of variational quantum circuits. Incorporating symmetries allows optimization of the parameter space by eliminating redundant degrees of freedom and focusing on physically relevant subspaces. This enhances optimization efficiency, reduces computational complexity, and improves the generalization capability of algorithms. In the context of quantum simulations, preserving symmetry ensures that results remain within the correct physical sector, which is critical for accurate modeling of quantum systems.

Among symmetry groups, the special unitary group \textit{SU(2)} holds a prominent place, naturally arising in quantum mechanics due to its connection with spin rotations. This symmetry is inherent in many Hamiltonians describing physical systems, such as the Heisenberg or Hubbard models, and plays a crucial role in stabilizing exotic quantum states, such as quantum spin liquids. The use of \textit{SU(2)}-equivariant quantum circuits enables the construction of ansatzes that automatically preserve spin quantum numbers, ensuring that simulations align with theoretical predictions and experimental data~\cite{Sachdev2011}.

\subsection{Mathematical Foundations of \textit{SU(2)} Symmetry}

The special unitary group \textit{SU(N)} is defined as the set of all \( N \times N \) unitary matrices with determinant 1:
\begin{equation}
    SU(N) = \{ U \in \mathbb{C}^{N \times N} \mid U U^{\dagger} = I, \det U = 1 \}.
\end{equation}
For \( N=2 \), elements of \textit{SU(2)} take the form:
\begin{equation}
    U = \begin{pmatrix} a & b \\ -b^* & a^* \end{pmatrix}, \quad \text{where} \quad a, b \in \mathbb{C}, \quad |a|^2 + |b|^2 = 1.
\end{equation}
This group has several fundamental properties:
\begin{itemize}
    \item Compactness, connectedness, and dimension 3 make it convenient for parameterization.
    \item It is a double cover of the rotation group \( SO(3) \), linking \textit{SU(2)} to three-dimensional rotations and spin systems.
    \item It can be parameterized using Pauli matrices, underscoring its central role in quantum mechanics and quantum computing.
\end{itemize}

The Lie algebra \( \mathfrak{su}(2) \) of the Lie group \textit{SU(2)} consists of \( 2 \times 2 \) skew-Hermitian matrices with trace 0, and its basis is formed by the generators \( T_x \), \( T_y \), and \( T_z \):
\begin{equation}
    \mathfrak{su}(2) = \text{span} \left\{ T_x = \frac{i}{2} \sigma_x, T_y = \frac{i}{2} \sigma_y, T_z = \frac{i}{2} \sigma_z \right\},
\end{equation}
where \( \sigma_x \), \( \sigma_y \), and \( \sigma_z \) are the Pauli matrices:
\begin{equation}
    \sigma_x = \begin{pmatrix} 0 & 1 \\ 1 & 0 \end{pmatrix}, \quad
    \sigma_y = \begin{pmatrix} 0 & -i \\ i & 0 \end{pmatrix}, \quad
    \sigma_z = \begin{pmatrix} 1 & 0 \\ 0 & -1 \end{pmatrix}.
\end{equation}
The generators satisfy the commutation relations:
\begin{equation}
    [T_x, T_y] = T_z, \quad [T_y, T_z] = T_x, \quad [T_z, T_x] = T_y,
\end{equation}
which reflect the structure of the spin algebra. Exponentiating these generators yields group elements:
\begin{equation}
    U(\theta, \hat{n}) = e^{-i \theta \hat{n} \cdot \mathbf{T}},
\end{equation}
where \( \theta \) is the rotation angle, \( \hat{n} \) is the unit vector of the rotation axis, and \( \mathbf{T} = (T_x, T_y, T_z) \). For example, a rotation around the \( z \)-axis by angle \( \theta \) takes the form:
\begin{equation}
    U_z(\theta) = e^{-i \theta T_z} = \begin{pmatrix} e^{-i \theta / 2} & 0 \\ 0 & e^{i \theta / 2} \end{pmatrix}.
\end{equation}
This formula is directly used in many quantum operators, such as \( R_z(\theta) \)~\cite{Nielsen2000}.

Additional details on Lie algebras, their structure, and their connection to symmetry groups are provided in Appendices~\ref{appendix:A1}, \ref{appendix:A2}, and~\ref{appendix:A3}.

\subsubsection{Irreducible Representations of the \textit{SU(2)} Group}

Irreducible representations of \textit{SU(2)} are classified by the quantum number \( j \), which can be integer or half-integer:
\begin{equation}
    j \in \{0, \tfrac{1}{2}, 1, \tfrac{3}{2}, 2, \dots \}.
\end{equation}
Each representation has dimension \( 2j+1 \), and the basis states are denoted \( |j, m\rangle \), where \( m = -j, -j+1, \dots, j \). For example:
\begin{itemize}
    \item For \( j = 1/2 \) (a particle with spin-\(\frac{1}{2}\)), the space is two-dimensional with basis:
    \begin{equation}
        | \uparrow \rangle = \begin{pmatrix} 1 \\ 0 \end{pmatrix}, \quad | \downarrow \rangle = \begin{pmatrix} 0 \\ 1 \end{pmatrix},
    \end{equation}
    where the generators \( \mathbf{S} = \frac{1}{2} \boldsymbol{\sigma} \) correspond to spin operators.
    \item For \( j = 1 \) (a particle with spin 1), the space is three-dimensional with basis \( |1, 1\rangle \), \( |1, 0\rangle \), \( |1, -1\rangle \), describing, for example, photon polarization or spin-1 chains.
\end{itemize}

In physical systems, these representations enable the classification of states by total spin~\cite{Auerbach1994}.

\subsubsection{Clebsch-Gordan Decomposition}

When combining multiple spin systems, the tensor product of their representations decomposes into irreducible components:
\begin{equation}
    V^{(j_1)} \otimes V^{(j_2)} = \bigoplus_{j=|j_1 - j_2|}^{j_1 + j_2} V^{(j)}.
\end{equation}
For example, for two spin-\(\frac{1}{2}\) particles (\( j_1 = j_2 = \frac{1}{2} \)), the four-dimensional space \( V^{(1/2)} \otimes V^{(1/2)} \) decomposes into \( V^{(1)} \oplus V^{(0)} \):
\begin{itemize}
    \item Triplet state (\( j = 1 \)): Symmetric states with total spin 1, such as:
    \begin{equation}
        |1, 0\rangle = \frac{1}{\sqrt{2}} (|\uparrow \downarrow\rangle + |\downarrow \uparrow\rangle),
    \end{equation}
    along with \( |1, 1\rangle = |\uparrow \uparrow\rangle \) and \( |1, -1\rangle = |\downarrow \downarrow\rangle \).
    \item Singlet state (\( j = 0 \)): Antisymmetric state with zero spin:
    \begin{equation}
        |0, 0\rangle = \frac{1}{\sqrt{2}} (|\uparrow \downarrow\rangle - |\downarrow \uparrow\rangle).
    \end{equation}
\end{itemize}

\textbf{Schur-Weyl duality} connects \textit{SU(2)} with the symmetric group \( S_n \). For the \( n \)-fold tensor product of the space \( V = \mathbb{C}^2 \), we have:
\begin{equation}
    V^{\otimes n} \cong \bigoplus_{\lambda} S^\lambda(\mathbb{C}^2) \otimes M_\lambda,
\end{equation}
where \( S^\lambda(\mathbb{C}^2) \) are irreducible representations of \textit{SU(2)}, and \( M_\lambda \) are representations of \( S_n \). For \( n=2 \), this yields the triplet (\( S=1 \)) and singlet (\( S=0 \)) states, consistent with the Clebsch-Gordan decomposition. In quantum circuits, this duality enables the construction of symmetric ansatzes, optimizing computations for systems with fixed spin~\cite{McArdle2020}. A more detailed overview is provided in Appendix~\ref{appendix:A4}.

\subsubsection{Spin Networks}

A \textit{spin network} is a graph in which each edge is labeled with an irreducible representation of \textit{SU(2)} (spin-\( j \)), and each vertex represents an \textit{SU(2)}-invariant interaction. Such structures are widely used in quantum gravity, tensor network methods in condensed matter physics, and variational quantum algorithms for strongly correlated electronic systems. In quantum computing, spin networks enable the systematic design of \textit{SU(2)}-equivariant quantum circuits, ensuring compliance with \textit{SU(2)} symmetry constraints~\cite{east2023needspinsu2equivariant}.

Let \( V \) and \( W \) be vector spaces on which representations of \textit{SU(2)} are realized. An \textbf{\textit{SU(2)}-equivariant mapping} is a linear transformation \( T: V \to W \) that commutes with the action of \textit{SU(2)} on \( V \) and \( W \):
\begin{equation}
    T (U \cdot v) = U \cdot T(v), \quad \forall U \in SU(2), \ \forall v \in V.
\end{equation}
This means the transformation preserves the symmetry of the input state space. In quantum circuits, this is achieved by restricting operations to those permitted by the \textit{SU(2)} Clebsch-Gordan decomposition.

Consider constructing an \textit{SU(2)}-equivariant mapping for a two-qubit system, where each qubit represents a spin-\(\frac{1}{2}\) particle. Instead of arbitrary two-qubit operators, only those that do not mix sectors with different spin values are selected. According to the Clebsch-Gordan decomposition, the tensor product of two spin-\(\frac{1}{2}\) particles can be written as \( V^{(1/2)} \otimes V^{(1/2)} = V^{(1)} \oplus V^{(0)} \), meaning the system can be in either a triplet state (total spin \( j = 1 \)), consisting of three symmetric states, or a singlet state (total spin \( j = 0 \)), which is antisymmetric. An \textit{SU(2)}-equivariant operation must act independently within each spin sector without mixing them. Accordingly, permitted transformations include applying the same \textit{SU(2)} rotation to both qubits and controlled interactions that preserve the total spin.

\subsubsection{Schur Transform}

The Schur transform is fundamental for \textit{SU(2)}-equivariant quantum circuits, as it enables a transition from the standard computational basis to a basis adapted to \textit{SU(2)} symmetry~\cite{Meyer2023}. Formally, the Schur transform is expressed as:
\begin{equation}
    \mathcal{U}_{\text{Schur}}: V^{\otimes n} \to \bigoplus_{j} S^j \otimes M_j,
\end{equation}
where \( S^j \) are subspaces of irreducible representations of \textit{SU(2)}, and \( M_j \) are multiplicity spaces. In quantum circuits, the transition to the Schur basis can be implemented using two-qubit and multi-qubit operators that permute states according to their symmetry properties.

Specifically, the Schur transform can be realized through a Fourier transform over the symmetric group \( S_n \), which separates subspaces associated with \textit{SU(2)} representations, controlled permutations that order basis vectors by total angular momentum, and local rotations that ensure precise correspondence between the standard and \textit{SU(2)}-equivariant bases.

\subsection{\textit{SU(2)} Symmetry in Physical Models}

The \textit{SU(2)} symmetry group plays a crucial role in systems with spin degrees of freedom, such as magnetic materials, quantum spin liquids, and topological phases~\cite{Balents2010, Ramirez1994}. The \textit{SU(2)} group consists of 2×2 unitary matrices with determinant 1 and is associated with spin-\(\frac{1}{2}\) and angular momentum. Its algebra is defined by the spin generators \( S^x, S^y, S^z \), which satisfy the commutation relations:
\[
[S^i, S^j] = i \epsilon^{ijk} S^k,
\]
where \(\epsilon^{ijk}\) is the antisymmetric Levi-Civita tensor, and the total spin \( S^2 = (S^x)^2 + (S^y)^2 + (S^z)^2 \) is an invariant quantity. Physically, \textit{SU(2)} symmetry reflects the invariance of a system under rotations in spin space. If the Hamiltonian commutes with the total spin operators (\([H, S^2] = 0\), \([H, S_z] = 0\)), the system exhibits global \textit{SU(2)} invariance, which is critical for studying magnetic, topological, and quantum-informational phenomena~\cite{Sachdev1992}.

Below, we examine key models that demonstrate \textit{SU(2)} symmetry.

\subsubsection{Isotropic Heisenberg Model}

The Heisenberg model describes isotropic spin interactions in crystalline structures and serves as a foundation for understanding magnetism~\cite{Haldane1983}. Its Hamiltonian for spin-\(\frac{1}{2}\) particles is given by:
\[
H = J \sum_{\langle i,j \rangle} \mathbf{S}_i \cdot \mathbf{S}_j,
\]
where \( J \) is the exchange interaction constant, \( \mathbf{S}_i \) is the vector of spin operators at site \( i \), and \( \langle i,j \rangle \) denotes nearest neighbors on the lattice.

The scalar product \( \mathbf{S}_i \cdot \mathbf{S}_j = S_i^x S_j^x + S_i^y S_j^y + S_i^z S_j^z \) ensures the Hamiltonian’s invariance under rotations in spin space, as confirmed by the commutation relations:
\[
[H, \mathbf{S}^2] = 0, \quad [H, S_z] = 0.
\]
Depending on the sign of \( J \), the model describes different magnetic states: for \( J < 0 \), spins align parallel, forming a ferromagnetic order, while for \( J > 0 \), they align antiparallel, corresponding to an antiferromagnetic order. In a one-dimensional (1D) chain, the antiferromagnetic model (\( J > 0 \)) exhibits critical behavior: long-range magnetic order is absent due to quantum fluctuations (Mermin-Wagner theorem), but correlations decay according to a power law, and the excitation spectrum is gapless, described by a conformal field theory with central charge \( c = 1 \). In two-dimensional (2D) and three-dimensional (3D) systems at low temperatures, spontaneous symmetry breaking can occur, leading to magnetic ordering, as observed in materials like Fe or Ni (ferromagnets) or MnO (antiferromagnets)~\cite{White1983, DiFrancesco1997}.

This model is fundamental for understanding phase transitions in magnetic systems and serves as a basis for numerical methods, such as quantum simulations on optical lattices with cold atoms.

\subsubsection{Heisenberg Model on the Kagome Lattice}

On the Kagome lattice (a two-dimensional structure with triangular elements), the Heisenberg model exhibits additional complexity due to geometric frustration~\cite{Mendels2016, Han2012}. The Hamiltonian is given by:
\[
H = J_1 \sum_{\langle i,j \rangle} \mathbf{S}_i \cdot \mathbf{S}_j + J_2 \sum_{\langle\langle i,k \rangle\rangle} \mathbf{S}_i \cdot \mathbf{S}_k,
\]
where \( J_1 > 0 \) is the antiferromagnetic interaction between nearest neighbors, and \( J_2 \) is the interaction between next-nearest neighbors.

The Kagome lattice with antiferromagnetic interactions exhibits geometric frustration, as its triangular topology prevents simultaneous minimization of the energy of all spin pairs. In the classical approximation, this leads to a highly degenerate ground state. In the quantum case for \( S = \frac{1}{2} \), quantum fluctuations can stabilize non-trivial phases, notably a quantum spin liquid (QSL)—a state without long-range magnetic order even at temperature \( T = 0 \). This state is characterized by fractionalized excitations, such as spinons (carriers of spin \( 1/2 \)) and visons (magnetic monopoles)~\cite{Mendels2016}.

Numerical methods, particularly the Density Matrix Renormalization Group (DMRG), indicate that for \( J_2 = 0 \), the ground state is a \( Z_2 \)-topological quantum spin liquid with short-range correlations and non-trivial topological entropy. Introducing additional terms, such as next-nearest-neighbor exchanges \( J_2 \) or anisotropic interactions (e.g., the Dzyaloshinskii-Moriya term), can lead to the emergence of magnetically ordered phases, such as \( q=0 \) or \( \sqrt{3} \times \sqrt{3} \) order.

Experimentally, QSL phases are realized in materials like herbertsmithite ZnCu\(_3\)(OH)\(_6\)Cl\(_2\), where neutron spectroscopy reveals a continuum of excitations without distinct magnetic peaks, indicating the presence of fractionalized spin states~\cite{Mendels2016, Han2012, Yan2011}. These phases hold potential for quantum computing due to their topological robustness against local perturbations, and their dynamics are investigated in quantum simulations with ultracold atoms.

\subsubsection{Hubbard Model}

The Hubbard model describes the interaction of electrons in narrow energy bands and is key to understanding strongly correlated systems~\cite{Hubbard1963}. Its Hamiltonian is given by:
\[
H = -t \sum_{\langle i,j \rangle, \sigma} (c_{i\sigma}^\dagger c_{j\sigma} + \text{h.c.}) + U \sum_i n_{i\uparrow} n_{i\downarrow},
\]
where \( t \) is the hopping amplitude between neighboring sites, \( U \) is the Coulomb interaction energy, \( c_{i\sigma}^\dagger \) is the creation operator for an electron with spin \( \sigma \), and \( n_{i\sigma} = c_{i\sigma}^\dagger c_{i\sigma} \) is the number operator for particles.

At half-filling (one electron per site) and strong interaction (\( U \gg t \)), charge fluctuations are suppressed, and in second-order perturbation theory, the model reduces to an effective spin Hamiltonian~\cite{Hubbard1963}:
\[
H_{\text{eff}} = J \sum_{\langle i,j \rangle} \mathbf{S}_i \cdot \mathbf{S}_j, \quad J = \frac{4t^2}{U}.
\]
This Hamiltonian exhibits \textit{SU(2)} symmetry in the spin sector, emerging as a derived property. In this regime, the system behaves as a Mott insulator, where electrons are localized, and spin degrees of freedom determine magnetic correlations. On the Kagome lattice, this can lead to the formation of a quantum spin liquid, while on square lattices, it results in antiferromagnetic order, as seen in the parent phases of cuprates (e.g., La\(_2\)CuO\(_4\)), which are foundational to high-temperature superconductivity~\cite{Anderson1987, Lee2006}. This model explains the metal-insulator transition and magnetic properties of real materials, and its \textit{SU(2)} symmetry is utilized in quantum simulations to model correlated states.

\subsubsection{AKLT Model}

The Affleck-Kennedy-Lieb-Tasaki (AKLT) model describes a spin chain with spin \( S = 1 \) and has the Hamiltonian~\cite{Affleck1987}:
\[
H = J \sum_i \left[ \mathbf{S}_i \cdot \mathbf{S}_{i+1} + \frac{1}{3} (\mathbf{S}_i \cdot \mathbf{S}_{i+1})^2 \right], \quad J > 0.
\]

This Hamiltonian preserves \textit{SU(2)} symmetry, and its ground state is a valence-bond solid (VBS) state, where spins form singlet pairs through projection onto the subspace with maximum spin between neighbors. Unlike the Heisenberg model for \( S = \frac{1}{2} \), which has a gapless spectrum, the AKLT chain exhibits a Haldane gap~\cite{Haldane1983}—an energy gap between the ground state and excitations, which protects topological edge states with spin \(\frac{1}{2}\). This phase is an example of a symmetry-protected topological (SPT) order, contrasting with trivial insulators and holding significance for quantum memory and information processing. Experimentally, the Haldane gap is observed in quasi-one-dimensional magnetic materials, such as CsNiCl\(_3\), and its topological properties are investigated in quantum simulations~\cite{Affleck1987, Haldane1983}.

\subsection{\textit{SU(2)}-Equivariant Quantum Circuits}

A quantum circuit is called \textit{SU(2)-equivariant} if it preserves the \textit{SU(2)} symmetry structure of the input state. Formally, an \( n \)-qubit quantum transformation \( T \) is equivariant if, for any unitary operator \( U \in SU(2) \), the following equality holds:
\begin{equation}
    U^{\otimes n} T = T U^{\otimes n}.
\end{equation}
This condition ensures that the quantum circuit commutes with global \textit{SU(2)} transformations.

\subsubsection{Two-Qubit \textit{SU(2)}-Equivariant Operators}

For two qubits, the tensor product space decomposes as:
\begin{equation}
    V^{(1/2)} \otimes V^{(1/2)} = V^{(1)} \oplus V^{(0)},
\end{equation}
where \( V^{(1)} \) (triplet states) is the symmetric subspace, and \( V^{(0)} \) (singlet state) is the antisymmetric subspace.

An \textit{SU(2)}-equivariant operator must act within these sectors, preserving the block structure of the representation.

The Schur transform transitions from the computational basis to a basis with fixed spin quantum numbers, where the matrix representation of the two-qubit Schur operator is:
\begin{equation}
    S_2 = \begin{pmatrix} 1 & 0 & 0 & 0 \\ 0 & \frac{1}{\sqrt{2}} & \frac{1}{\sqrt{2}} & 0 \\ 0 & 0 & 0 & 1 \\ 0 & \frac{1}{\sqrt{2}} & -\frac{1}{\sqrt{2}} & 0 \end{pmatrix}.
\end{equation}
This matrix ensures alignment of spin basis states with their total angular momentum representations. After transforming the system into the spin basis, a fundamental two-qubit equivariant operator can be constructed by applying a phase to the singlet state (\( J = 0, J_z = 0 \)). Such an operator is represented as:
\begin{equation}
    P_2(\theta) = \begin{pmatrix} 1 & 0 & 0 & 0 \\ 0 & 1 & 0 & 0 \\ 0 & 0 & 1 & 0 \\ 0 & 0 & 0 & e^{i\theta} \end{pmatrix}.
\end{equation}
This transformation isolates the singlet subspace (\( J=0 \)) and applies a phase rotation only within it, ensuring that the operation remains block-diagonal and preserves \textit{SU(2)} symmetry~\cite{east2023needspinsu2equivariant}.

From a geometric perspective, such a construction can be represented as a spin network. An \textit{SU(2)}-equivariant operator corresponds to a vertex transformation that preserves angular momentum rules. Its structure ensures that different irreducible representations remain separated, manifested in the block-diagonal form of the transformation matrix.

\textit{SU(2)}-equivariant quantum circuits for two qubits are built from the following basic components:
\begin{itemize}
    \item \textbf{Schur operator} (\( S_2 \)) – transforms the computational basis into the spin basis.
    \item \textbf{Controlled phase} (\( P_2(\theta) \)) – applies a phase rotation only in the spin-0 sector.
    \item \textbf{Inverse Schur transform} (\( S_2^{\dagger} \)) – returns the system to the computational basis.
\end{itemize}
Thus, the general structure of an \textit{SU(2)}-equivariant operator is represented as:
\begin{equation}
    V_2(\theta) = S_2 P_2(\theta) S_2^{\dagger}.
\end{equation}

Extending \textit{SU(2)}-equivariant operators to three qubits requires considering the decomposition:
\begin{equation}
    V^{(1/2)} \otimes V^{(1/2)} \otimes V^{(1/2)} = V^{(3/2)} \oplus V^{(1/2)} \oplus V^{(1/2)}.
\end{equation}
This separates the state space into a fully symmetric spin-\( 3/2 \) subspace and two mixed spin-\( 1/2 \) subspaces. The corresponding three-qubit Schur transform is:
\begin{equation}
    S_3 = \begin{pmatrix} 1 & 0 & 0 & 0 & 0 & 0 & 0 & 0 \\
    0 & \frac{1}{\sqrt{3}} & \frac{1}{\sqrt{3}} & 0 & \frac{1}{\sqrt{3}} & 0 & 0 & 0 \\
    0 & 0 & 0 & \frac{1}{\sqrt{3}} & 0 & \frac{1}{\sqrt{3}} & \frac{1}{\sqrt{3}} & 0 \\
    0 & 0 & 0 & 0 & 0 & 0 & 0 & 1 \\
    0 & \sqrt{\frac{2}{3}} & -\frac{1}{\sqrt{6}} & 0 & -\frac{1}{\sqrt{6}} & 0 & 0 & 0 \\
    0 & 0 & 0 & \frac{1}{\sqrt{6}} & 0 & \frac{1}{\sqrt{6}} & -\sqrt{\frac{2}{3}} & 0 \\
    0 & 0 & \frac{1}{\sqrt{2}} & 0 & -\frac{1}{\sqrt{2}} & 0 & 0 & 0 \\
    0 & 0 & 0 & -\frac{1}{\sqrt{2}} & 0 & \frac{1}{\sqrt{2}} & 0 & 0 \end{pmatrix}.
\end{equation}
To maintain equivariance, the parameterized unitary transformation \( P_3(\theta) \) acts in the spin basis:
\begin{equation}
    P_3(\theta) = \begin{pmatrix} I_4 & 0 \\ 0 & U_2(\theta) \end{pmatrix},
\end{equation}
where \( U_2(\theta) \) is a two-dimensional unitary operator that mixes states with spin-\( 1/2 \).

For the general case of \( k \)-qubit systems, \textit{SU(2)}-equivariant circuits have the structure:
\begin{equation}
    V_k(\theta) = S_k P_k(\theta) S_k^{\dagger},
\end{equation}
where \( S_k \) is the Schur transform for \( k \) qubits, and \( P_k(\theta) \) parameterizes transformations acting within each irreducible subspace. The unitary operator \( P_k(\theta) \) has a block structure corresponding to the decomposition into irreducible representations:
\begin{equation}
    P_k(\theta) = \bigoplus_{i} (U_i(\theta) \otimes I_{d_i}),
\end{equation}
where each \( U_i(\theta) \) is a unitary operator with dimension equal to the multiplicity of the corresponding irreducible representation, and \( d_i = 2J + 1 \) is the dimension of the \( i \)-th irreducible representation~\cite{east2023needspinsu2equivariant}.    
\chapter{Methodology}

\section{Structure and Properties of the \textit{SU(2)} Ansatz}
\label{sec:su2_circuit}

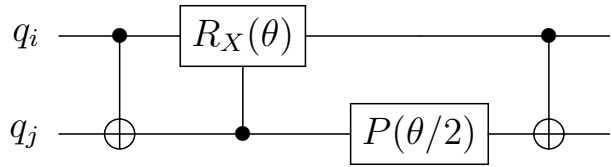
\begin{figure}[h]
    \centering
    \scalebox{1.2}{
        \Qcircuit @C=1.2em @R=1.0em @!R {
            \lstick{q_i} & \ctrl{1} & \gate{R_X(\theta)} & \qw & \ctrl{1} & \qw \\
            \lstick{q_j} & \targ & \ctrl{-1} & \gate{P(\theta/2)} & \targ & \qw \\
        }
    }
    \caption{\textit{SU(2)}-equivariant block, the fundamental element of the ansatz}
    \label{fig:su2_block}
\end{figure}

The development of an efficient ansatz that preserves \textit{SU(2)} symmetry is a key task for achieving potential advantages in quantum computing. The foundation of this construction is the \textit{SU(2)}-equivariant block, depicted in Figure~\ref{fig:su2_block}; an analytical proof that it satisfies the \textit{SU(2)}-equivariance condition is provided in Appendix~\ref{appendix:B1}. This block serves as the primary building block for creating more complex quantum circuit architectures for varying numbers of qubits.

Using the \textit{SU(2)}-equivariant block as a foundation, various quantum circuits can be constructed, tailored to specific requirements. By arranging these blocks in different configurations, ansatzes with diverse structures, such as linear, brickwall, cyclic, or more complex architectures, can be created. Additionally, the flexibility in design allows for varying the order of control qubits within blocks, enabling the generation of different entanglement structures.

Below, we present four examples of ansatzes, each demonstrating a unique arrangement of \textit{SU(2)} blocks.

\subsubsection{Ansatz with Linear and Brickwall Structure of \textit{SU(2)} Blocks}

The linear structure arranges \textit{SU(2)} blocks sequentially along a chain of qubits, connecting each pair of neighboring qubits with an \textit{SU(2)} block, while the brickwall structure alternates the placement of \textit{SU(2)} blocks between neighboring qubits in each layer.
\newpage
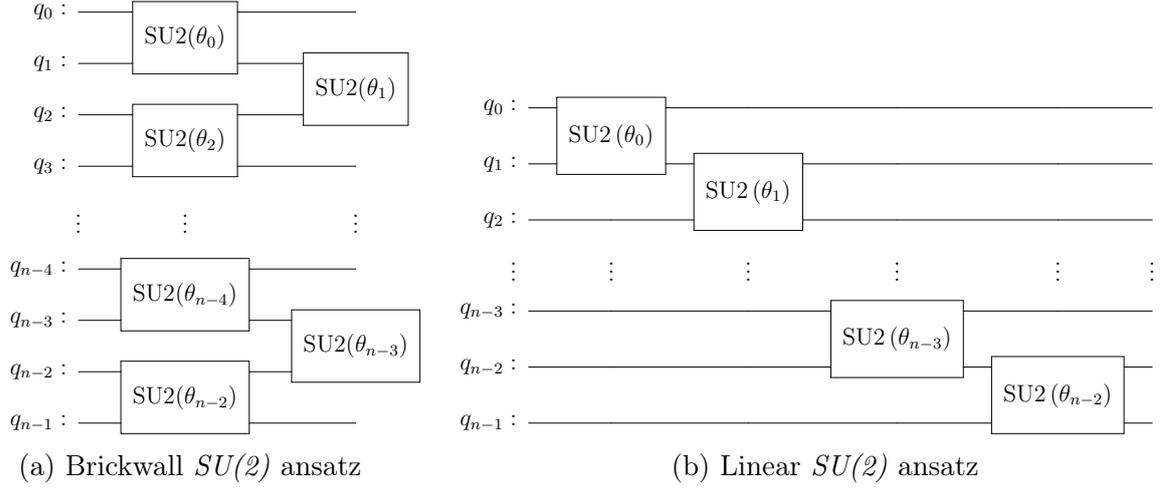
\begin{figure}[h]
    \centering
    \noindent
    \begin{subfigure}[b]{0.4\textwidth}
        \scalebox{0.75}{
        \Qcircuit @C=1.8em @R=1.3em @!R {
	\nghost{{q}_{0} :  } & \lstick{{q}_{0} :  } & \multigate{1}{\mathrm{SU2}(\theta_0)} & \qw \\
	\nghost{{q}_{1} :  } & \lstick{{q}_{1} :  } & \ghost{\mathrm{SU2}(\theta_0)} & \multigate{1}{\mathrm{SU2}(\theta_1)} \\
	\nghost{{q}_{2} :  } & \lstick{{q}_{2} :  } & \multigate{1}{\mathrm{SU2}(\theta_2)} & \ghost{\mathrm{SU2}(\theta_1)} \\
	\nghost{{q}_{3} :  } & \lstick{{q}_{3} :  } & \ghost{\mathrm{SU2}(\theta_2)} & \qw \\
	\nghost{\vdots} & \vdots & \vdots & \vdots \\
	\nghost{{q}_{n-4} :  } & \lstick{{q}_{n-4} :  } & \multigate{1}{\mathrm{SU2}(\theta_{n-4})} & \qw \\
	\nghost{{q}_{n-3} :  } & \lstick{{q}_{n-3} :  } & \ghost{\mathrm{SU2}(\theta_{n-4})} & \multigate{1}{\mathrm{SU2}(\theta_{n-3})} \\
	\nghost{{q}_{n-2} :  } & \lstick{{q}_{n-2} :  } & \multigate{1}{\mathrm{SU2}(\theta_{n-2})} & \ghost{\mathrm{SU2}(\theta_{n-3})} \\
	\nghost{{q}_{n-1} :  } & \lstick{{q}_{n-1} :  } & \ghost{\mathrm{SU2}(\theta_{n-2})} & \qw \\
        }
        }
        \subcaption{Brickwall \textit{SU(2)} ansatz}
        \label{fig:su2_brickwall}
    \end{subfigure}
    \hfill
    \begin{subfigure}[b]{0.55\textwidth}
        \scalebox{0.75}{
       	\Qcircuit @C=1.2em @R=1.5em { 
		& \lstick{{q}_{0} : } & \multigate{1}{\mathrm{SU2}\,(\theta_0)} & \qw & \qw & \qw & \qw \\
	& \lstick{{q}_{1} : } & \ghost{\mathrm{SU2}\,(\theta_0)} & \multigate{1}{\mathrm{SU2}\,(\theta_1)} & \qw & \qw & \qw \\
	& \lstick{{q}_{2} : } & \qw & \ghost{\mathrm{SU2}\,(\theta_1)} & \qw & \qw & \qw \\
	& \lstick{\vdots} & \vdots & \vdots & \vdots & \vdots & \vdots \\
	& \lstick{{q}_{n-3} : } & \qw & \qw & \multigate{1}{\mathrm{SU2}\,(\theta_{n-3})} & \qw & \qw \\
	& \lstick{{q}_{n-2} : } & \qw & \qw & \ghost{\mathrm{SU2}\,(\theta_{n-3})} & \multigate{1}{\mathrm{SU2}\,(\theta_{n-2})} & \qw \\
	& \lstick{{q}_{n-1} : } & \qw & \qw & \qw & \ghost{\mathrm{SU2}\,(\theta_{n-2})} & \qw \\
				}
        }
        \subcaption{Linear \textit{SU(2)} ansatz}
        \label{fig:su2_linear}
    \end{subfigure}

    \caption{Comparison of \textit{SU(2)}-equivariant ansatz structures}
    \label{fig:su2_blocks}
\end{figure}

\subsubsection{Ansatz with Linear and Brickwall Structure of \textit{SU(2)} Blocks with Variable Control Qubit Order}

By varying the order of control qubits in the central \textit{SU(2)} block, a richer entanglement structure can be achieved.
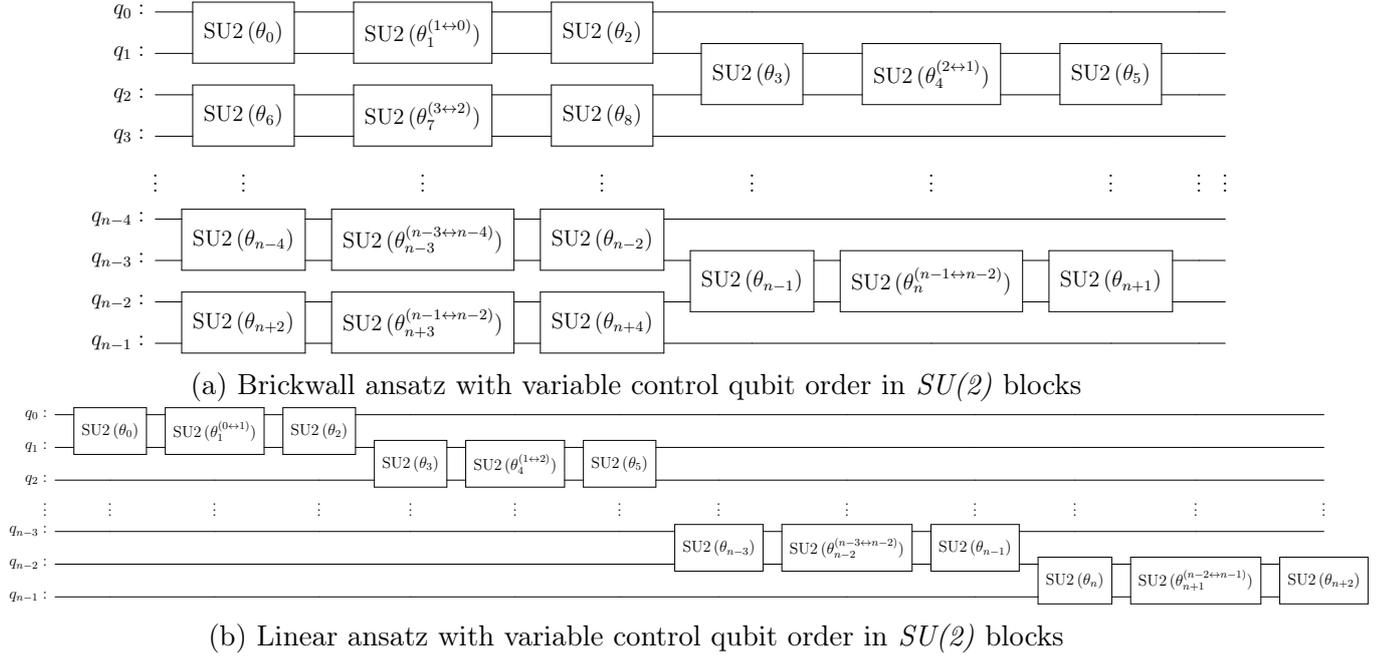
\begin{figure}[h!]
    \centering
    \begin{subfigure}[b]{1\textwidth}
        \centering
        \scalebox{0.7}{
        \Qcircuit @C=1.2em @R=1.0em @!R { \\
	\nghost{{q}_{0} :  } & \lstick{{q}_{0} :  } & \multigate{1}{\mathrm{SU2}\,(\theta_0)} & \multigate{1}{\mathrm{SU2}\,(\theta_1^{(1\leftrightarrow0)})} & \multigate{1}{\mathrm{SU2}\,(\theta_2)} & \qw & \qw & \qw & \qw & \qw\\
	\nghost{{q}_{1} :  } & \lstick{{q}_{1} :  } & \ghost{\mathrm{SU2}\,(\theta_0)} & \ghost{\mathrm{SU2}\,(\theta_1^{(0\leftrightarrow1)})} & \ghost{\mathrm{SU2}\,(\theta_2)} & \multigate{1}{\mathrm{SU2}\,(\theta_3)} & \multigate{1}{\mathrm{SU2}\,(\theta_4^{(2\leftrightarrow1)})} & \multigate{1}{\mathrm{SU2}\,(\theta_5)} & \qw & \qw\\
	\nghost{{q}_{2} :  } & \lstick{{q}_{2} :  } & \multigate{1}{\mathrm{SU2}\,(\theta_6)} & \multigate{1}{\mathrm{SU2}\,(\theta_7^{(3\leftrightarrow2)})} & \multigate{1}{\mathrm{SU2}\,(\theta_8)} & \ghost{\mathrm{SU2}\,(\theta_3)} & \ghost{\mathrm{SU2}\,(\theta_4^{(1\leftrightarrow2)})} & \ghost{\mathrm{SU2}\,(\theta_5)} & \qw & \qw\\
	\nghost{{q}_{3} :  } & \lstick{{q}_{3} :  } & \ghost{\mathrm{SU2}\,(\theta_6)} & \ghost{\mathrm{SU2}\,(\theta_7^{(2\leftrightarrow3)})} & \ghost{\mathrm{SU2}\,(\theta_8)} & \qw & \qw & \qw & \qw & \qw\\
	\nghost{\vdots} & \vdots & \vdots & \vdots & \vdots & \vdots & \vdots & \vdots & \vdots & \vdots\\
	\nghost{{q}_{n-4} :  } & \lstick{{q}_{n-4} :  } & \multigate{1}{\mathrm{SU2}\,(\theta_{n-4})} & \multigate{1}{\mathrm{SU2}\,(\theta_{n-3}^{(n-3\leftrightarrow n-4)})} & \multigate{1}{\mathrm{SU2}\,(\theta_{n-2})} & \qw & \qw & \qw & \qw & \qw\\
	\nghost{{q}_{n-3} :  } & \lstick{{q}_{n-3} :  } & \ghost{\mathrm{SU2}\,(\theta_{n-4})} & \ghost{\mathrm{SU2}\,(\theta_{n-3}^{(n-4\leftrightarrow n-3)})} & \ghost{\mathrm{SU2}\,(\theta_{n-2})} & \multigate{1}{\mathrm{SU2}\,(\theta_{n-1})} & \multigate{1}{\mathrm{SU2}\,(\theta_{n}^{(n-1\leftrightarrow n-2)})} & \multigate{1}{\mathrm{SU2}\,(\theta_{n+1})} & \qw & \qw\\
	\nghost{{q}_{n-2} :  } & \lstick{{q}_{n-2} :  } & \multigate{1}{\mathrm{SU2}\,(\theta_{n+2})} & \multigate{1}{\mathrm{SU2}\,(\theta_{n+3}^{(n-1\leftrightarrow n-2)})} & \multigate{1}{\mathrm{SU2}\,(\theta_{n+4})} & \ghost{\mathrm{SU2}\,(\theta_{n-1})} & \ghost{\mathrm{SU2}\,(\theta_{n}^{(n-2\leftrightarrow n-1)})} & \ghost{\mathrm{SU2}\,(\theta_{n+1})} & \qw & \qw\\
	\nghost{{q}_{n-1} :  } & \lstick{{q}_{n-1} :  } & \ghost{\mathrm{SU2}\,(\theta_{n+2})} & \ghost{\mathrm{SU2}\,(\theta_{n+3}^{(n-2\leftrightarrow n-1)})} & \ghost{\mathrm{SU2}\,(\theta_{n+4})} & \qw & \qw & \qw & \qw & \qw\\
				}}
        \caption{Brickwall ansatz with variable control qubit order in \textit{SU(2)} blocks}
        \label{fig:complex_ansatz_brick}
    \end{subfigure}

    \begin{subfigure}[b]{1\textwidth}
        \centering
        \scalebox{0.5}{
	\Qcircuit @C=1.2em @R=1.2em { 
	& \lstick{{q}_{0} : } & \multigate{1}{\mathrm{SU2}\,(\theta_0)} & \multigate{1}{\mathrm{SU2}\,(\theta_1^{(0\leftrightarrow1)})} & \multigate{1}{\mathrm{SU2}\,(\theta_2)} & \qw & \qw & \qw & \qw & \qw & \qw & \qw & \qw & \qw \\
	& \lstick{{q}_{1} : } & \ghost{\mathrm{SU2}\,(\theta_0)} & \ghost{\mathrm{SU2}\,(\theta_1^{(0\leftrightarrow1)})} & \ghost{\mathrm{SU2}\,(\theta_2)} & \multigate{1}{\mathrm{SU2}\,(\theta_3)} & \multigate{1}{\mathrm{SU2}\,(\theta_4^{(1\leftrightarrow2)})} & \multigate{1}{\mathrm{SU2}\,(\theta_5)} & \qw & \qw & \qw & \qw & \qw & \qw \\
	& \lstick{{q}_{2} : } & \qw & \qw & \qw & \ghost{\mathrm{SU2}\,(\theta_3)} & \ghost{\mathrm{SU2}\,(\theta_4^{(1\leftrightarrow2)})} & \ghost{\mathrm{SU2}\,(\theta_5)} & \qw & \qw & \qw & \qw & \qw & \qw \\
	& \lstick{\vdots} & \vdots & \vdots & \vdots & \vdots & \vdots & \vdots & \vdots & \vdots & \vdots & \vdots & \vdots & \vdots \\
	& \lstick{{q}_{n-3} : } & \qw & \qw & \qw & \qw & \qw & \qw & \multigate{1}{\mathrm{SU2}\,(\theta_{n-3})} & \multigate{1}{\mathrm{SU2}\,(\theta_{n-2}^{(n-3\leftrightarrow n-2)})} & \multigate{1}{\mathrm{SU2}\,(\theta_{n-1})} & \qw & \qw & \qw \\
	& \lstick{{q}_{n-2} : } & \qw & \qw & \qw & \qw & \qw & \qw & \ghost{\mathrm{SU2}\,(\theta_{n-3})} & \ghost{\mathrm{SU2}\,(\theta_{n-2}^{(n-3\leftrightarrow n-2)})} & \ghost{\mathrm{SU2}\,(\theta_{n-1})} & 
    \multigate{1}{\mathrm{SU2}\,(\theta_{n})} & \multigate{1}{\mathrm{SU2}\,(\theta_{n+1}^{(n-2\leftrightarrow n-1)})} & \multigate{1}{\mathrm{SU2}\,(\theta_{n+2})} \\
	& \lstick{{q}_{n-1} : } & \qw & \qw & \qw & \qw & \qw & \qw & \qw & \qw & \qw & \ghost{\mathrm{SU2}\,(\theta_{n})} & \ghost{\mathrm{SU2}\,(\theta_{n+1}^{(n-2\leftrightarrow n-1)})} & \ghost{\mathrm{SU2}\,(\theta_{n+2})} \\
				}
        }
        \caption{Linear ansatz with variable control qubit order in \textit{SU(2)} blocks}
        \label{fig:complex_ansatz_linear}
    \end{subfigure}

    \caption[Comparison of ansatzes with different \textit{SU(2)} block topologies: brickwall and linear]{Comparison of ansatzes with different \textit{SU(2)} block topologies: brickwall (top) and linear (bottom). Both architectures account for qubit permutations in the central blocks $(i\leftrightarrow j)$}
    \label{fig:general_su2_ansatze}
\end{figure}

\newpage

\subsubsection{Comparison of Ansatzes}

To compare the effectiveness of the four different \textit{SU(2)} ansatzes, their ability to generate entanglement and their expressibility can be evaluated. To assess the entanglement capability, for each of the four \textit{SU(2)} ansatzes, \( n \) random parameter sets are generated, and for each set, the Meyer--Wallach measure is computed and averaged. To measure expressibility, the average Kullback--Leibler (KL) divergence between the distribution generated by the ansatz and the uniform Haar distribution is similarly computed (see~\ref{exrp_and_ent}). The obtained averaged values are used as quality metrics according to the chosen criterion.

The results of this analysis are presented in the plots below, showing the dependence of the averaged Meyer--Wallach measure and the averaged KL divergence on the number of layers in the ansatz for systems with \( 6, 8, 10 \), and \( 12 \) qubits. The evaluation used \( n=1000 \) parameter sets.

\begin{figure}[h]
    \centering
    \includegraphics[width=1\textwidth]{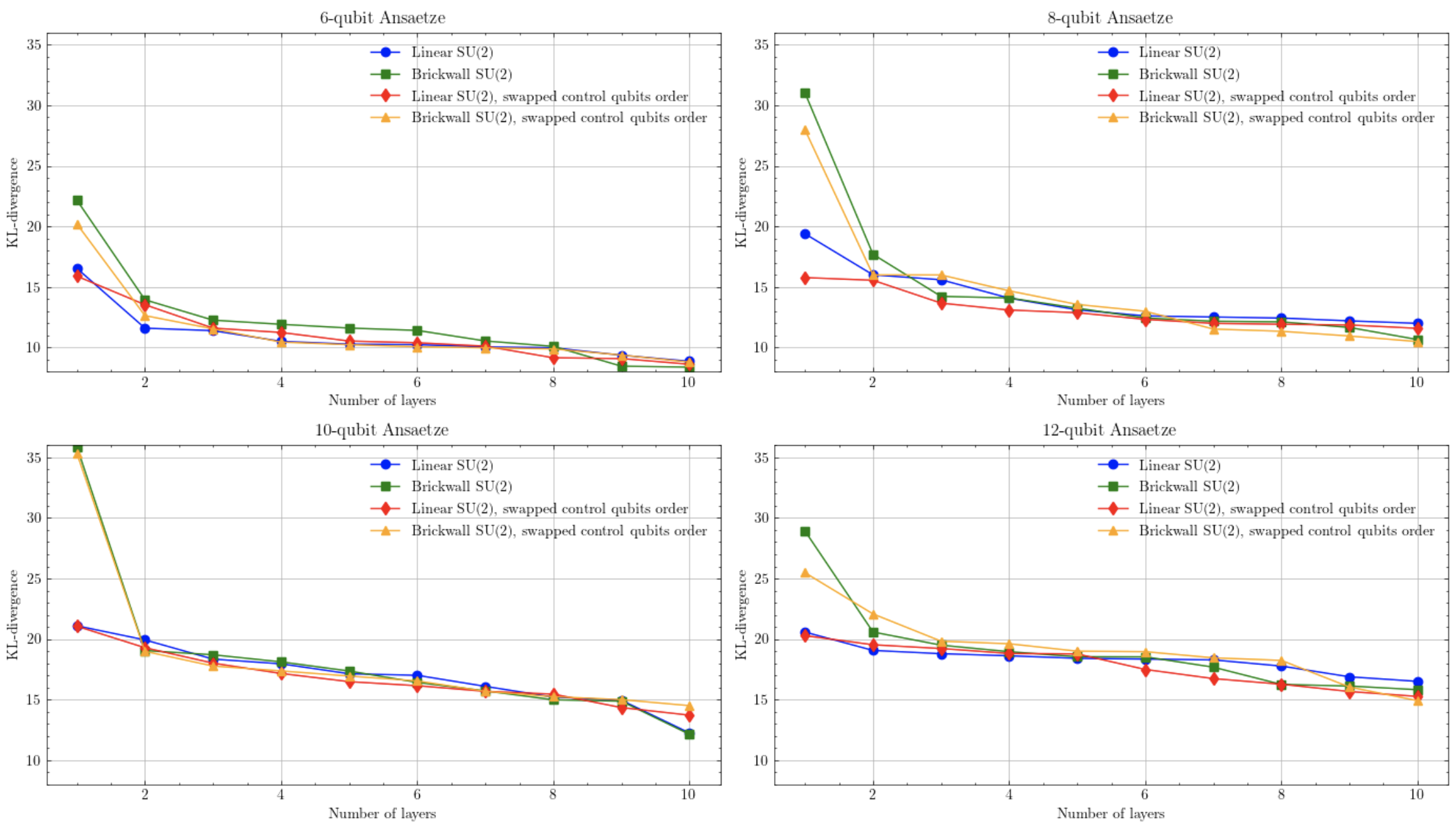}
    \caption{Dependence of the KL divergence on the number of layers for four \textit{SU(2)} ansatzes (linear, brickwall, linear and brickwall with variable control qubit order)}
    \label{fig:expressibility}
\end{figure}

From the plots, it is evident that all ansatzes exhibit similar KL divergence values for a fixed number of qubits. The divergence decreases as the number of layers increases, indicating improved expressibility. However, no ansatz demonstrates a significant advantage in expressibility over the others.

\begin{figure}[h]
    \centering
    \includegraphics[width=0.9\textwidth]{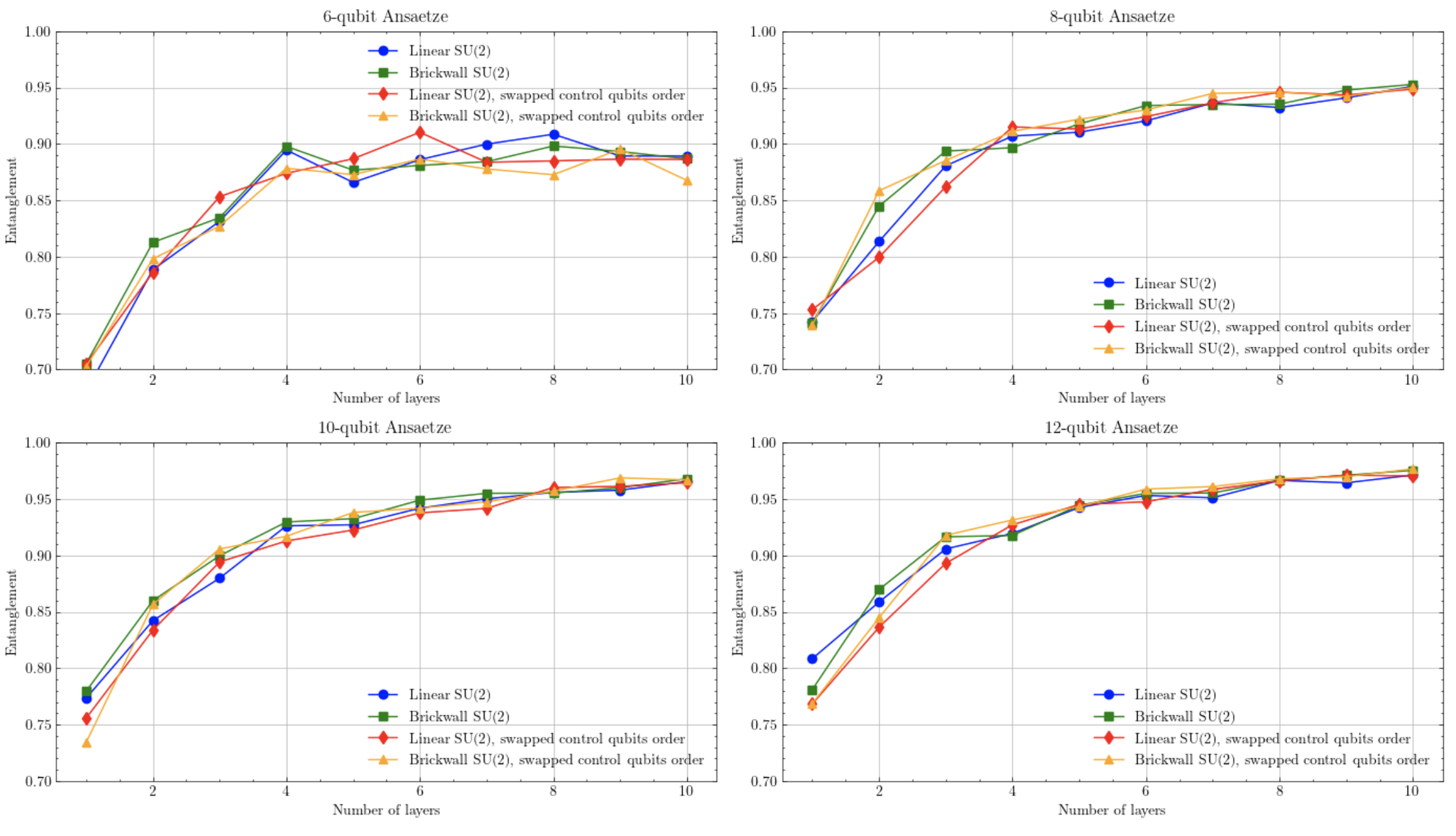}
    \caption{Dependence of the entanglement capability on the number of layers for four \textit{SU(2)} ansatzes (linear, brickwall, linear and brickwall with variable control qubit order)}
    \label{fig:entangling_capability}
\end{figure}

A similar analysis of entanglement (fig.\ref{fig:entangling_capability}) shows that, as with expressibility, different \textit{SU(2)} ansatz topologies have comparable effectiveness in generating entangled states.

This suggests that focusing on more complex ansatzes with variable control qubit order is not justified, as they require significantly more two-qubit operations compared to standard linear or brickwall ansatzes, without providing advantages in either entanglement or expressibility.

\subsection{Comparison with General Ansatzes}

Based on the previous analysis, we select two \textit{SU(2)} ansatzes (with linear and brickwall structures) without variable control qubit order. To complete the analysis, it is worth comparing their properties with general ansatzes that have similar structures. The architectures of these ansatzes are presented below.

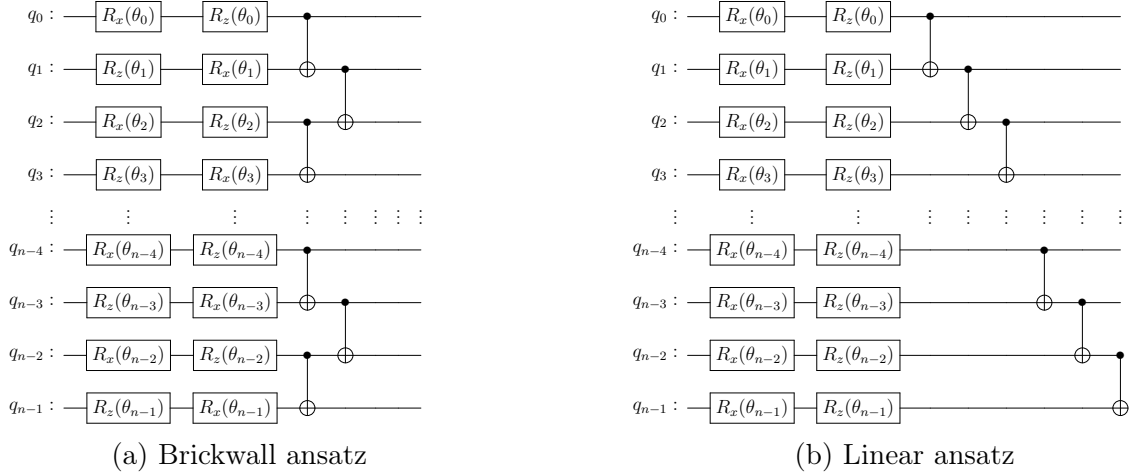
\begin{figure}[h]
    \centering

    \begin{subfigure}[c]{0.45\textwidth}
        \centering
        \scalebox{0.6}{
        \Qcircuit @C=1.2em @R=1.2em {
            \lstick{q_0 :} & \gate{R_x(\theta_{0})} & \gate{R_z(\theta_{0})} & \ctrl{1} & \qw      & \qw      & \qw      & \qw \\
            \lstick{q_1 :} & \gate{R_z(\theta_{1})} & \gate{R_x(\theta_{1})} & \targ{} & \ctrl{1} & \qw      & \qw      & \qw \\
            \lstick{q_2 :} & \gate{R_x(\theta_{2})} & \gate{R_z(\theta_{2})} & \ctrl{1} & \targ{}  & \qw      & \qw      & \qw \\
            \lstick{q_3 :} & \gate{R_z(\theta_{3})} & \gate{R_x(\theta_{3})} & \targ{} & \qw      & \qw      & \qw      & \qw \\
            \lstick{\vdots} & \vdots & \vdots & \vdots & \vdots & \vdots & \vdots & \vdots \\
            \lstick{q_{n-4} :} & \gate{R_x(\theta_{n-4})} & \gate{R_z(\theta_{n-4})} & \ctrl{1} & \qw & \qw & \qw & \qw \\
            \lstick{q_{n-3} :} & \gate{R_z(\theta_{n-3})} & \gate{R_x(\theta_{n-3})} & \targ{} & \ctrl{1} & \qw & \qw & \qw \\
            \lstick{q_{n-2} :} & \gate{R_x(\theta_{n-2})} & \gate{R_z(\theta_{n-2})} & \ctrl{1} & \targ{} & \qw & \qw & \qw \\
            \lstick{q_{n-1} :} & \gate{R_z(\theta_{n-1})} & \gate{R_x(\theta_{n-1})} & \targ{} & \qw & \qw & \qw & \qw \\
        }}
        \subcaption{Brickwall ansatz}
        \label{fig:general_brickwall}
    \end{subfigure}
    \hfill
    \begin{subfigure}[c]{0.45\textwidth}
        \centering
        \scalebox{0.6}{
        \Qcircuit @C=1.2em @R=1.2em {
            \lstick{q_0 :} & \gate{R_x(\theta_{0})} & \gate{R_z(\theta_{0})} & \ctrl{1} & \qw      & \qw      & \qw      & \qw      & \qw \\
            \lstick{q_1 :} & \gate{R_x(\theta_{1})} & \gate{R_z(\theta_{1})} & \targ{} & \ctrl{1} & \qw      & \qw      & \qw      & \qw \\
            \lstick{q_2 :} & \gate{R_x(\theta_{2})} & \gate{R_z(\theta_{2})} & \qw     & \targ{}  & \ctrl{1} & \qw      & \qw      & \qw \\
            \lstick{q_3 :} & \gate{R_x(\theta_{3})} & \gate{R_z(\theta_{3})} & \qw     & \qw      & \targ{}  & \qw      & \qw      & \qw \\
            \lstick{\vdots} & \vdots & \vdots & \vdots & \vdots & \vdots & \vdots & \vdots & \vdots \\
            \lstick{q_{n-4} :} & \gate{R_x(\theta_{n-4})} & \gate{R_z(\theta_{n-4})} & \qw & \qw & \qw & \ctrl{1} & \qw & \qw \\
            \lstick{q_{n-3} :} & \gate{R_x(\theta_{n-3})} & \gate{R_z(\theta_{n-3})} & \qw & \qw & \qw & \targ{} & \ctrl{1} & \qw \\
            \lstick{q_{n-2} :} & \gate{R_x(\theta_{n-2})} & \gate{R_z(\theta_{n-2})} & \qw & \qw & \qw & \qw & \targ{} & \ctrl{1} \\
            \lstick{q_{n-1} :} & \gate{R_x(\theta_{n-1})} & \gate{R_z(\theta_{n-1})} & \qw & \qw & \qw & \qw & \qw & \targ{}
        }}
        \subcaption{Linear ansatz}
        \label{fig:general_linear}
    \end{subfigure}

    \caption{General ansatzes with brickwall and linear architectures}
    \label{fig:line_brick_arch}
\end{figure}

Comparing expressibility between \textit{SU(2)} and general ansatzes is not meaningful, as \textit{SU(2)} ansatzes are inherently limited to generating \textit{SU(2)}-equivariant states, whereas general ansatzes can span the entire state space. However, comparing entanglement capability is relevant, as shown in the plot below.

\begin{figure}[h]
    \centering
    \includegraphics[width=1\textwidth]{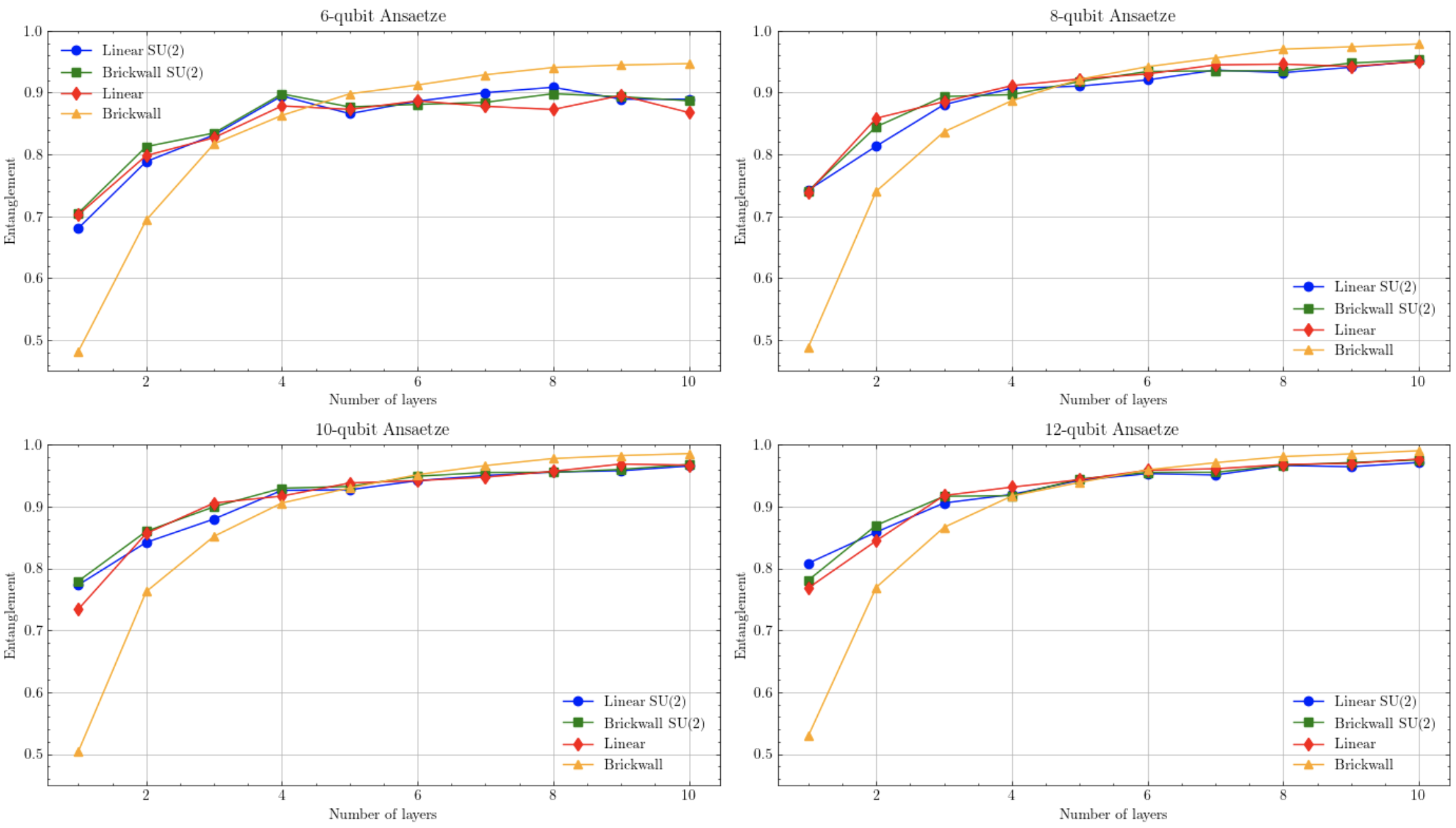}
    \caption[Averaged Meyer--Wallach measure for \textit{SU(2)} ansatzes (linear, brickwall) and general linear and brickwall ansatzes]{Averaged Meyer--Wallach measure for 1000 parameter sets for selected \textit{SU(2)} ansatzes (linear, brickwall) and general linear and brickwall ansatzes for 6, 8, 10, and 12 qubits}
    \label{fig:entanglement_comparison}
\end{figure}

The entanglement analysis shows that all selected ansatzes achieve high entanglement (\( 0.9 \)--\( 1.0 \)) for different numbers of qubits. The \textit{SU(2)} ansatz exhibits faster entanglement growth in early layers, particularly for \( 8 \), \( 10 \), and \( 12 \) qubits, reaching values of \( 0.9 \) within \( 2 \)--\( 3 \) layers, whereas linear and brickwall ansatzes stabilize at \( 0.9 \)--\( 1.0 \) after \( 6 \)--\( 8 \) layers. General linear and brickwall ansatzes achieve higher entanglement values (\( >0.95 \)) with a large number of layers.

Despite the same number of layers in the compared ansatzes, it should be noted that they involve different numbers of two-qubit operators. For a fairer comparison, if circuit depth or the number of multi-qubit operators is critical, the ansatzes should be transpiled to a unified operator basis (dependent on the specific quantum device) and further compared, taking into account hardware constraints.

\subsection{Using Quantum Machine Learning to Improve Trotterization Techniques}

Quantum machine learning (QML) opens a promising avenue for enhancing the accuracy of quantum system simulations. In particular, within the framework of Trotterization methods, applying QML can help reduce errors associated with evolution discretization and ensure better preservation of physical symmetries.

Despite the presence of \textit{SU(2)} symmetry in physical systems, the need to apply Trotterization to represent the Hamiltonian can break this symmetry. Specifically, for the isotropic Heisenberg model, the standard Trotterization approach introduces asymmetry into the model, as the evolution operator

\[ U_{stand} = 
\exp\left(-i H_{XX} \frac{\Delta t}{\hbar}\right) \exp\left(-i H_{YY} \frac{\Delta t}{\hbar}\right) \exp\left(-i H_{ZZ} \frac{\Delta t}{\hbar}\right)
\]
is not \textit{SU(2)}-invariant. This is because the operator depends on the order of application of the Hamiltonian terms, and \( H_{XX} \), \( H_{YY} \), and \( H_{ZZ} \) do not commute with each other, i.e., \( [H_{XX}, H_{YY}] \neq 0 \), \( [H_{YY}, H_{ZZ}] \neq 0 \), \( [H_{XX}, H_{ZZ}] \neq 0 \).

In contrast, applying an \textit{optimized approach} that accurately implements the operator 
\[ U_{opt} = 
\exp\left(-i \frac{\Delta t}{\hbar} (H_{XX} + H_{YY} + H_{ZZ})\right),
\]
preserves \textit{SU(2)} symmetry. In this case, the Hamiltonian can be represented as \( H = H_1 + H_2 + \dots + H_n \), where each \( H_j \) is \textit{SU(2)}-invariant (e.g., \( H_j = J (X_i X_j + Y_i Y_j + Z_i Z_j) \) for a pair of qubits \( i, j \)). The evolution operator is written as
\[
U = e^{-i H t / \hbar} = e^{-i H_1 t / \hbar} \otimes \cdots \otimes e^{-i H_n t / \hbar},
\]
and since each factor \( e^{-i H_j t / \hbar} \) is \textit{SU(2)}-invariant, the entire operator \( U \) also preserves this property. Formally, this can be expressed through the \textit{SU(2)}-equivariance condition: for any \( V \in \text{SU}(2) \),
\[
V^{\otimes n} e^{-i H_j t / \hbar} = e^{-i H_j t / \hbar} V^{\otimes n},
\]
ensuring symmetry preservation for each component \( H_j \), and thus for the full operator \( U \).

An interesting application is the use of an \textit{SU(2)} ansatz to approximate the Trotterized state via a variational algorithm. When using the optimized Trotterization approach, which preserves \textit{SU(2)} symmetry, the behavior of the \textit{SU(2)} ansatz is quite predictable: during training, the ansatz effectively converges to the approximated state.

The behavior of the \textit{SU(2)} ansatz when approximating asymmetric states obtained through the standard Trotterization scheme is non-trivial. In this case, the target state loses \textit{SU(2)} symmetry due to errors, with symmetry breaking during the Trotter approximation. Applying the \textit{SU(2)} ansatz in this context forces the variational algorithm to find the closest \textit{SU(2)}-symmetric state to the asymmetric target.

To study the state, we introduce a loss function \( \mathcal{L}(\theta) \), which characterizes the distance between the trained and target states:
\[
\mathcal{L}(\theta) = 1 - F\left(\ket{\psi_{\text{trained}}}, \ket{\psi_{\text{Trotter}}}\right),
\]
where the fidelity \( F \) is computed as:
\[
F\left(\ket{\psi_{\text{trained}}},\ket{\psi_{\text{Trotter}}}\right) = |\langle \psi_{\text{Trotter}} | \psi_{\text{trained}}(\theta) \rangle|^2.
\]
The parameters \( \theta \) are optimized using the NFT algorithm~\cite{Nakanishi_2020}.

\begin{figure}[h!]
    \centering
    \begin{subfigure}[b]{0.48\textwidth}
        \centering
        \includegraphics[width=\textwidth]{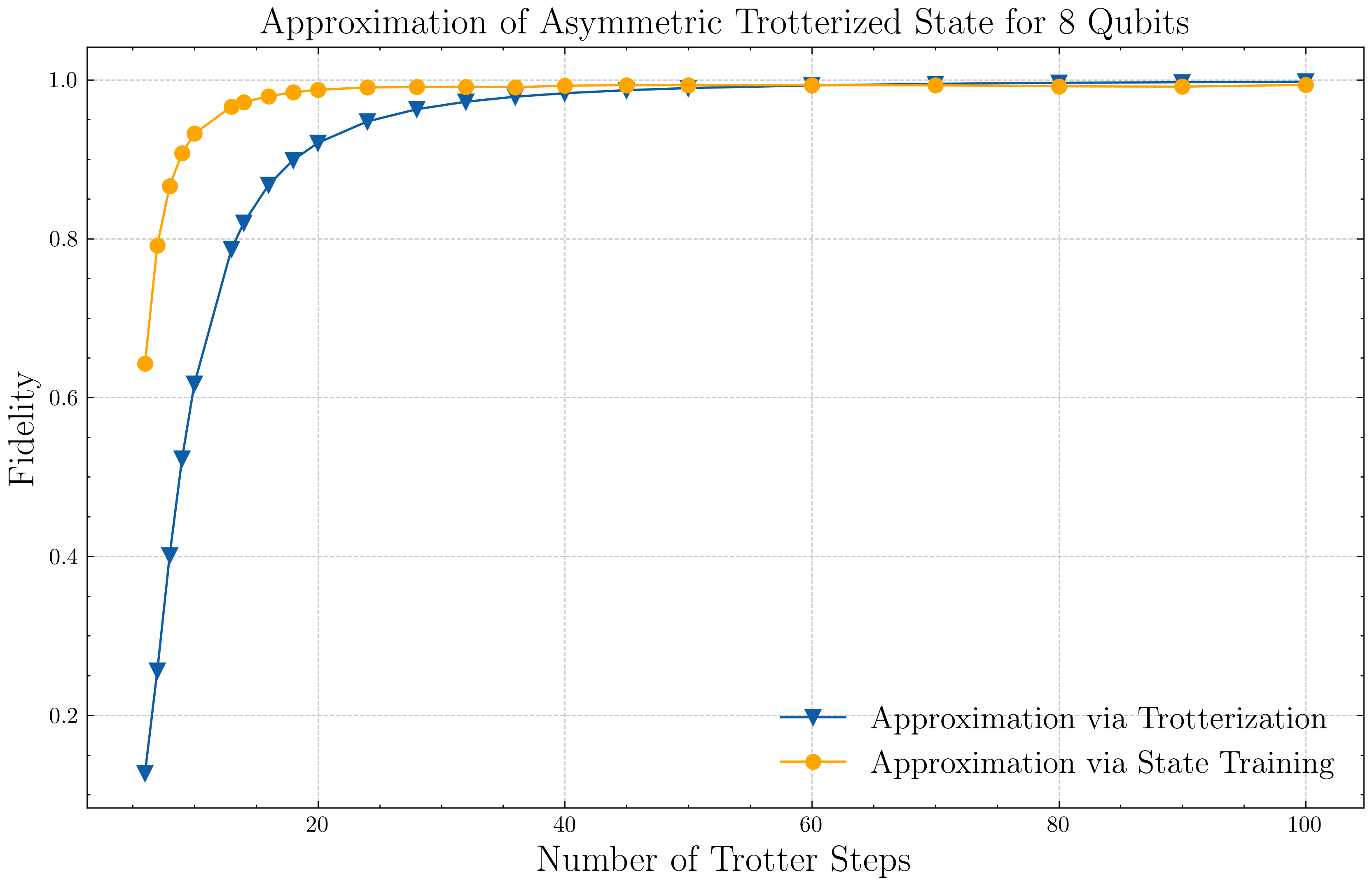}
        \caption{8 qubits}
        \label{fig:trot_unsym_8qb}
    \end{subfigure}
    \hfill
    \begin{subfigure}[b]{0.48\textwidth}
        \centering
        \includegraphics[width=\textwidth]{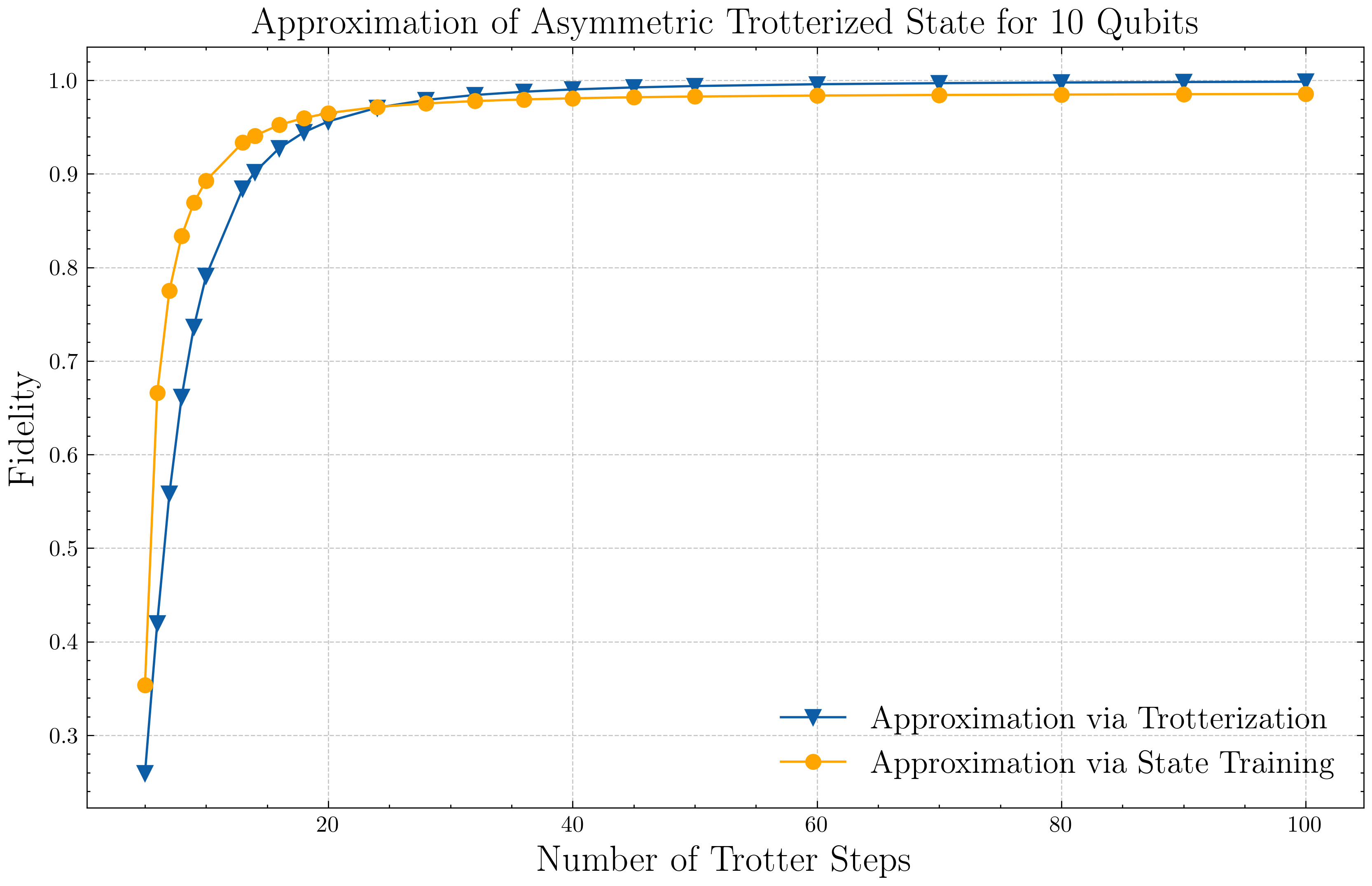}
        \caption{10 qubits}
        \label{fig:trot_unsym_10qb}
    \end{subfigure}

    \vspace{1em}
    \begin{subfigure}[b]{0.5\textwidth}
        \centering
        \includegraphics[width=\textwidth]{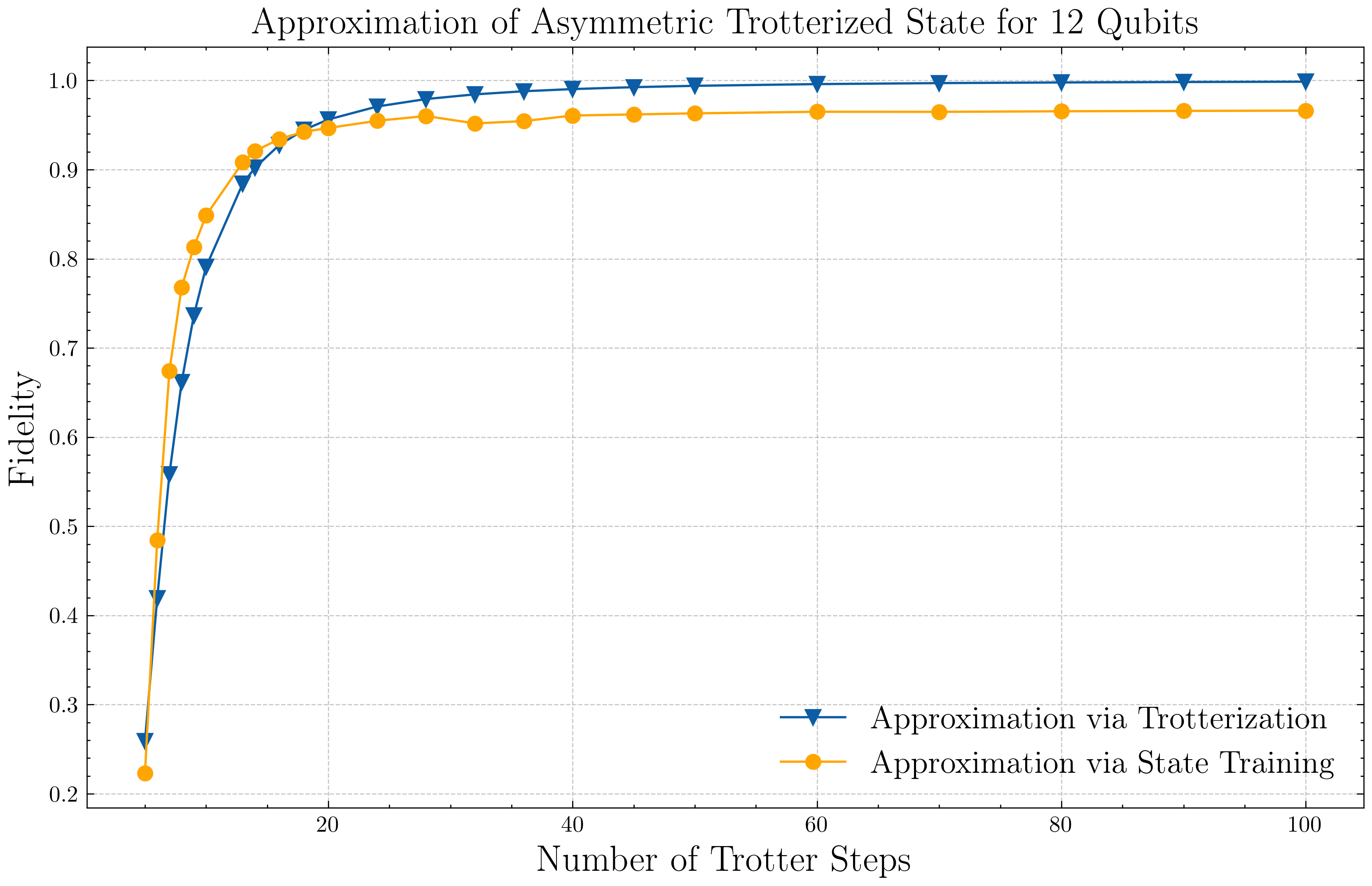}
        \caption{12 qubits}
        \label{fig:trot_unsym_12qb}
    \end{subfigure}

    \caption[Comparison of fidelity to the exact \textit{SU(2)}-symmetric state using standard Trotterization and \textit{SU(2)} ansatz training]{Comparison of fidelity to the exact \textit{SU(2)}-symmetric state using standard Trotterization (blue line) and \textit{SU(2)} ansatz training (orange line)}
    \label{fig:trotter_learning_comparison}
\end{figure}

On Figure~\ref{fig:trotter_learning_comparison}, a comparison of the fidelity to the exact symmetric state is presented for two approaches:
\begin{itemize}
    \item \textbf{Standard Trotterization} — evolution using the Trotter decomposition, which does not preserve \textit{SU(2)} symmetry;
    \item \textbf{\textit{SU(2)} Ansatz Training} — variational approximation of the Trotterized state \( |\psi_T\rangle \) using an \textit{SU(2)}-equivariant ansatz, which reduces to optimizing the loss \( 1 - \mathcal{F}(|\psi_{\mathrm{ansatz}}\rangle, |\psi_T\rangle) \).
\end{itemize}
\smallskip

\noindent
In all simulations, the system begins its evolution from a state that is a tensor product of singlet pairs:
\[
\ket{\psi(0)} = \bigotimes_{i=1}^{n/2} \frac{1}{\sqrt{2}} \left( \ket{01}_{2i-1,2i} - \ket{10}_{2i-1,2i} \right),
\]
where each pair \( \ket{01} - \ket{10} \) is a two-qubit singlet. This initial state is not an eigenstate of the Heisenberg Hamiltonian, allowing the study of non-trivial evolution dynamics.

The results show that for a small number of Trotter steps (i.e., with significant errors and symmetry violations), the fidelity to the exact symmetric state is significantly higher when training the ansatz compared to directly using the Trotterized evolution. 

This indicates that the \textit{SU(2)} ansatz `projects' the asymmetric Trotterized state onto the closest \textit{SU(2)}-symmetric state, which turns out to be closer to the exact solution. Thus, even for imperfect approximations, the ansatz effectively corrects symmetry errors.

After demonstrating the effective training of the \textit{SU(2)} ansatz on asymmetric Trotterized states, a natural question arises: does such an ansatz have an advantage in learning capability compared to general ansatzes without symmetry constraints? To this end, we compare the effectiveness of approximating the target (symmetrically Trotterized) state for different types of ansatzes using a variational algorithm. Similar to the previous example, we consider the evolution of a 12-qubit Heisenberg system with an initial singlet state.

\begin{figure}[h]
    \centering
    \includegraphics[width=0.9\textwidth]{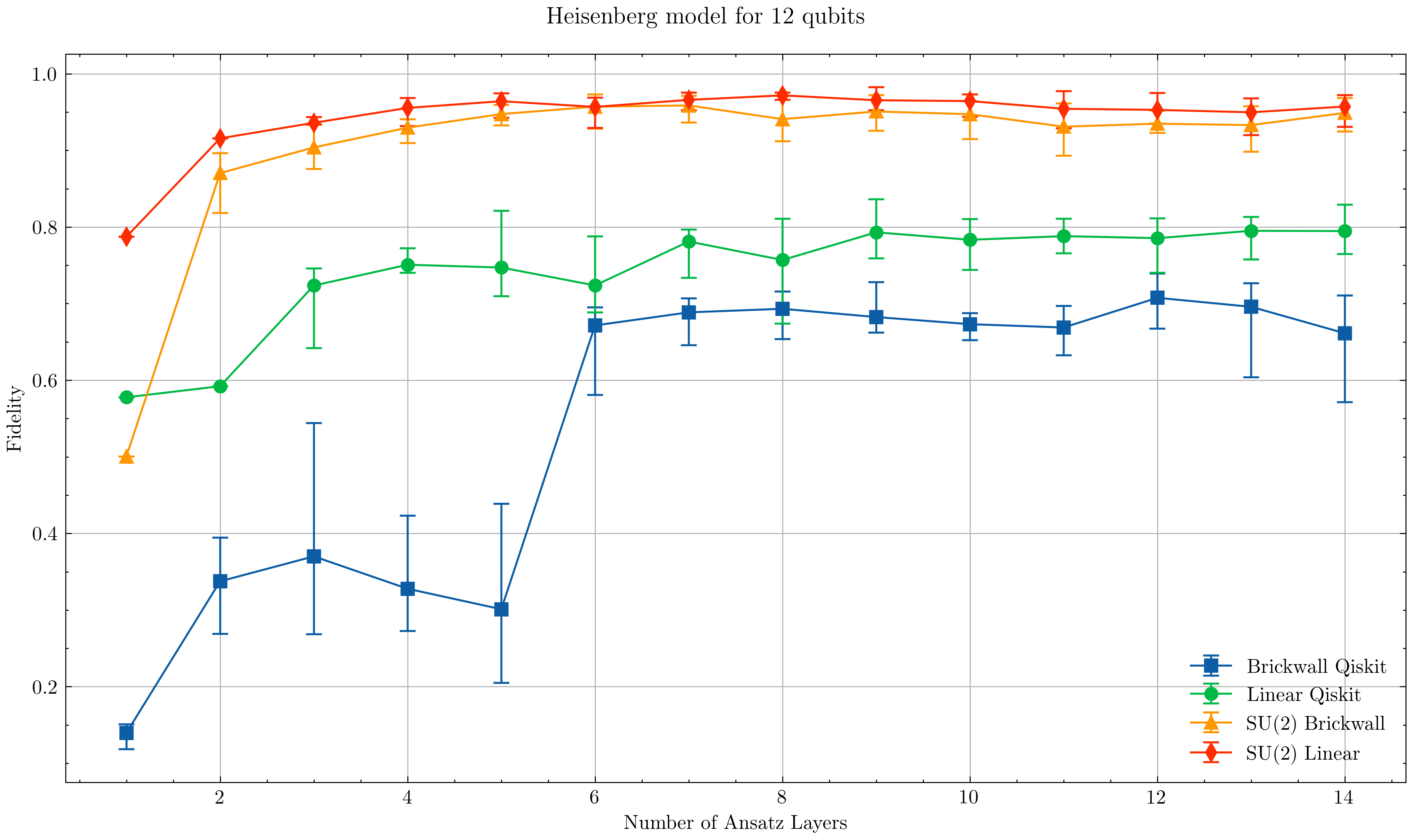}
    \caption[Comparison of learning capability of different variational ansatzes in approximating the symmetric Trotterized state of the Heisenberg model for 12 qubits]
{
    Comparison of learning capability of different ansatzes in approximating symmetric Trotterized states of the Heisenberg model for 12 qubits. The plot shows the mean value and the corresponding range of values (min/max).
    }

    \label{fig:fidelity_su2_vs_qiskit}
\end{figure}

As seen in Figure~\ref{fig:fidelity_su2_vs_qiskit}, ansatzes with built-in \textit{SU(2)} symmetry (red and orange curves) significantly outperform in learning the \textit{SU(2)}-symmetric state. They achieve high accuracy with just \( 2 \)--\( 3 \) layers and exhibit stable behavior as the number of layers increases. In contrast, general (non-specialized) ansatzes (blue and green curves) perform worse even with more layers and show greater variability during training. This confirms that leveraging symmetry not only reduces the dimensionality of the parameter space but also improves the optimization landscape, ensuring better trainability.

\newpage


\section{Parameter Extrapolation of the Ansatz}
\label{sec:extrapolation_parameters}

The analysis of the previous plots naturally raises the question: is it possible, given information about the system state after \( k \) Trotter steps, to predict its state after \( k+1 \), \( k+2 \), and so on? Since we are considering the evolution of a quantum system with a fixed time \( t \), increasing the number of Trotter steps \( r \) (with \( \delta t = t/r \to 0 \)) improves the accuracy of the approximation of the evolution operator. In the limit \( r \to \infty \), the Trotterized operator converges to the exact unitary evolution operator, and the corresponding states coincide:
\[
\lim_{r \to \infty} \left\| \ket{\psi_{\text{Trotter}}^{(r)}} - \ket{\psi(t)} \right\| = 0.
\]
This implies that the amplitudes of the wavefunction \( \ket{\psi_j^{(r)}} \) should gradually approach the amplitudes of the exact state as \( r \) increases.

However, direct application of extrapolation to the wavefunction vector is practically infeasible. The size of the state vector grows exponentially with the number of qubits: \( 2^n \) amplitudes for an \( n \)-qubit system. Already for \( n = 10 \), this means predicting over 1000 parameters, and for \( n = 12 \), over 4000. For larger systems, the full wavefunction becomes inaccessible both for computation and storage. Moreover, extrapolating each amplitude individually does not guarantee that the result will be a physically normalized quantum state.

At this stage, it is appropriate to revisit quantum machine learning methods. If each state after \( r \) Trotter steps is approximated by an ansatz \( U(\boldsymbol{\theta}^{(r)}) \ket{0}^{\otimes n} \), where the parameters \( \boldsymbol{\theta}^{(r)} \) are obtained through optimization, then instead of extrapolating the state itself, we can attempt to extrapolate the parameters \( \boldsymbol{\theta}^{(r)} \).

This offers several advantages:
\begin{itemize}
    \item The dimensionality of the parameter vector is typically polynomial with respect to the number of qubits;
    \item Extrapolation occurs in a parameter space with physical meaning (preserving symmetry, controlling smoothness, etc.);
    \item The resulting state is inherently physically admissible, as it is generated by the unitary evolution of the ansatz.
\end{itemize}

After approximating the Trotterized states using the \textit{SU(2)} ansatz for each number of Trotter steps \( r \), we can study the behavior of the optimized parameters \( \boldsymbol{\theta}^{(r)} \) as a function of \( r \). This allows us to assess the smoothness of parameter changes and the potential for their extrapolation to predict future states.

\begin{figure}[h] 
\centering 
\includegraphics[width=0.85\textwidth]{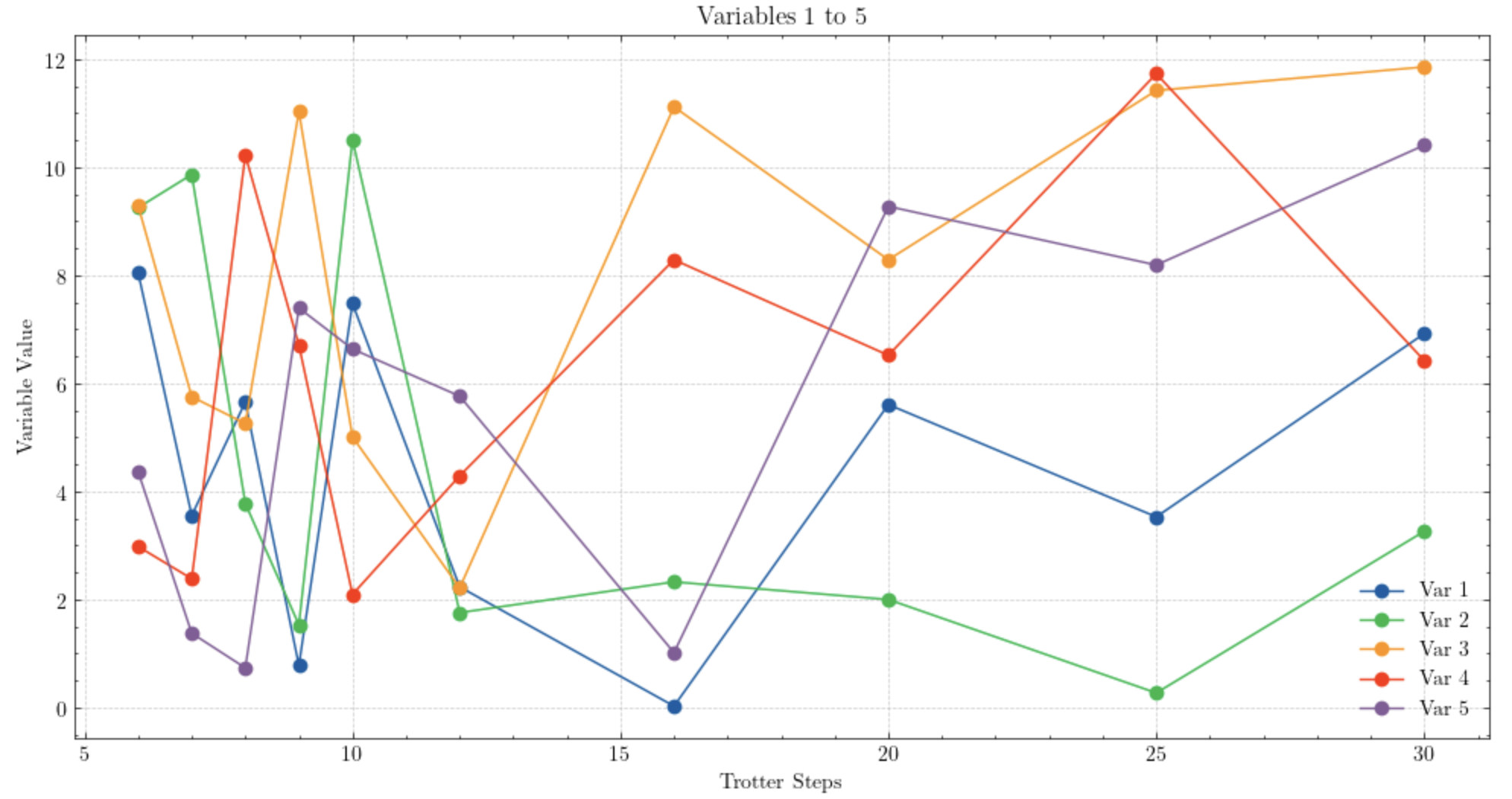} 
\caption{Dynamics of selected parameters \( \theta_i^{(r)} \) of the \textit{SU(2)} ansatz as a function of the number \( r \) of Trotter steps} 
\label{fig:param_dynamics} 
\end{figure}

Figure~\ref{fig:param_dynamics} illustrates the dynamics of only a few ansatz parameters after reducing each to a period of \( 4\pi \), accounting for the periodicity of the corresponding quantum operators. For convenience, this section considers only states obtained using \textit{optimized symmetric Trotterization}, which preserves \textit{SU(2)} symmetry. This enables better control over the quality of learning, as the target state is already \textit{SU(2)}-equivariant, and the ansatz should accurately reproduce it.

As expected, if the ansatz truly learns the physical evolution, the parameters \( \boldsymbol{\theta}^{(r)} \) should exhibit a certain smooth and monotonic behavior as \( r \) increases. However, as seen in the plot, a complex, oscillatory structure of changes is observed, with no clear trend toward parameter convergence.

It is worth emphasizing that this parameter behavior is not a result of poor approximation quality: the fidelity between the ansatz and the target Trotterized state exceeds 0.99 in all cases.

The emergence of such parameter dependence may be due to overparameterization of the Hilbert space. This means that multiple parameter configurations can generate the same quantum state—i.e., the parameter set is not unique. In such a case, the optimization landscape has a high degree of multiplicity: many different points in the parameter space correspond to the same physical state (with identical fidelity to the target state).

As a result, even if the Trotterized states for different \( r \) are very close in the Hilbert space (having nearly identical amplitudes), the corresponding optimal parameters \( \boldsymbol{\theta}^{(r)} \) may lie far apart in the parameter space. This results in discontinuities in the parameter trajectory: although the quantum states vary continuously with increasing \( r \), the corresponding parameters may exhibit abrupt changes. This behavior arises because the optimization process is solely designed to minimize the fidelity loss, without enforcing continuity in parameter space. As a consequence, the optimization may converge to distinct local minima that yield nearly identical quantum states but differ significantly in parameter values.

\begin{figure}[h] 
\centering 
\includegraphics[width=0.85\textwidth]{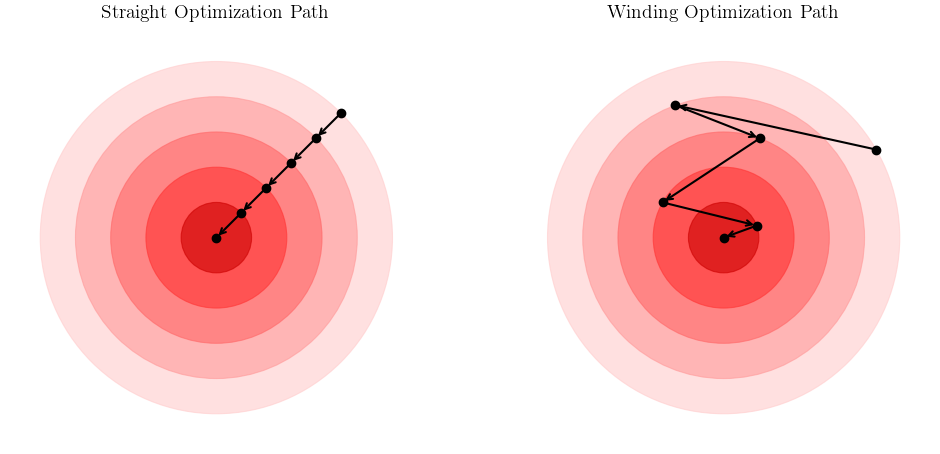} 
\caption{Schematic representation of the optimization trajectories of the ansatz in the parameter space} 
\label{fig:parameter_landscape_paths}  
\end{figure}

Figure~\ref{fig:parameter_landscape_paths} schematically depicts the structure of the parameter space of the \textit{SU(2)} ansatz. Each concentric circle corresponds to a fixed number \( r \) of Trotter steps, with inner circles representing more accurate approximations of the evolution state. On the other hand, the colored regions symbolize different parameter regions that, despite differences in values, generate the same or nearly the same quantum state. This can be interpreted as regions with degenerate loss function values.

The right side of the figure shows the situation that arises without additional regularization of the weights: due to the ambiguity of the parameter space, even quantum states that are very close in the Hilbert space may correspond to significantly distant parameter configurations. In this case, optimization jumps between distant minima, leading to abrupt changes in \( \boldsymbol{\theta}^{(r)} \) values as \( r \) changes, even though the corresponding states remain nearly identical.

The left side of the figure depicts an ideal scenario in which the optimization of the ansatz parameters occurs smoothly: as the number \( r \) of Trotter steps increases, the parameters \( \boldsymbol{\theta}^{(r)} \) monotonically approach the center, minimizing the path in the parameter space.

\subsection{Parameter Regularization via Modification of the Loss Function}

To achieve smoother behavior of the parameters \( \boldsymbol{\theta}^{(r)} \) with an increasing number \( r \) of Trotter steps, an approach involving the modification of the loss function by adding a regularization term was investigated. Specifically, the following form of the loss function was used:

\begin{equation} 
\mathcal{L} = \alpha \cdot \left(1 - F\left( \ket{\psi_{\text{ansatz}}^{(r)}}, \ket{\psi_{\text{target}}^{(r)}} \right)\right) + \beta \cdot \widetilde{D}(\boldsymbol{\theta}^{(r)}, \boldsymbol{\theta}^{(r-1)}), 
\end{equation}

where: 
\begin{itemize}
    \item \( F(\cdot,\cdot) \) — the fidelity between the state generated by the ansatz and the Trotterized state;
    \item \( \widetilde{D} \) — a modified Euclidean distance between the parameter vectors \( \boldsymbol{\theta}^{(r)} \) and \( \boldsymbol{\theta}^{(r-1)} \);
    \item \( \alpha, \beta \) — weighting coefficients (in the experiment, equal values were used: \( \alpha = \beta = 0.5 \)).
\end{itemize}

To mitigate the excessive influence of large Euclidean distance values (since its values are unbounded above), a modification in the form of a smoothing function was applied:
\begin{equation} 
\widetilde{D}(\boldsymbol{\theta}^{(r)}, \boldsymbol{\theta}^{(r-1)}) = \sigma\left(\gamma \cdot \left| \boldsymbol{\theta}^{(r)} - \boldsymbol{\theta}^{(r-1)} \right|^2 - c\right), 
\end{equation}
where \( \sigma(x) = \frac{1}{1 + e^{-x}} \) is the sigmoid function, and \( \gamma \), \( c \) are smoothing hyperparameters. This transformation ensures saturation at large distances (values approach 1) and sensitivity in the region of small changes (values close to 0).

This modification of the loss function allows explicit control over the change in parameters between consecutive Trotter steps and encourages the ansatz to move more smoothly in the parameter space.

\medskip

\noindent\textbf{Limitations of the Approach.} Despite the method's effectiveness in reducing discontinuities and oscillations in the parameters, numerical experiments showed that such regularization reduces the ansatz's ability to accurately approximate the target state. Specifically, the fidelity between the trained ansatz and the target Trotterized state decreased on average from approximately 0.99 to 0.92–0.94 when regularization was applied.

Another approach to controlling parameter smoothness is the use of dynamic weights in the loss function. Instead of a fixed coefficient for the regularization term, the weight of the term accounting for the distance between parameters was chosen to be proportional to the fidelity between the ansatz and the target Trotterized state. This allows focusing on state approximation in the early stages of training and only after achieving acceptable accuracy minimizing the distance between parameters:
\begin{equation} 
\mathcal{L} = \alpha \cdot (1 - F) + F \cdot \widetilde{D}, 
\end{equation}
where \( F \) is the fidelity between the trained and target states, and \( \widetilde{D} \) is the modified distance between parameters.

This approach improved the quality of approximation—the fidelity remained at a high level, comparable to the non-regularized case. However, experiments showed that dynamic weighting introduces instability into the optimization structure: the regularization component only becomes active at later stages of training, leading to less controlled weight dynamics in the early steps. Consequently, the parameters may still remain irregular or too scattered in the optimization landscape.

\subsection{Regularization Based on Dynamic Mode Decomposition (DMD)}

An additional approach implemented and explored in this work is the use of the \textit{Dynamic Mode Decomposition (DMD)} method (a detailed description of the method is provided in Appendix~\ref{appendix:DMD}) to improve the quality of training variational algorithms. The core idea of the method is to predict the parameters of quantum circuits for a larger number of Trotter steps using DMD extrapolation based on already trained parameters.

Algorithmically, the approach consists of the following main stages:
\begin{itemize}
    \item \textbf{Stage 1.} Training the quantum ansatz for a small number of Trotter steps (e.g., 5), where the parameters at each subsequent step are initialized based on the trained parameters of the previous step.
    \item \textbf{Stage 2.} Using DMD to extrapolate the trained weights and predict the parameters for the next (yet untrained) Trotter step.
    \item \textbf{Stage 3.} Using the DMD-predicted parameters as initial values (initialization) for further training of the ansatz.
\end{itemize}

A modified loss function was used, combining fidelity and the Euclidean distance between the proposed and DMD-predicted parameters:
\[
\mathcal{L} = \alpha \cdot (1 - F) + F \cdot D(\boldsymbol{\theta}_{\text{proposed}}, \boldsymbol{\theta}_{\text{DMD}}),
\]
where \( F \) is the fidelity, and \( D(\cdot,\cdot) \) is the Euclidean distance between parameters.

However, this form of the loss function proved insufficiently effective for regularization, leading to parameter oscillations during training and poor extrapolation quality.

To address these challenges, several modifications to the approach were proposed:

\begin{enumerate}
    \item \textbf{Local Alignment of the Next Step.} 
    
    In this approach, the DMD prediction was used to determine the parameters for only the next step (\( k+2 \)). This improved stability but did not ensure smoothness in the parameter trajectory. Numerical issues also arose due to DMD instability, which were resolved through adaptive rank truncation of matrices, limiting parameter changes in SPSA, and controlling overflows.
    
    \item \textbf{Trajectory Alignment with Future Steps.}

    Multiple future steps were considered (e.g., from \( k+1 \) to \( k+10 \)). While this allowed better control over the parameter trajectory, the approach was computationally expensive and did not significantly improve extrapolation accuracy for larger steps.

    \item \textbf{Full Trajectory Alignment (Past and Future).}

    This approach ensured internal consistency of parameters on the training data, but extrapolation to new steps was unsatisfactory due to sensitivity to general patterns.

    \item \textbf{Localized Trajectory Alignment (Local Window).}

    The most effective method proved to be one using a local window around the trained step (from \( k-m \) to \( k+n \), optimally \( m=n=7 \)). This stabilized the parameter dynamics and ensured high training quality.
\end{enumerate}

Although the above-described regularization methods based on Dynamic Mode Decomposition (DMD) aimed to improve parameter smoothness and extrapolation stability, numerical experiments showed that their use during training led to worse results compared to the simplest regularization strategy—namely, using previously trained parameters as initialization for the next optimization step.

Specifically, despite their more complex structure and sophisticated prediction logic, methods involving DMD extrapolation during training were sensitive to numerical instabilities and demonstrated lower quality in approximating the final quantum state. Surprisingly, the best compromise between target state approximation accuracy and parameter smoothness was achieved using the simplest regularization model, which involves a fixed coefficient for the regularization term:
\begin{equation} 
\mathcal{L} = \alpha \cdot (1 - F) + \beta \cdot \widetilde{D}, 
\end{equation}
where \( F \) is the fidelity between the trained and target states, and \( \widetilde{D} \) is the Euclidean distance between parameters (smoothed using a sigmoid transformation). Optimal results were obtained with the coefficient \( \beta \) in the range of 0.5–0.7.

\begin{figure}[h] 
\centering 
\includegraphics[width=0.85\textwidth]{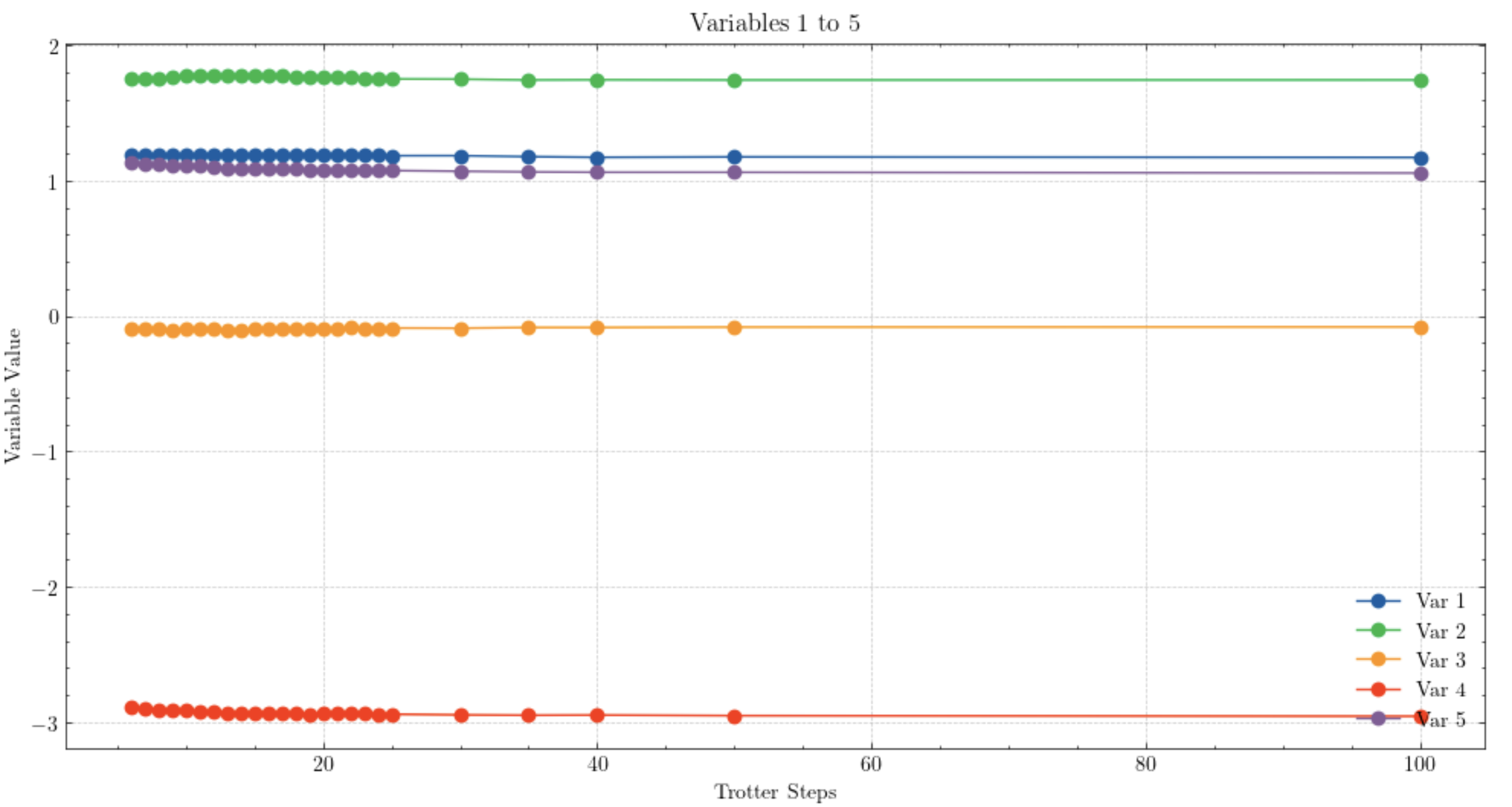} 
\caption{Dynamics of selected parameters after training with regularization \( \theta_i^{(r)} \) of the \textit{SU(2)} ansatz as a function of the number \( r \) of Trotter steps} 
\label{fig:param_dynamics_regul} 
\end{figure}

In this configuration, the ansatz demonstrated consistently high fidelity values (in the range of 0.97–0.99) and significantly better smoothness in parameter changes between consecutive Trotter steps (see Figure~\ref{fig:param_dynamics_regul}). Therefore, despite the conceptually interesting approaches involving DMD, it was decided to adopt the simpler regularization strategy due to its stability, high efficiency, and better practical quality of the results obtained.

\subsection{Approaches Used for Extrapolation}

After completing the training process of the \textit{SU(2)} ansatz for each number \( r \) of Trotter steps, a natural question arises: is it possible to predict the ansatz parameters \( \boldsymbol{\theta}^{(r)} \) for new (larger) values of \( r \) without performing repeated optimization training? Such an approach would allow efficient approximation of states at large \( r \), avoiding the computationally expensive training process.

\medskip

\textbf{Core Idea:} Each ansatz parameter is treated as a function of \( r \), and an approximating function is fitted for each parameter based on known points in the training set \( \{r_1, \dots, r_m\} \).

\subsubsection*{1. Linear and Polynomial Regression}

The first approach involves using classical models of linear and polynomial regression:
\[
\theta_i^{(r)} \approx a_0 + a_1 r + a_2 r^2 + \dots + a_k r^k,
\]
where \( k = 1 \) for linear regression, and \( k > 1 \) for polynomial regression.

\begin{figure}[h]
    \centering
    \begin{subfigure}[t]{0.48\textwidth}
        \centering
        \includegraphics[width=\textwidth]{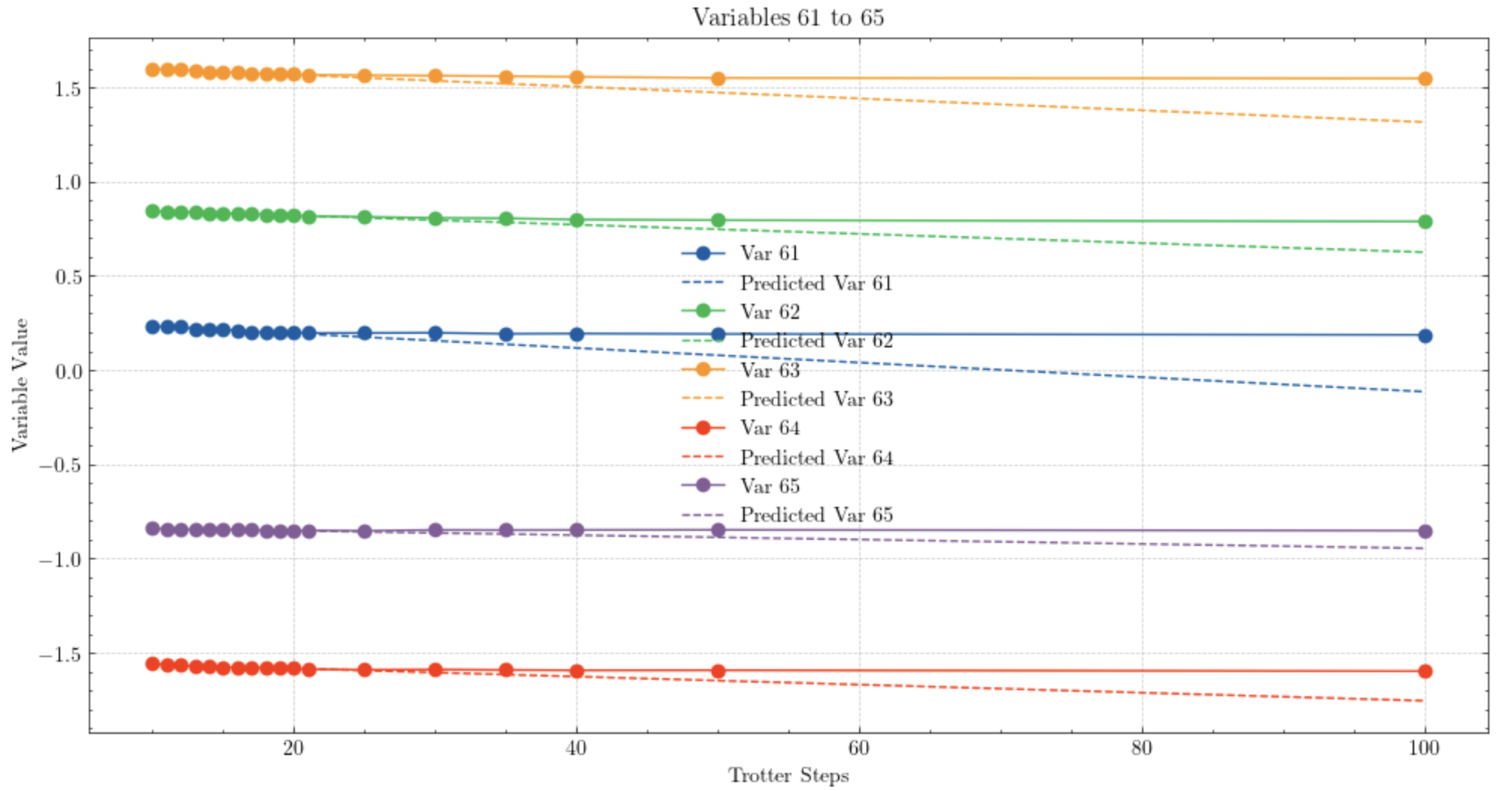}
        \caption{Polynomial approximation of parameters \( \theta_i^{(r)} \). Solid lines represent training values, dashed lines represent the approximation.}
        \label{fig:poly_param_fit}
    \end{subfigure}
    \hfill
    \begin{subfigure}[t]{0.48\textwidth}
        \centering
        \includegraphics[width=\textwidth]{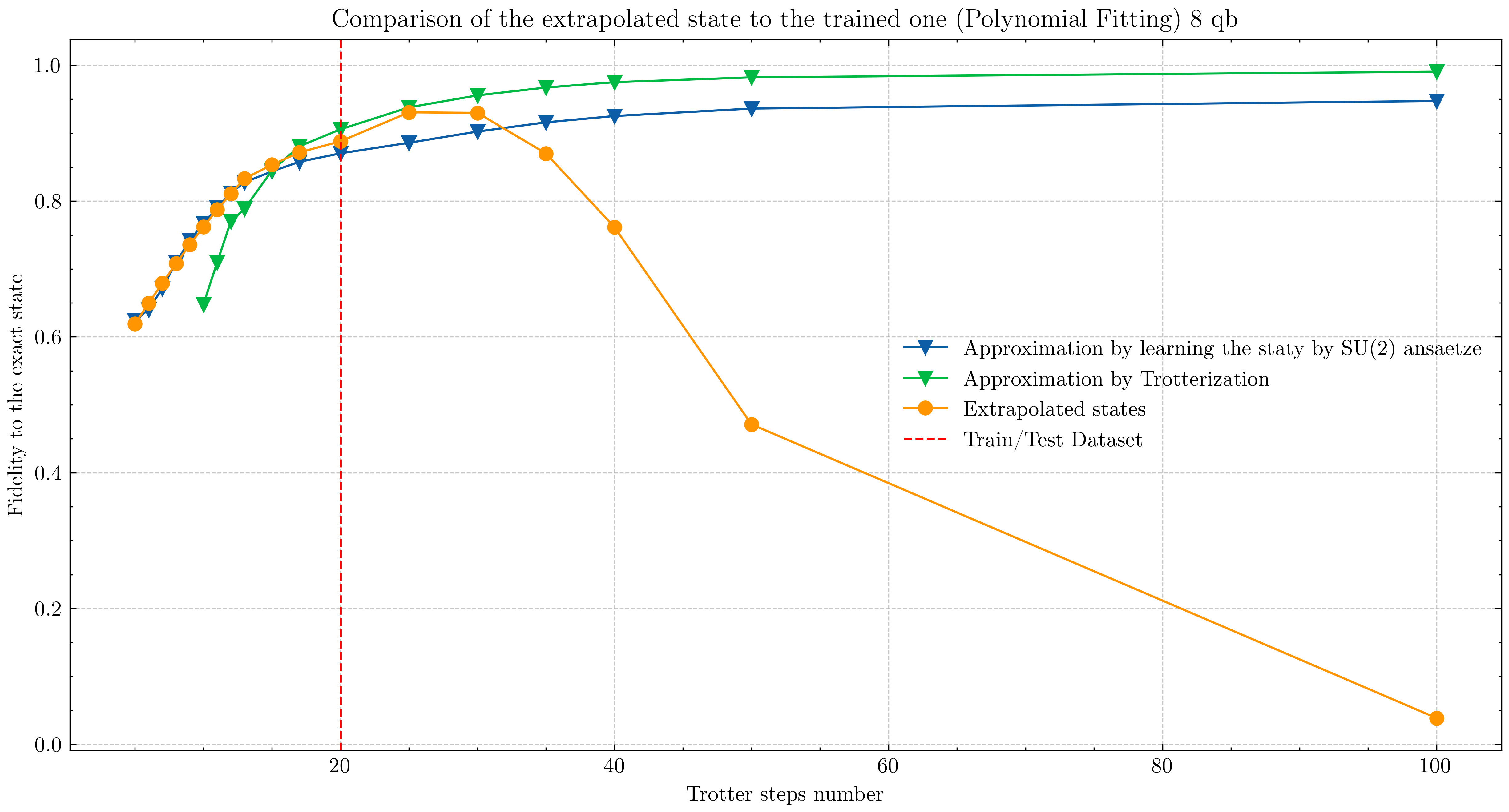}
        \caption{Fidelity of reconstructed states. A decline is observed at large \( r \).}
        \label{fig:poly_fid}
    \end{subfigure}
    \caption{Evaluation of polynomial extrapolation of parameters and its impact on quantum state reconstruction}
    \label{fig:poly_extrapol_analysis}
\end{figure}

Figure~\ref{fig:poly_param_fit} shows an example of polynomial approximation for several ansatz parameters. As can be seen, the approximation locally reproduces the parameter values well within the training region but degrades rapidly outside it. This is particularly evident in Figure~\ref{fig:poly_fid}, which presents the fidelity of the reconstructed states. At large \( r \), the fidelity begins to decline, which is an undesirable effect since, ideally, we expect it to plateau as \( r \to \infty \).

\subsubsection*{2. Nonlinear Approximation: Exponential and Logarithmic Models}

In the second approach, parameter approximation was performed using functions that better capture asymptotic behavior:
\begin{align*}
\theta_i^{(r)} &\approx a + b e^{-c r} \quad \text{(exponential model)},\\
\theta_i^{(r)} &\approx a + b \log(r + c) \quad \text{(logarithmic model)}.
\end{align*}

\begin{figure}[h]
    \centering
    \begin{subfigure}[t]{0.52\textwidth}
        \centering
        \includegraphics[width=\textwidth]{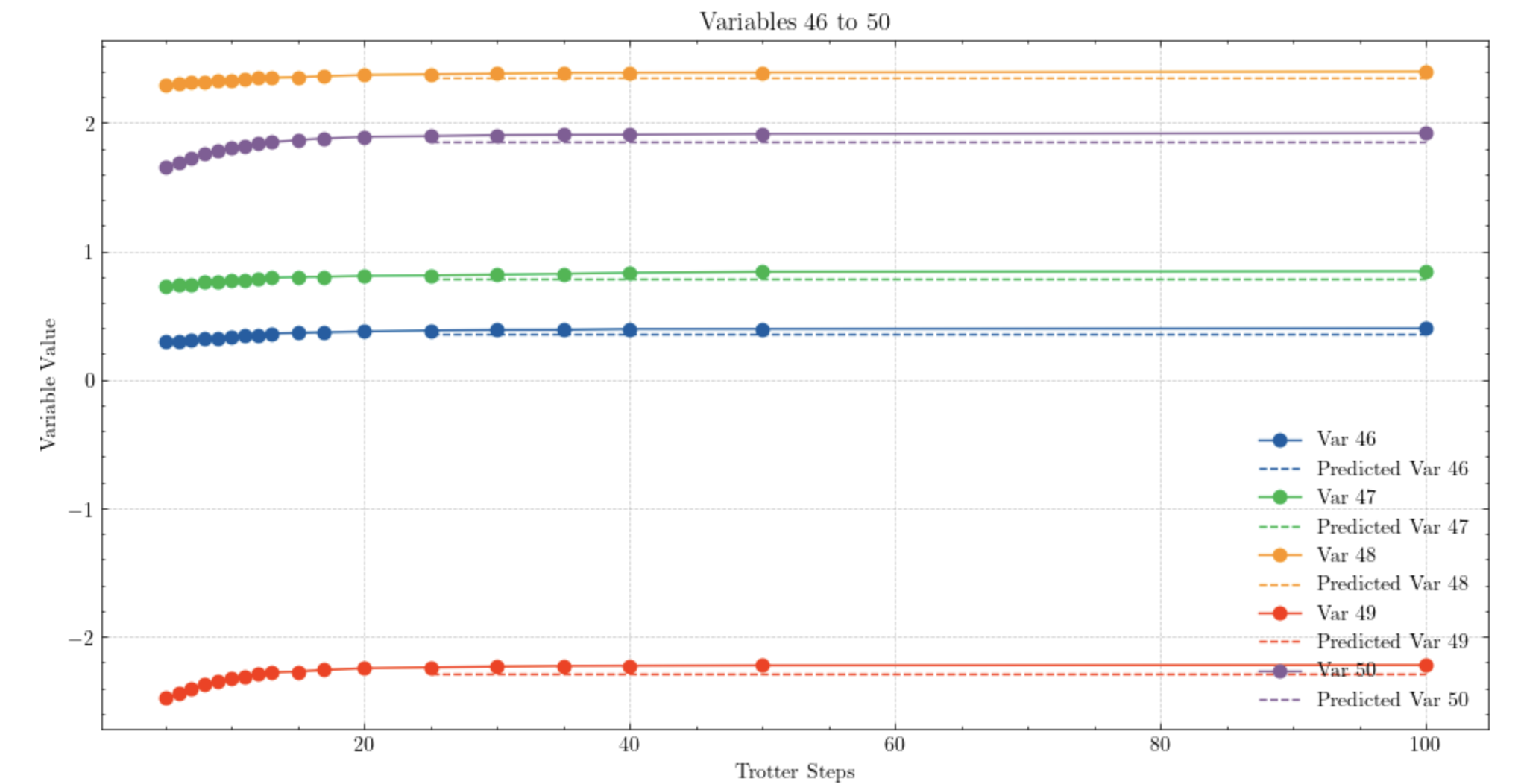}
        \caption{Exponential approximation of parameters \( \theta_i^{(r)} \). Smooth behavior and improved approximation.}
        \label{fig:exp_param_fit}
    \end{subfigure}
    \hfill
    \begin{subfigure}[t]{0.44\textwidth}
        \centering
        \includegraphics[width=\textwidth]{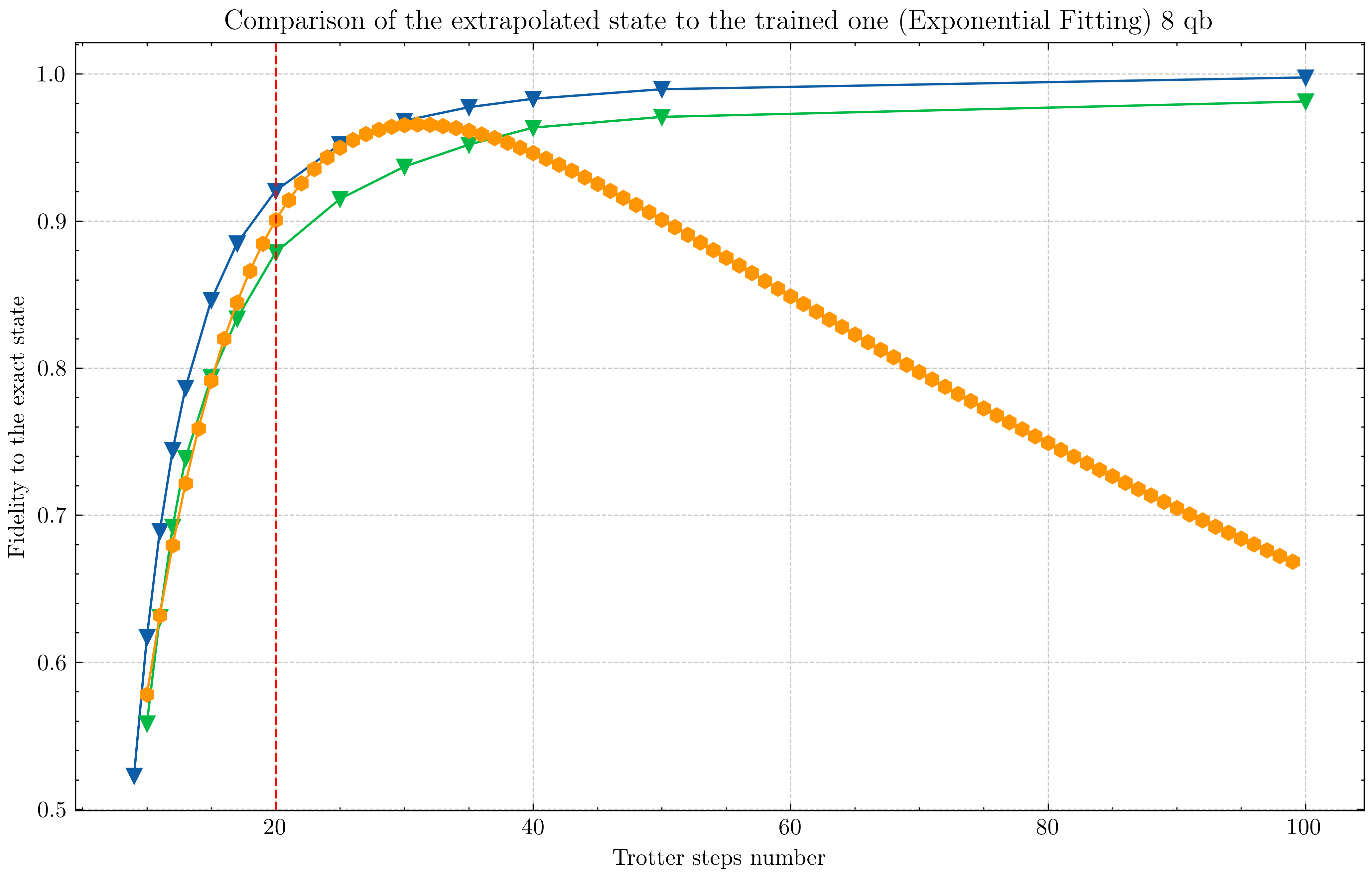}
        \caption{Fidelity of extrapolated states.}
        \label{fig:exp_fid}
    \end{subfigure}
    \caption{Analysis of exponential approximation of ansatz parameters and its impact on state reconstruction quality}
    \label{fig:exp_joint}
\end{figure}

The results show somewhat better behavior. Figure~\ref{fig:exp_param_fit} demonstrates that the smoothing of fluctuations is more pronounced. However, as seen in Figure~\ref{fig:exp_fid}, an undesirable decline in fidelity occurs at large \( r \), indicating the limitations of such models.

\medskip
All the extrapolation methods considered earlier—linear, polynomial, and exponential—were based on treating each ansatz parameter independently as a separate variable. Despite good parameter approximation on the training data, these models exhibit a degradation in quality on the test set, especially at large values of \( r \). This indicates the sensitivity of the quantum state to parameter changes and the potentially incorrect assumption of their independence. Thus, simply approximating each parameter individually is an inadequate approach for long-term extrapolation. To account for internal correlations between parameters, it is necessary to consider the system's dynamics as a unified whole.

\medskip
To overcome these limitations, we apply the Dynamic Mode Decomposition (DMD) method, which allows us to treat the change in ansatz parameters as a coherent collective evolution. Unlike previous approaches, DMD analyzes the entire set of parameters as a single dynamic system and builds a forecast of their behavior based on observed dynamics.

\newpage
\subsubsection*{3. Application of Dynamic Mode Decomposition (DMD) for Extrapolation}

The procedure for forming a forecast using DMD consists of the following steps:

\begin{enumerate}
    \item \textbf{Formation of Snapshot Matrices}. Two matrices are constructed from the ansatz parameters obtained for consecutive values of the number of Trotter steps:
    \[
    \mathbf{X} = \left[\boldsymbol{\theta}^{(r_1)}, \boldsymbol{\theta}^{(r_2)}, \dots, \boldsymbol{\theta}^{(r_{m-1})}\right], \quad
    \mathbf{X}' = \left[\boldsymbol{\theta}^{(r_2)}, \boldsymbol{\theta}^{(r_3)}, \dots, \boldsymbol{\theta}^{(r_{m})}\right].
    \]
    
    \item \textbf{Construction of a Reduced Linear Operator}. Perform the singular value decomposition (SVD) of the matrix \( \mathbf{X} \):
    \[
    \mathbf{X} = \mathbf{U} \mathbf{\Sigma} \mathbf{V}^*,
    \]
    after which the reduced operator is defined by:
    \[
    \tilde{\mathbf{A}} = \mathbf{U}^* \mathbf{X}' \mathbf{V} \mathbf{\Sigma}^{-1}.
    \]
    
    \item \textbf{Dynamics Forecasting}. Compute the eigenvalues \( \lambda_j \) and eigenvectors \( \mathbf{w}_j \) of the operator \( \tilde{\mathbf{A}} \), as well as the eigenmodes \( \mathbf{\phi}_j = \mathbf{U} \mathbf{w}_j \). The forecast for the parameters at future values of the number of Trotter steps is constructed as:
    \[
    \boldsymbol{\theta}^{(r_k)} \approx \sum_{j=1}^{r} \mathbf{\phi}_j \lambda_j^{k-1} b_j = \mathbf{\Phi} \Lambda^{k-1} \mathbf{b},
    \]
    where the coefficient vector \( \mathbf{b} \) is determined by projecting the initial state onto the space of eigenmodes.
\end{enumerate}

\medskip

Thus, DMD enables the prediction of the coherent evolution of all ansatz parameters, taking into account their collective dynamics, making it significantly more effective for long-term extrapolation.             
\chapter{Results}
\label{ch:resuts}

This chapter presents the results of applying the Dynamic Mode Decomposition (DMD) method for extrapolating parameters of variational quantum circuits for the isotropic Heisenberg model. The analysis was conducted for systems of different sizes: \( N = 8, 10 \), and \( 12 \) qubits. Despite the limited number of Trotter steps in the training set (\( r \leq 18 \)), DMD enables effective prediction of parameters for larger values of \( r \), significantly improving the quality of quantum state approximation.

\medskip

\begin{figure}[h!]
    \centering

    \begin{subfigure}[t]{0.85\textwidth}
        \centering
        \includegraphics[width=0.85\textwidth]{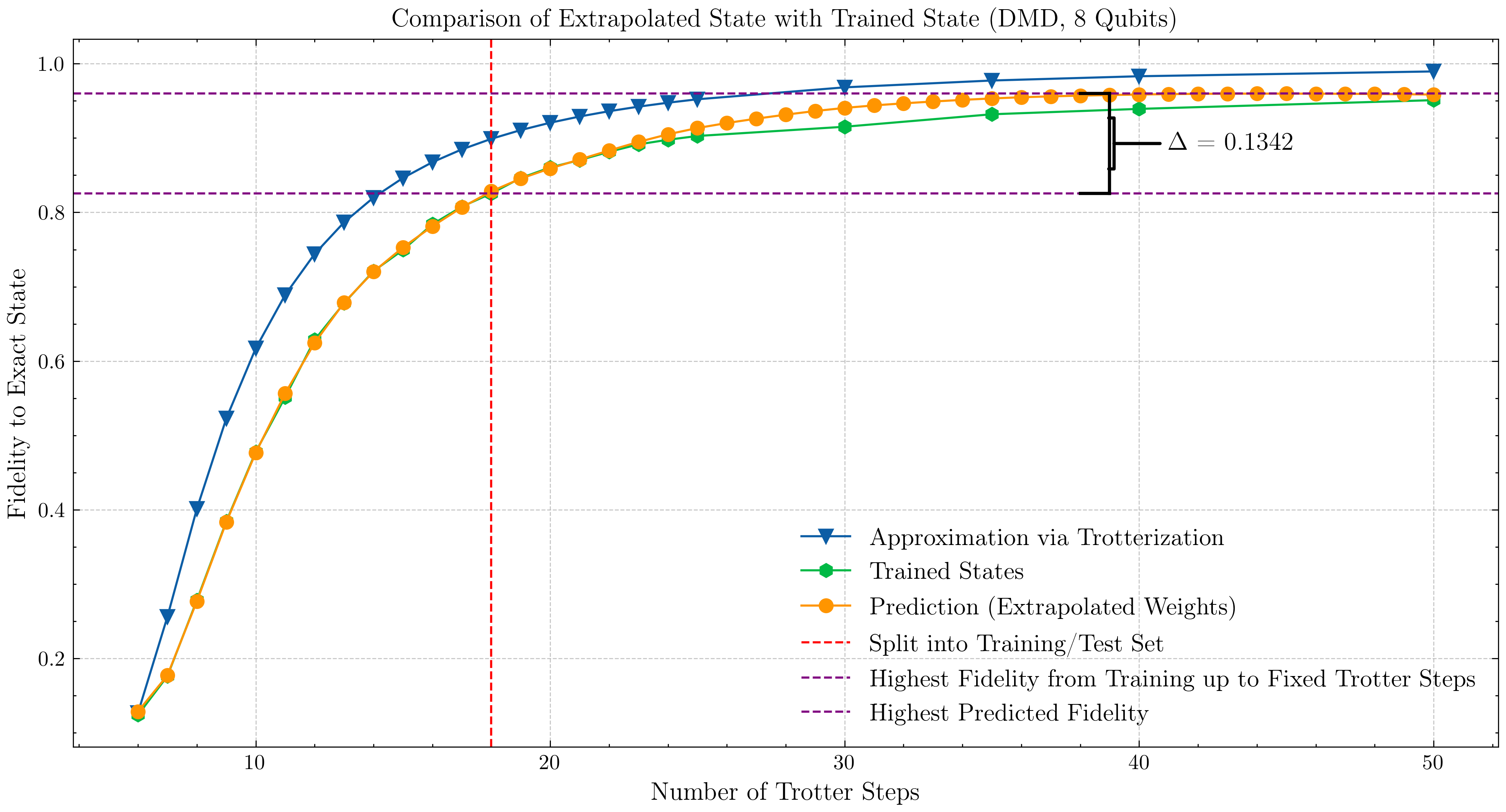}
        \caption{DMD extrapolation (8 qubits), fidelity increase: \(\Delta = 0.1342\)}
        \label{fig:dmd_8qb}
    \end{subfigure}
    \vspace{0.5cm}

    \begin{subfigure}[t]{0.85\textwidth}
        \centering
        \includegraphics[width=0.85\textwidth]{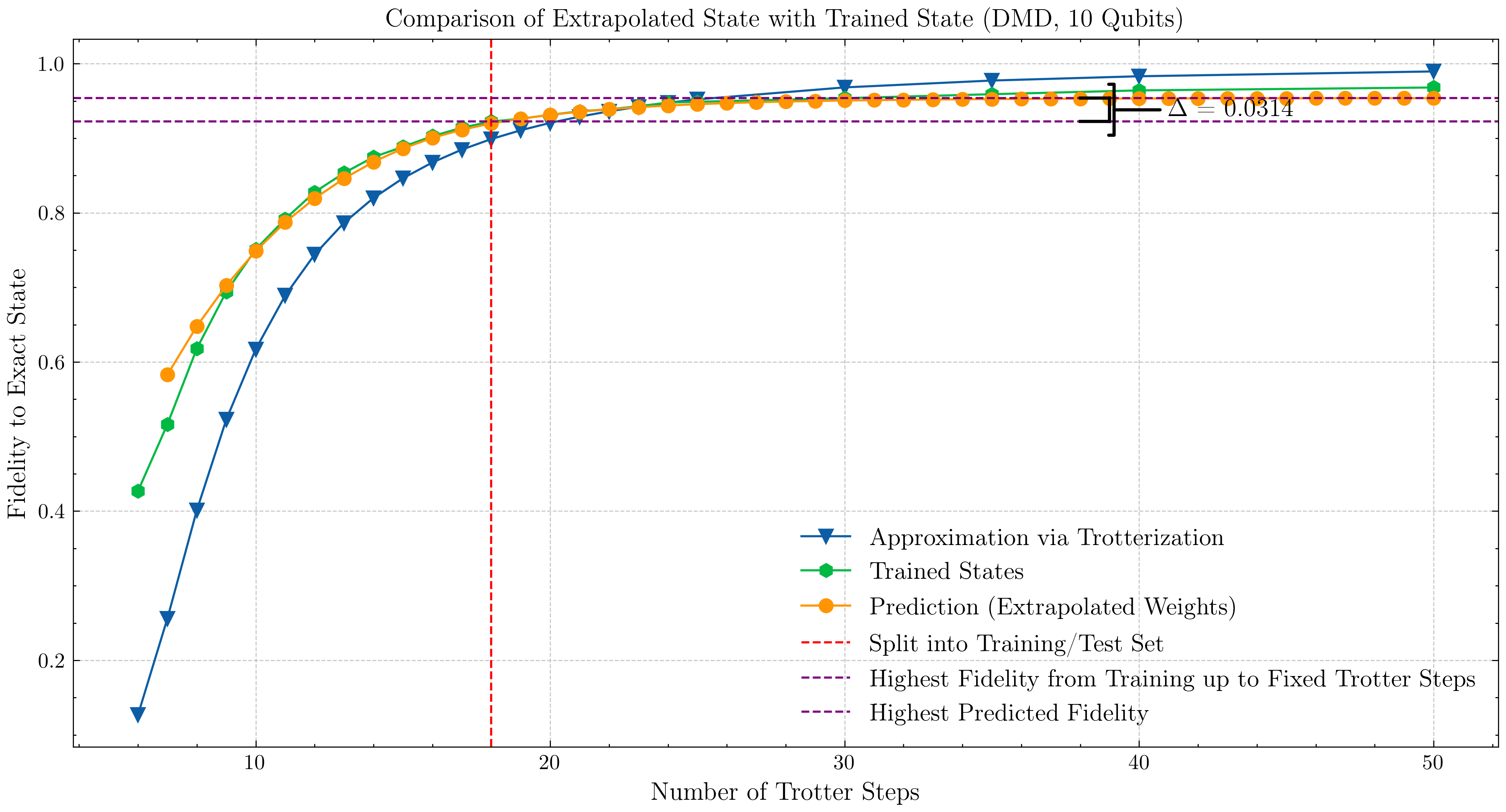}
        \caption{DMD extrapolation (10 qubits), fidelity increase: \(\Delta = 0.0314\)}
        \label{fig:dmd_10qb}
    \end{subfigure}
    \vspace{0.5cm}

    \begin{subfigure}[t]{0.85\textwidth}
        \centering
        \includegraphics[width=0.85\textwidth]{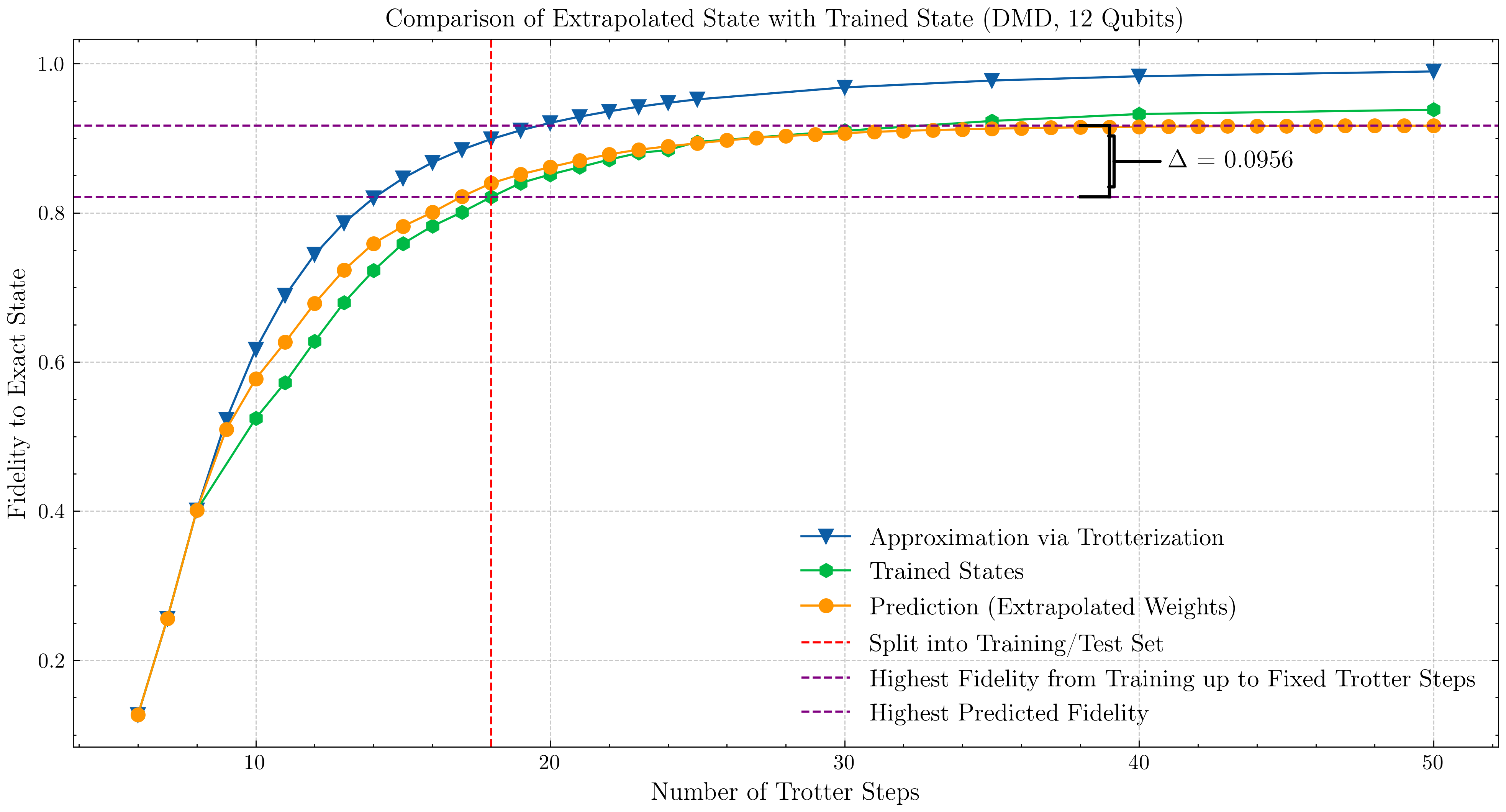}
        \caption{DMD extrapolation (12 qubits), fidelity increase: \(\Delta = 0.0956\)}
        \label{fig:dmd_12qb}
    \end{subfigure}

    \caption{Comparison of fidelity for trained and extrapolated states for \( 8, 10 \), and \( 12 \) qubits}
    \label{fig:dmd_comparison}
\end{figure}

\noindent\textbf{Summary of Results.} Figures~\ref{fig:dmd_8qb}--\ref{fig:dmd_12qb} demonstrate the effectiveness of the DMD-based approach in the task of extrapolating variational circuit parameters. In all experiments, the fidelity to the exact evolved state was evaluated relative to the wave function obtained through direct diagonalization of the Hamiltonian.

Table~\ref{tab:dmd_vs_training_resources} presents the results for three systems with \( N = 8, 10 \), and \( 12 \) qubits at \( r = 18 \) Trotter steps. It compares the accuracy for:
\begin{itemize}
    \item direct Trotterization,
    \item ansatz training,
    \item parameter extrapolation using the DMD method,
\end{itemize}
and also indicates the number of two-qubit operators (CX) in the corresponding circuits. All circuits were transpiled to the hardware basis \( \{\mathrm{R}_x, \mathrm{R}_y, \mathrm{CX}\} \) for an objective comparison of implementation complexity.

\begin{table}[h]
\centering
\small
\begin{tabular}{|c|c|c|c|c|c|}
\hline
\textbf{Qubits} & \textbf{Trotter} & \textbf{Training} & \textbf{DMD} & \textbf{CX (Trotter)} & \textbf{CX (Ansatz)} \\ \hline
8  & 0.89  & 0.83  & \textbf{0.98}  & 420 & \textbf{224}\\ \hline
10 & 0.93  & 0.92  & \textbf{0.95}  & 486 & \textbf{288} \\ \hline
12 & 0.87  & 0.83 & \textbf{0.90}   & 594 & \textbf{352}\\ \hline
\end{tabular}
\caption{Comparison of fidelity and quantum circuit complexity. All circuits were transpiled to the basis \( \{\mathrm{R}_x, \mathrm{R}_y, \mathrm{CX}\} \).}
\label{tab:dmd_vs_training_resources}
\end{table}

\noindent As evident from the table, the DMD method achieves \textbf{higher fidelity} compared to both direct ansatz training and classical Trotterization. Additionally, the ansatz contains significantly fewer two-qubit operators, reducing the quantum circuit depth and making it more suitable for implementation on NISQ devices.
\newpage
\medskip

\newpage
\textbf{Universality of the Approach.} It is noted that all results were obtained at a fixed evolution time \( t \). However, the DMD method is independent of the specific value of \( t \) and is \textbf{general}: changes in time affect only the numerical values of the parameters, not the structure of the Trotterized circuit or the extrapolation algorithm. This makes DMD extrapolation a universal tool for predicting quantum evolution across a wide range of problems.                 
\chapter{Conclusions}

In this work, an approach for the efficient simulation of quantum systems with internal \textit{SU(2)} symmetry was developed and implemented using variational quantum algorithms. The main results can be summarized as follows:

\begin{itemize}
    \item \textit{SU(2)}-equivariant ansatzes were constructed, which, by their structure, preserve the physical symmetry of the studied quantum systems. Detailed analysis showed that such ansatzes provide high expressivity and the ability to generate multipartite entanglement with a small number of layers, making them effective for use on NISQ devices.
    
    \item It was demonstrated that employing an \textit{SU(2)} ansatz in simulating time evolution via Trotterization enables the approximation of Trotterized states with higher accuracy compared to general variational circuits.
    
    \item Complex behavior of the optimized ansatz parameters as a function of the number of Trotter steps \( r \) was observed. This is attributed to the multiplicity of minima in the loss function within the parameter space, leading to irregular dynamics even with high approximation accuracy.
    
    \item A method to address the irregular behavior of parameters was proposed through regularization of the loss function, which ensures smoother and more monotonic parameter changes without significant loss of approximation accuracy.
    
    \item Extrapolation of ansatz parameters based on obtained training data was performed using the Dynamic Mode Decomposition (DMD) method. This approach enables the prediction of parameters for larger values of \( r \) without requiring additional variational training.
    
    \item It was established that using DMD-extrapolated parameters significantly improves the quality of approximation to the exact evolved state. For example, with a fixed number of \( r=18 \) Trotter steps, the increase in fidelity to the exact state ranges from 7--14\%.
    
    \item It was shown that the proposed approach is general and does not depend on a fixed evolution time---time acts solely as a parameter in the Trotterized circuit, altering the numerical values of the parameters without changing the method's structure.
\end{itemize}

The obtained results position the proposed approach as an alternative \textbf{hybrid Trotterization method}, which combines the advantages of a variationally optimized ansatz and the classical construction of Trotterized evolution. By leveraging \textit{SU(2)}-equivariant ansatzes and parameter extrapolation methods via DMD, high accuracy in approximating the exact evolved state is achieved with limited quantum circuit depth. Notably, preserving the system's internal symmetry significantly enhances the quality of approximation, making the proposed approach promising for modeling strongly correlated quantum systems with \textit{SU(2)} symmetry.             

\appendix

\renewcommand{\thechapter}{A}
\chapter{Fundamentals of Lie Groups and Lie Algebras}
\label{sec:appA}

This appendix compiles the basic definitions and statements of the theory of Lie groups and Lie algebras used in the construction and analysis of the algorithms in this work. These concepts are drawn from sources~\cite{Knapp2002,Hall2013,Fuchs1995,Humphreys1972}.

\section{Concepts of Lie Groups and Lie Algebras}
\label{appendix:A1}

\addtocontents{toc}{\protect\setcounter{tocdepth}{1}} 

\subsection{Lie Groups}
A \textit{group} $(G, \cdot)$ is a set $G$ equipped with a binary operation $\cdot: G \times G \to G$, which assigns to each pair of elements $g, h \in G$ an element $g \cdot h \in G$ and satisfies the following axioms:
\begin{itemize}
    \item[(i)] \textbf{Associativity}: For all $g, h, k \in G$, the equality $(g \cdot h) \cdot k = g \cdot (h \cdot k)$ holds;
    \item[(ii)] \textbf{Existence of an identity element}: There exists an element $e \in G$ such that $e \cdot g = g \cdot e = g$ for all $g \in G$;
    \item[(iii)] \textbf{Existence of an inverse element}: For each $g \in G$, there exists an element $g^{-1} \in G$ such that $g \cdot g^{-1} = g^{-1} \cdot g = e$.
\end{itemize}
A group $(G, \cdot)$ is called abelian (or commutative) if the group operation is commutative, i.e., $g \cdot h = h \cdot g$ for all $g, h \in G$.

Groups are a fundamental tool for describing symmetries in mathematics and physics. For example, the group $SO(3)$ represents rotations in three-dimensional space, which are essential transformations in classical and quantum mechanics.

\smallskip

A \textit{Lie group} $(G, \cdot)$ is a group that is also a smooth manifold, with the group operations of multiplication
\begin{equation}
    \cdot\:: G \times G \to G, \quad (g, h) \mapsto g \cdot h,
\end{equation}
and inversion
\begin{equation}
    \iota: G \to G, \quad g \mapsto \iota(g) = g^{-1}
\end{equation}
being smooth mappings~\cite{Hall2013, Fuchs1995}.

The structure of a smooth manifold allows the application of differential geometry methods to analyze continuous symmetries within the group. Lie groups play a central role in quantum mechanics, describing symmetries of physical systems, such as rotations, phase transformations, or spin states.

Classical \emph{examples of Lie groups} include matrix groups of size $n \times n$ over the field of real or complex numbers with the standard matrix multiplication operation and the topology of the space $\mathbb{R}^{n \times n}$ or $\mathbb{C}^{n \times n}$, respectively, such as:
\begin{itemize}
    \item[(a)] $GL(n, \mathbb{R})$: The \textit{general linear} group of invertible real $n \times n$ matrices;
    \item[(b)] $SL(n, \mathbb{R})$: The \textit{special linear} group of real matrices $U \in GL(n, \mathbb{R})$ with the condition $\det U = 1$;
    \item[(c)] $U(n)$: The group of \textit{unitary} $n \times n$ matrices, i.e., matrices $U \in \mathbb{C}^{n \times n}$ satisfying $U U^{\dagger} = I$;
    \item[(d)] $SU(n)$: The \textit{special unitary} group of matrices $U \in U(n)$ with the condition $\det U = 1$;
    \item[(e)] $SO(n)$: The \textit{special orthogonal} group of $n \times n$ matrices preserving the Euclidean metric; it includes matrices $U \in \mathbb{R}^{n \times n}$ such that $U U^\top = I$ and $\det U = 1$ (the real analogue of the group $SU(n)$).
\end{itemize}

In quantum computing, the groups $SU(2)$ and $SU(4)$ describe unitary operators that implement single- and two-qubit quantum gates~\cite{Nielsen2000}.

\subsection{Lie Algebras}

Associated with every Lie group $G$ is its \textit{Lie algebra} $\mathfrak{g}$, which is the tangent space at the identity element $e \in G$, equipped with an operation $[\cdot, \cdot]: \mathfrak{g} \times \mathfrak{g} \to \mathfrak{g}$, called the \textit{Lie bracket}. This operation has the following properties:
\begin{enumerate}
    \item[(i)] \textbf{Antisymmetry}: $[X, Y] = -[Y, X]$;
    \item[(ii)] \textbf{Bilinearity}: $[a X + b Y, Z] = a [X, Z] + b [Y, Z]$ for $a, b \in \mathbb{R}$;
    \item[(iii)] \textbf{Jacobi identity}: $[X, [Y, Z]] + [Y, [Z, X]] + [Z, [X, Y]] = 0$.
\end{enumerate}

A Lie algebra is a linear vector space, and thus concepts such as linear combinations and bases are defined for it. A \textit{Lie subalgebra} $\mathfrak{h}$ of a Lie algebra is a linear subspace of $\mathfrak{g}$ that is closed under the Lie bracket, i.e., $[\mathfrak{h}, \mathfrak{h}] \subset \mathfrak{h}$.

For matrix Lie groups, the Lie bracket is the commutator
\begin{equation}
    [X, Y] = X Y - Y X, \quad X, Y \in \mathfrak{g}.
\end{equation}
Properties (i) and (ii) are evident, and the Jacobi identity can be verified through direct computation.

The Lie algebra $\mathfrak{g}$ describes the local structure of the Lie group and generates its infinitesimal transformations via the exponential map $\exp: \mathfrak{g} \to G$, where $\exp(tX) = e^{t X} \in G$ for $X \in \mathfrak{g}$ and $t \in \mathbb{R}$. In physics, elements of the Lie algebra $\mathfrak{g}$ correspond to generators of groups and symmetries, such as angular momentum or spin operators.

\subsection{Lie Algebra $\mathfrak{su}(2)$ of the Group $SU(2)$}

The special unitary group $SU(2)$ consists of $2 \times 2$ unitary matrices with determinant $1$:
\begin{equation}
    SU(2) = \{ U \in \mathbb{C}^{2 \times 2} \mid U U^{\dagger} = I, \det U = 1 \}.
\end{equation}
Its Lie algebra $\mathfrak{su}(2)$ is the set of skew-Hermitian matrices with zero trace:
\begin{equation}
    \mathfrak{su}(2) = \{ X \in \mathbb{C}^{2 \times 2} \mid X^{\dagger} = -X, \text{Tr}(X) = 0 \}.
\end{equation}
Indeed, the eigenvalues of any matrix $U \in SU(2)$ lie on the unit circle in the complex plane and are $e^{i\theta}$ and $e^{-i\theta}$. The corresponding eigenvectors $\mathbf{v}_1$ and $\mathbf{v}_2$ can be chosen to form an orthonormal basis of $\mathbb{C}^2$, and then the matrix $V = [\mathbf{v}_1, \mathbf{v}_2]$ is unitary. In this basis, we have:
\begin{equation*}
    U = V \begin{pmatrix} e^{i\theta} & 0 \\ 0 & e^{-i\theta} \end{pmatrix} V^\dagger.
\end{equation*}
Direct computations show that the derivative at zero of the trajectory
\begin{equation*}
    U(t) = V \begin{pmatrix} e^{it\theta} & 0 \\ 0 & e^{-it\theta} \end{pmatrix} V^\dagger \in SU(2)
\end{equation*}
is
\begin{equation*}
    U'(0) = V \begin{pmatrix} i\theta & 0 \\ 0 & -i\theta \end{pmatrix} V^\dagger,
\end{equation*}
which is a skew-Hermitian matrix with zero trace. Clearly, every skew-Hermitian matrix with zero trace has such a representation and is thus an element of the Lie algebra $\mathfrak{su}(2)$.

The basis elements of $\mathfrak{su}(2)$ are the generators:
\begin{equation}\label{eq:su2-T}
    T_x = \frac{1}{2} \begin{pmatrix} 0 & i \\ i & 0 \end{pmatrix}, \quad
    T_y = \frac{1}{2} \begin{pmatrix} 0 & -1 \\ 1 & 0 \end{pmatrix}, \quad
    T_z = \frac{1}{2} \begin{pmatrix} i & 0 \\ 0 & -i \end{pmatrix},
\end{equation}
which can be expressed in terms of the Pauli matrices as $T_j = \frac{i}{2} \sigma_j$. They satisfy the commutation relations:
\begin{equation}\label{eq:commut-su2}
    [T_x, T_y] = T_z, \quad [T_y, T_z] = T_x, \quad [T_z, T_x] = T_y.
\end{equation}
These relations are analogous to those describing angular momentum or spin operators in quantum mechanics.

An element $U \in SU(2)$ can be parameterized using the exponential map:
\begin{equation}
    U = e^{\theta\, \hat{n} \cdot \mathbf{T}} = \cos\left(\frac{\theta}{2}\right) I + i \sin\left(\frac{\theta}{2}\right) (\hat{n} \cdot \boldsymbol{\sigma}),
\end{equation}
where $\theta$ is the rotation angle, $\hat{n}$ is a unit vector, and $\boldsymbol{\sigma} = (\sigma_x, \sigma_y, \sigma_z)$.

This form is widely used in quantum computing to describe single-qubit rotations on the Bloch sphere.

\subsection{Lie Algebra $\mathfrak{so}(3)$ of the Group $SO(3)$}

The group $SO(3)$ is the group of rotations in $\mathbb{R}^3$, and its Lie algebra $\mathfrak{so}(3)$ is the subspace of skew-symmetric matrices within the space of all real matrices $\mathbb{R}^{n \times n}$. It is generated by the elements
\begin{equation*}
    E = \begin{pmatrix}
        0 & 1 & 0 \\ -1 & 0 & 0 \\ 0 & 0 & 0
    \end{pmatrix}, 
    \quad 
    F = \begin{pmatrix}
        0 & 0 & 0 \\ 0 & 0 & 1 \\ 0 & -1 & 0
    \end{pmatrix},
    \quad
    G = \begin{pmatrix}
        0 & 0 & 1 \\ 0 & 0 & 0 \\ -1 & 0 & 0
    \end{pmatrix}.
\end{equation*}
The commutation relations
\begin{equation}
    [E, F] = G, \quad [F, G] = E, \quad [G, E] = F
\end{equation}
are identical to those in~\eqref{eq:commut-su2}, and thus the Lie algebra $\mathfrak{so}(3)$ is \textit{isomorphic} to the Lie algebra $\mathfrak{su}(2)$, which we denote as $\mathfrak{so}(3) \cong \mathfrak{su}(2)$.
\section{Structure of Lie Algebras}
\label{appendix:A2}

\subsection{Abelian Lie Algebras}

A Lie algebra $\mathfrak{g}$ is called \textit{abelian} (or \textit{commutative}) if all its elements commute:
\begin{equation}
    \forall X, Y \in \mathfrak{g}: \quad [X, Y] = 0.
\end{equation}
Abelian Lie algebras are the simplest, and their structure is equivalent to that of linear vector spaces. For example, the algebra $\mathbb{R}^n$ with the operation $[X, Y] = 0$ is an abelian Lie algebra corresponding to the torus $\mathbb{T}^n$ as a Lie group.

The \textit{center} $Z(\mathfrak{g})$ of a Lie algebra $\mathfrak{g}$ is the subset of elements that commute with all other elements of the algebra~\cite{Humphreys1972, Hall2013}:
\begin{equation}
    Z(\mathfrak{g}) = \{ X \in \mathfrak{g} \mid [X, Y] = 0, \ \forall Y \in \mathfrak{g} \}.
\end{equation}
The center $Z(\mathfrak{g})$ is an (abelian) subalgebra of $\mathfrak{g}$: the fact that $Z(\mathfrak{g})$ is a linear subspace follows from the linearity of the Lie bracket, and its closure under the Lie bracket follows from the definition of the center.

\smallskip

\emph{Example}: The center $Z(\mathfrak{su}(2))$ of the Lie algebra $\mathfrak{su}(2)$ is trivial, i.e., $Z(\mathfrak{su}(2)) = \{0\}$, since no non-zero element $c_x T_x + c_y T_y + c_z T_z$ can commute with all basis elements $T_x$, $T_y$, and $T_z$.

\smallskip

A general Lie algebra may contain other abelian subalgebras; the maximal commutative subalgebra of semisimple elements of a Lie algebra $\mathfrak{g}$ is called its \textit{Cartan subalgebra} (see Section~\ref{A:Cartan-subalgebra}).

\subsection{Ideals of Lie Algebras}

An \textit{ideal} $\mathfrak{h}$ of a Lie algebra $\mathfrak{g}$ is a subalgebra of $\mathfrak{g}$ that satisfies the property
\begin{equation}
    \forall X \in \mathfrak{g},\ \forall Y \in \mathfrak{h}: \qquad [X, Y] \in \mathfrak{h}.
\end{equation}

Each element $X \in \mathfrak{g}$ generates an \textit{adjoint linear map} $\mathrm{ad}_X : \mathfrak{g} \to \mathfrak{g}$ defined by $\mathrm{ad}_X(Y) = [X, Y]$; by definition, an ideal $\mathfrak{h}$ is an invariant subspace of the linear operator $\mathrm{ad}_X$ for each $X \in \mathfrak{g}$, i.e., $\mathrm{ad}_X (\mathfrak{h}) \subset \mathfrak{h}$.

An ideal $\mathfrak{h}$ of a Lie algebra $\mathfrak{g}$ is called \textit{solvable} if, for some finite $k$, we have $\mathfrak{h}^{(k)} = \{0\}$; here, $\mathfrak{h}^{(0)} \supset \mathfrak{h}^{(1)} \supset \mathfrak{h}^{(2)} \supset \dots$ is the derived chain of nested ideals:
\begin{equation}\label{eq:derived-chain}
    \mathfrak{h} = \mathfrak{h}^{(0)}, \quad \mathfrak{h}^{(k+1)} = [\mathfrak{h}^{(k)}, \mathfrak{h}^{(k)}] \quad \text{for} \quad k = 0, 1, \dots
\end{equation}
The maximal solvable ideal of a Lie algebra $\mathfrak{g}$ is called its \textit{radical}. The radical of a Lie algebra is uniquely defined and is the union of all its solvable ideals; it is a ``simple'' structural component of the algebra, as discussed below.

\smallskip

For example, the center $Z(\mathfrak{g})$ of a Lie algebra $\mathfrak{g}$ is a solvable ideal. Any abelian ideal of a Lie algebra $\mathfrak{g}$ is solvable. If a Lie algebra contains a non-trivial solvable ideal, then in the corresponding derived chain~\eqref{eq:derived-chain}, the last non-zero element is an abelian ideal. Thus, the following three properties are equivalent:
\begin{itemize}
    \item The radical of the Lie algebra $\mathfrak{g}$ is non-trivial;
    \item The Lie algebra $\mathfrak{g}$ contains a non-trivial solvable ideal;
    \item The Lie algebra $\mathfrak{g}$ contains a non-trivial abelian ideal.
\end{itemize}

\subsection{Simple and Semisimple Lie Algebras}

\textbf{Definition 1.} A Lie algebra $\mathfrak{g}$ is \textit{simple} if it does not contain non-trivial ideals and is not abelian.

An example of a simple Lie algebra is $\mathfrak{su}(2)$: it is not abelian, and if $G = c_x T_x + c_y T_y + c_z T_z$ is an element of a non-trivial ideal $\mathfrak{h}$ with, for example, $c_z \neq 0$, then the equality $[[G, T_y], T_z] = c_z T_y$ implies that $T_y \in \mathfrak{h}$; hence, the basis elements $T_x$, $T_y$, and $T_z$ belong to $\mathfrak{h}$, and thus $\mathfrak{h} = \mathfrak{g}$ (see~\cite[Proposition~3.12]{Hall2013}).

\smallskip
\noindent\textbf{Definition 2.} A Lie algebra $\mathfrak{g}$ is \textit{semisimple} if it does not contain non-zero solvable ideals.

\smallskip
Every simple Lie algebra is also semisimple. Simple Lie algebras are the ``building blocks'' for semisimple Lie algebras, as every semisimple Lie algebra can be expressed as a direct sum of simple Lie algebras. An example of a semisimple Lie algebra that is not simple is $\mathfrak{so}(4)$, since
\begin{equation}\label{eq:so4-decomposition}
    \mathfrak{so}(4) \cong \mathfrak{so}(3) \oplus \mathfrak{so}(3) \cong \mathfrak{su}(2) \oplus \mathfrak{su}(2).
\end{equation}
Indeed, let
\begin{equation*}
    E_x = \begin{pmatrix}
        0 & 1 & 0 & 0 \\ -1 & 0 & 0 & 0 \\ 0 & 0 & 0 & 1 \\ 0 & 0 & -1 & 0
    \end{pmatrix},    
    \qquad
    E_y = \begin{pmatrix}
        0 & 0 & 0 & \ 1 \\ 0 & 0 & 1 & \ 0 \\ 0 & -1 & 0 & \ 0 \\ -1 & 0 & 0 & \ 0
    \end{pmatrix},
    \qquad
    E_z = \begin{pmatrix}
        0 & 0 & \ 1 & 0 \\ 0 & 0 & \ 0 & -1 \\ -1 & 0 & \ 0 & 0 \\ 0 & 1 & \ 0 & 0
    \end{pmatrix},
\end{equation*}
and let
\begin{equation*}
    A = \begin{pmatrix}
        0 & a & b & c \\ -a & 0 & d & e \\ -b & -d & 0 & f \\ -c & -e & -f & 0
    \end{pmatrix}
\end{equation*}
be an arbitrary element of $\mathfrak{so}(4)$. Direct computations show that
\begin{equation*}
    [E_x, A] = (e - b) E_y + (c + d) E_z,
\end{equation*}
from which we obtain the equalities
\begin{equation*}
    [E_x, E_y] = 2 E_z, \qquad [E_z, E_x] = 2 E_y.
\end{equation*}
Similar reasoning yields the equality $[E_y, E_z] = 2 E_x$, and thus the subspace $\mathfrak{h}$ generated by the elements $E_x$, $E_y$, $E_z$ forms a subalgebra of $\mathfrak{so}(4)$, isomorphic to the simple Lie algebra $\mathfrak{su}(2)$. Moreover, this subalgebra forms an ideal, and its complement with respect to the Lie bracket is also isomorphic to the Lie algebra $\mathfrak{su}(2)$ (generated by elements similar to $E_x$, $E_y$, and $E_z$, but with a different distribution of signs $\pm$), which gives the decomposition~\eqref{eq:so4-decomposition}.

Thus, the Lie algebra $\mathfrak{so}(4)$ is not simple. From the decomposition~\eqref{eq:so4-decomposition}, it follows that this algebra contains no other ideals, and hence it is semisimple.

\subsection{Killing Form and Semisimplicity Criterion}

Semisimple Lie algebras have important properties, such as the existence of a non-degenerate bilinear form (the Killing form), which allows their classification using Dynkin diagrams. The Killing form is invariant under automorphisms of the algebra (i.e., isomorphisms preserving the Lie bracket) and is an effective tool for determining orthogonality in the Cartan decomposition and classifying Lie algebras.

The \textit{Killing form} $B$ on a Lie algebra $\mathfrak{g}$ is a symmetric bilinear form defined by the equality
\begin{equation}\label{eq:Killing}
    B(X, Y) = \mathrm{Tr}(\mathrm{ad}_X \circ \mathrm{ad}_Y), \quad X, Y \in \mathfrak{g},
\end{equation}
where the adjoint representation operator $\mathrm{ad}_X : \mathfrak{g} \to \mathfrak{g}$ is a linear transformation acting as $\mathrm{ad}_X(Y) = [X, Y]$, and $\mathrm{Tr}$ is the matrix trace of a linear operator. In a fixed basis $X_1, X_2, \dots, X_N$ of the Lie algebra $\mathfrak{g}$, the operators $\mathrm{ad}_X$ and $\mathrm{ad}_Y$ correspond to $N \times N$ matrices, and the composition $\mathrm{ad}_X \circ \mathrm{ad}_Y$ corresponds to their product, the trace of which computes the Killing form~\eqref{eq:Killing}.

\smallskip

\textbf{Example.} Consider the Lie algebra $\mathfrak{su}(2)$ with generators $T_x$, $T_y$, $T_z$. In this basis, the matrices of the linear transformations $\mathrm{ad}_{T_j}$ are
\begin{equation*}
    \mathrm{ad}_{T_x} = \begin{pmatrix}
        0 & 0 & 0 \\ 0 & 0 & -1 \\ 0 & 1 & 0
    \end{pmatrix}, \quad
    \mathrm{ad}_{T_y} = \begin{pmatrix}
        \hphantom{-}0 & 0 & 1 \\ \hphantom{-}0 & 0 & 0 \\ -1 & 0 & 0
    \end{pmatrix}, \quad
    \mathrm{ad}_{T_z} = \begin{pmatrix}
        0 & -1 & 0 \\ 1 & 0 & 0 \\ 0 & 0 & 0
    \end{pmatrix}.
\end{equation*}
The Killing form $B$ in this basis is given by the matrix $-2 I_3$, i.e., if
\begin{equation*}
    X = c_x T_x + c_y T_y + c_z T_z, \qquad Y = d_x T_x + d_y T_y + d_z T_z,
\end{equation*}
then
\begin{equation*}
    B(X, Y) = -2 (c_x d_x + c_y d_y + c_z d_z) = 4 \mathrm{Tr} (X Y).
\end{equation*}

\medskip

\textbf{Theorem (Cartan’s Criterion).} A Lie algebra $\mathfrak{g}$ is semisimple if and only if its Killing form $B$ is non-degenerate, i.e., $B(X, Y) = 0$ for all $Y \in \mathfrak{g}$ only if $X = 0$.

\smallskip

Indeed, if the Killing form is degenerate, then the linear subspace $S$ of elements $X \in \mathfrak{g}$ such that $B(X, Y) = 0$ for all $Y \in \mathfrak{g}$ is an ideal of the algebra. This follows from the identity
\begin{equation*}
    B([X, Z], Y) + B(X, [Y, Z]) = 0,
\end{equation*}
which expresses the invariance of the Killing form with respect to the action of the adjoint map $\mathrm{ad}$; if we take $X \in S$ and $Z \in \mathfrak{g}$ in this identity, we obtain $[X, Z] \in S$. It can be further shown~\cite[Theorem~1.45]{Knapp2002} that the ideal $S$ is solvable, and thus the Lie algebra $\mathfrak{g}$ is not semisimple.

Conversely, if the algebra $\mathfrak{g}$ is not semisimple, it contains a non-trivial abelian ideal $\mathfrak{p}$ — the last non-zero ideal in the derived chain of inclusions $\mathfrak{p}^{(0)} \supset \mathfrak{p}^{(1)} \supset \dots$~\eqref{eq:derived-chain}. Let $\mathfrak{h}$ be any complement of the linear subspace $\mathfrak{p}$ in the linear space $\mathfrak{g}$. By construction, for each $X \in \mathfrak{p}$, we have $\mathrm{ad}_X(\mathfrak{p}) = 0$ and $\mathrm{ad}_X(\mathfrak{h}) \subset \mathfrak{p}$, and for any $Y \in \mathfrak{h}$, the inclusion $\mathrm{ad}_Y(\mathfrak{p}) \subset \mathfrak{p}$ holds. Thus,
\begin{equation*}
    \mathrm{ad}_X \circ \mathrm{ad}_Y (\mathfrak{p}) = \{0\}, \qquad \mathrm{ad}_X \circ \mathrm{ad}_Y(\mathfrak{h}) \subset \mathfrak{p},
\end{equation*}
and hence the matrix $\mathrm{ad}_X \circ \mathrm{ad}_Y$ has a zero diagonal in a basis consistent with the decomposition $\mathfrak{g} = \mathfrak{p} \dotplus \mathfrak{h}$, and $B(X, Y) = 0$.

\smallskip

\textbf{Example.} As shown above, in the Lie algebra $\mathfrak{su}(2)$, the Killing form is given by $B(X, Y) = 4 \text{Tr}(X Y)$ and is non-degenerate, and thus, by Cartan’s criterion, the Lie algebra $\mathfrak{su}(2)$ is semisimple.         
\section{Cartan Decomposition and Related Structures}
\label{appendix:A3}

\subsection{Cartan--Killing Theorem}

For semisimple Lie algebras, the Killing form defines an inner product that allows the algebra to be decomposed into orthogonal components. The Cartan--Killing theorem (also known as the Cartan decomposition) is a key tool for studying the structure of Lie algebras, particularly in the context of distinguishing their compact and non-compact subalgebras.

\smallskip

\textbf{Theorem 2.} Let $\mathfrak{g}$ be a semisimple Lie algebra over the field $\mathbb{R}$. Then there exists a decomposition of the linear space
\begin{equation}\label{eq:Cartan-Killing}
    \mathfrak{g} = \mathfrak{k} \oplus \mathfrak{p},
\end{equation}
orthogonal with respect to the Killing form (i.e., $B(\mathfrak{k}, \mathfrak{p}) = 0$), where
\begin{itemize}
    \item[(i)] $\mathfrak{k}$ is the maximal compact Lie subalgebra;
    \item[(ii)] $\mathfrak{p}$ is the orthogonal complement of $\mathfrak{k}$ with respect to the Killing form.
\end{itemize}
In this case, the Killing form is negative definite on $\mathfrak{k}$ and positive definite on $\mathfrak{p}$; additionally, the relations $[\mathfrak{k}, \mathfrak{k}] \subseteq \mathfrak{k}$, $[\mathfrak{k}, \mathfrak{p}] \subseteq \mathfrak{p}$, and $[\mathfrak{p}, \mathfrak{p}] \subseteq \mathfrak{k}$ hold.

\smallskip

In item (i), a compact Lie algebra is understood as the Lie algebra of a compact Lie group. The decomposition~\eqref{eq:Cartan-Killing} is analogous to the decomposition of matrices into antisymmetric ($\mathfrak{k}$) and symmetric ($\mathfrak{p}$) parts for algebras such as $\mathfrak{so}(n)$ and $\mathfrak{su}(n)$, and in the context of Lie groups, it is a generalization of the polar decomposition of a matrix, representing it as a product of a unitary and a non-negative Hermitian component. The Cartan decomposition also plays a significant role in the theory of symmetric spaces and the classification of Lie algebras, and in the context of quantum computing, it is applied to analyze symmetries in quantum circuits.

\smallskip

\textbf{Example:} For the Lie algebra $\mathfrak{so}(3)$ of the orthogonal group $SO(3)$, which describes rotations in three-dimensional space, the Cartan decomposition is trivial: $\mathfrak{so}(3) = \mathfrak{so}(3) \oplus \{0\}$, since the entire algebra is compact.

The Lie algebra $\mathfrak{sl}(2, \mathbb{R})$ of the Lie group $SL(2, \mathbb{R})$ consists of all real $2 \times 2$ matrices with zero trace and is generated by the matrices
\begin{equation*}
    H = \begin{pmatrix} 1 & \hphantom{-}0 \\ 0 & -1 \end{pmatrix}, \quad
    E_+ = \begin{pmatrix} 0 & 1 \\ 0 & 0 \end{pmatrix}, \quad
    E_- = \begin{pmatrix} \hphantom{-}0 & 0 \\ -1 & 0 \end{pmatrix}
\end{equation*}
with commutation relations
\begin{equation*}
    [H, E_+] = 2 E_+, \qquad [H, E_-] = -2 E_-, \qquad [E_+, E_-] = -H.
\end{equation*}
It follows that $\mathfrak{sl}(2, \mathbb{R})$ is semisimple: if $G = a H + b_+ E_+ + b_- E_-$ is an element of a non-trivial ideal $\mathfrak{h}$ and, for example, $a \neq 0$, then the equality $[[G, E_\pm], H] = -4 a E_\pm$ together with $[E_+, E_-] = H$ shows that all basis elements belong to $\mathfrak{h}$, and thus $\mathfrak{h} = \mathfrak{sl}(2, \mathbb{R})$; if, however, $b_\pm \neq 0$, then $[[G, H], E_-] = 2 b_\pm H$ belongs to the ideal, along with $E_+$ and $E_-$, and again we obtain $\mathfrak{h} = \mathfrak{sl}(2, \mathbb{R})$. The Cartan decomposition of the algebra $\mathfrak{sl}(2, \mathbb{R})$ has the form
\begin{equation}
    \mathfrak{sl}(2, \mathbb{R}) = \mathfrak{so}(2) \oplus \mathfrak{p},
\end{equation}
where $\mathfrak{so}(2)$ is the compact Lie subalgebra of the rotation group $SO(2)$, i.e., the subspace of skew-symmetric matrices, and $\mathfrak{p}$ is the subspace of symmetric matrices with zero trace.

\subsection{Cartan Subalgebra and KHK Decomposition of a Lie Group}\label{A:Cartan-subalgebra}

\textbf{Definition 1.} An element $X$ of a Lie algebra $\mathfrak{g}$ is called \textit{semisimple} if the operator $\mathrm{ad}_X$ is diagonalizable. A \textit{Cartan subalgebra} $\mathfrak{h}$ of a Lie algebra $\mathfrak{g}$ is the maximal commutative subalgebra of $\mathfrak{g}$ consisting of its semisimple elements, i.e., the maximal subset $\mathfrak{h}$ of semisimple elements for which
\begin{equation}
    \forall X, Y \in \mathfrak{h}: \qquad [X, Y] = 0.
\end{equation}
From the definition, it follows that for a Cartan subalgebra $\mathfrak{h}$, the corresponding set of operators $\{\mathrm{ad}_X \mid X \in \mathfrak{h}\}$ can be simultaneously diagonalized.

\smallskip

\noindent\textbf{Example:} In the Lie algebra $\mathfrak{su}(2)$ of $2 \times 2$ skew-Hermitian matrices with zero trace, the standard Cartan subalgebra $\mathfrak{h}$ is generated by the element $T_z$ from~\eqref{eq:su2-T}. Its dimension is 1, reflecting the maximal commutativity in this algebra of dimension 3. Note that every non-zero element of $\mathfrak{su}(2)$ generates a one-dimensional abelian subalgebra of this algebra, so the Cartan subalgebra is not uniquely defined. However, all Cartan subalgebras of a Lie algebra $\mathfrak{g}$ have the same dimension and are conjugate to one another, i.e., for any pair of Cartan subalgebras $\mathfrak{h}_1$, $\mathfrak{h}_2$, there exists an element $X \in G$ of the corresponding Lie group such that $\mathfrak{h}_2 = X \mathfrak{h}_1 X^{-1}$.

\smallskip

The Cartan subalgebra plays a key role in studying the structure of a Lie algebra, allowing it to be decomposed into subspaces defined by the commutation properties of its elements. It is also closely related to the so-called $KHK$ decomposition of the associated semisimple Lie group $G = \exp(i \mathfrak{g})$.

Denote by $\mathfrak{h} \subset \mathfrak{p}$ the maximal abelian subspace of the space $\mathfrak{p}$ from the Cartan decomposition~\eqref{eq:Cartan-Killing}, and by $H = \exp(i \mathfrak{h})$ the corresponding abelian subgroup in $G$. Let $K$ be the maximal compact subgroup corresponding to the subalgebra $\mathfrak{k}$ of the Cartan decomposition~\eqref{eq:Cartan-Killing}; then the so-called $KHK$ decomposition
\[
G = KHK
\]
of the group $G$ holds, meaning that every element $g \in G$ can be expressed as
\begin{equation}
    g = k_1 h k_2,
\end{equation}
where $k_1, k_2 \in K$, $h \in H$. For a Hermitian Hamiltonian $H$, the corresponding $KHK$ decomposition of the unitary operator $e^{i t H}$ can be written as
\begin{equation}
    e^{i t H} = k e^{i t H_0} k^\dagger.
\end{equation}

\subsection{Root Space Decomposition of a Lie Algebra}

The roots of a Lie algebra are a tool for classifying and analyzing the structure of semisimple Lie algebras through subspaces associated with Cartan subalgebras. Let $\mathfrak{g}$ be a semisimple Lie algebra with a Cartan subalgebra $\mathfrak{h}$, and let $\mathfrak{h}^*$ be the dual space of linear functionals on $\mathfrak{h}$. For each $\alpha \in \mathfrak{h}^*$, consider the subspace
\begin{equation}
    \mathfrak{g}_\alpha = \{ X \in \mathfrak{g} \mid [H, X] = \alpha(H) X, \forall H \in \mathfrak{h} \}.
\end{equation}

\noindent
\textbf{Definition 3.} A linear functional $\alpha \in \mathfrak{h}^*$ is a \textit{root} if the corresponding space $\mathfrak{g}_\alpha$ is non-trivial; then $\mathfrak{g}_\alpha$ is called a \textit{root space}. Note that $\mathfrak{h}$ is a root space with the zero root $\alpha_0$: $\alpha_0(H) = 0$.

\smallskip

\noindent
\textbf{Theorem (Root Space Decomposition)} For a semisimple Lie algebra $\mathfrak{g}$ with a Cartan subalgebra $\mathfrak{h}$, there exists a decomposition
\begin{equation}
    \mathfrak{g} = \mathfrak{h} \oplus \sum_{\alpha \in \Delta} \mathfrak{g}_\alpha,
\end{equation}
where $\Delta$ is the set of non-zero roots.

\smallskip

Note that $[\mathfrak{g}_\alpha, \mathfrak{g}_\beta] \subset \mathfrak{g}_{\alpha - \beta}$; in particular, $[\mathfrak{g}_\alpha, \mathfrak{g}_\alpha] \subset \mathfrak{h}$.

\smallskip

\noindent\textbf{Example:} Consider the Lie algebra $\mathfrak{su}(2)$ with the Cartan subalgebra $\mathfrak{h} = \text{span}\{T_z\}$. For an element $X = a T_x + b T_y + c T_z$ from a root space $\mathfrak{g}_\alpha$, we have $[T_z, X] = \alpha(T_z) X = -b T_x + a T_y$. From this, we determine two roots $\alpha_\pm$, defined by the equalities $\alpha_\pm(T_z) = \pm i$, and the corresponding root spaces are generated by the elements $T_x \mp i T_y$, which correspond to the raising and lowering spin operators.

The root system forms the basis for the classification of Lie algebras using Dynkin diagrams, which is fundamental in the theory of representations of Lie algebras and Lie groups.      
\section{Schur--Weyl Duality and Representations of the Symmetric Group}
\label{appendix:A4}

Schur--Weyl duality connects the actions of the unitary group $SU(N)$ and the symmetric group $S_n$ on tensor products of vector spaces and is of significant importance in quantum mechanics, quantum computing, and combinatorics~\cite{Sagan2001, Goodman2009}.

\subsection{Schur--Weyl Theorem}

\textbf{Theorem 1 (Schur--Weyl Duality).} Let $V = \mathbb{C}^N$ be a vector space of dimension $N$, and let $V^{\otimes n}$ be its $n$-fold tensor product. Then
\begin{equation}\label{eq:Schur}
    V^{\otimes n} \cong \bigoplus_{\lambda} S^\lambda(\mathbb{C}^N) \otimes M_\lambda,
\end{equation}
where
\begin{itemize}
    \item $S^\lambda(\mathbb{C}^N)$ are irreducible representations of $SU(N)$, indexed by Young diagrams $\lambda = (\lambda_1, \lambda_2, \dots, \lambda_k)$, $\lambda_1 \geq \lambda_2 \geq \dots \geq \lambda_k$, $\sum \lambda_i = n$, $k \leq N$;
    \item $M_\lambda$ are irreducible representations of $S_n$, corresponding to the same diagram $\lambda$;
    \item the direct sum is taken over all partitions of $n$ compatible with $N$.
\end{itemize}
The operators of $SU(N)$ act on $V^{\otimes n}$ diagonally on each tensor factor, while $S_n$ acts by permuting the tensor factors. The algebra of $SU(N)$-equivariant endomorphisms of the space $V^{\otimes n}$ is generated by the action of the group $S_n$, and conversely, the $S_n$-equivariant endomorphisms of the space $V^{\otimes n}$ form an algebra generated by the action of the group $SU(N)$, establishing a duality between these groups.

The decomposition~\eqref{eq:Schur} allows for the classification of quantum states by symmetry, which is a key aspect in the study of $SU(N)$-equivariant quantum circuits.

\subsubsection{Example 1: $SU(2)$ with Two Qubits}
For $V = \mathbb{C}^2$ and $n = 2$, we have $V^{\otimes 2}$ of dimension 4. The decomposition is:
\begin{equation}
    V^{\otimes 2} \cong S^{(2)}(\mathbb{C}^2) \otimes M_{(2)} \oplus S^{(1,1)}(\mathbb{C}^2) \otimes M_{(1,1)},
\end{equation}
where:
\begin{itemize}
    \item $\lambda = (2)$: triplet ($S = 1$), $S^{(2)}(\mathbb{C}^2)$ of dimension 3, $M_{(2)}$ is the trivial representation of $S_2$ (dimension 1);
    \item $\lambda = (1,1)$: singlet ($S = 0$), $S^{(1,1)}(\mathbb{C}^2)$ of dimension 1, $M_{(1,1)}$ is the sign representation of $S_2$ (dimension 1).
\end{itemize}
The basis of the triplet: $\{|\uparrow\uparrow\rangle, \frac{1}{\sqrt{2}}(|\uparrow\downarrow\rangle + |\downarrow\uparrow\rangle), |\downarrow\downarrow\rangle\}$, singlet: $\frac{1}{\sqrt{2}}(|\uparrow\downarrow\rangle - |\downarrow\uparrow\rangle)$.

\subsubsection{Example 2: $SU(3)$ with Two Particles}
For $V = \mathbb{C}^3$ and $n = 2$, we have $V^{\otimes 2}$ of dimension 9. The decomposition is:
\begin{equation}
    V^{\otimes 2} \cong S^{(2)}(\mathbb{C}^3) \otimes M_{(2)} \oplus S^{(1,1)}(\mathbb{C}^3) \otimes M_{(1,1)},
\end{equation}
where:
\begin{itemize}
    \item $S^{(2)}(\mathbb{C}^3)$: symmetric representation (sextet), dimension 6;
    \item $S^{(1,1)}(\mathbb{C}^3)$: antisymmetric representation (triplet), dimension 3.
\end{itemize}

\subsection{Representations of the Symmetric Group}

The irreducible representations of the group $S_n$ are indexed by Young diagrams $\lambda = (\lambda_1, \lambda_2, \dots)$, where $\sum \lambda_i = n$. A Young tableau is a diagram $\lambda$ filled with the numbers $\{1, 2, \dots, n\}$ in increasing order along rows and columns. The number of such tableaux $d_\lambda$ is the dimension of $M_\lambda$.

\noindent
\textbf{Example:} For the group $S_3$, we have
\begin{itemize}
    \item $\lambda = (3)$: $d_{(3)} = 1$, trivial representation $\pi(\sigma) = 1$;
    \item $\lambda = (2,1)$: $d_{(2,1)} = 2$, Young tableaux: \qquad
    $
        \begin{matrix} 1 & 2 \\ 3 & \end{matrix}, \qquad \begin{matrix} 1 & 3 \\ 2 & \end{matrix}
    $
    \item $\lambda = (1,1,1)$: $d_{(1,1,1)} = 1$, representation is the sign of the permutation, $\pi(\sigma) = \mbox{sign}(\sigma)$.
\end{itemize}

\setcounter{chapter}{1} 
\renewcommand{\thechapter}{B}
\chapter{Proof of \textit{SU(2)}-Equivariance}

\label{appendix:B1}

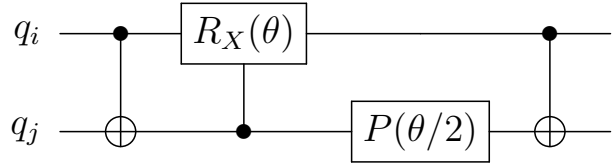
\begin{figure}[h]
    \centering
    \scalebox{1.2}{
        \Qcircuit @C=1.2em @R=1.0em @!R {
            \lstick{q_i} & \ctrl{1} & \gate{R_X(\theta)} & \qw & \ctrl{1} & \qw \\
            \lstick{q_j} & \targ & \ctrl{-1} & \gate{P(\theta/2)} & \targ & \qw \\
        }
    }
    \caption{\textit{SU(2)}-equivariant block}
    \label{fig:single_block}
\end{figure}

In this section, we analytically verify that the block $U$ of our ansatz (see Fig.~\ref{fig:single_block}) is \textit{SU(2)-equivariant}.
To do this, we need to check whether it commutes with every global transformation from \textit{SU(2)}, i.e., whether
\begin{equation} \label{eq:equivariance_condition}
V^{\otimes 2} U = U V^{\otimes 2}
\end{equation}
for all \( V \in \textit{SU(2)} \).
A general matrix \( V \in \textit{SU(2)} \) can be written as
\begin{equation}
    V = \begin{bmatrix}
        e^{i \alpha} \cos \beta & e^{i \phi} \sin \beta \\
        -e^{-i \phi} \sin \beta & e^{-i \alpha} \cos \beta
    \end{bmatrix},
\end{equation}
where \( \alpha, \beta, \phi \in \mathbb{R} \); this matrix is unitary (\( V^{\dagger} V = I \)) and has determinant 1.

\section{Matrix Representations of Quantum Operators}

We write the operators used in the ansatz explicitly, taking into account the order of controlled qubits:

\textbf{CNOT Operator} (\(\operatorname{CNOT}_{01}\)):
\begin{equation}
    \operatorname{CNOT} = \begin{bmatrix}
        1 & 0 & 0 & 0 \\
        0 & 1 & 0 & 0 \\
        0 & 0 & 0 & 1 \\
        0 & 0 & 1 & 0
    \end{bmatrix}.
\end{equation}

\textbf{Phase Shift Operator $P$} (\(P_1\)):
\begin{equation}
    P = \begin{bmatrix}
        1 & 0 & 0 & 0 \\
        0 & 1 & 0 & 0 \\
        0 & 0 & \cos \left( \frac{\theta}{2} \right) + i \sin \left( \frac{\theta}{2} \right) & 0 \\
        0 & 0 & 0 & \cos \left( \frac{\theta}{2} \right) + i \sin \left( \frac{\theta}{2} \right)
    \end{bmatrix}.
\end{equation}

\textbf{\(\operatorname{CRX}(\theta)\) Operator} (\(\operatorname{CRX}(\theta)_{10}\)):
\begin{equation}
    \operatorname{CRX}(\theta) = \begin{bmatrix}
        1 & 0 & 0 & 0 \\
        0 & \cos \left( \frac{\theta}{2} \right) & 0 & -i \sin \left( \frac{\theta}{2} \right) \\
        0 & 0 & 1 & 0 \\
        0 & -i \sin \left( \frac{\theta}{2} \right) & 0 & \cos \left( \frac{\theta}{2} \right)
    \end{bmatrix}.
\end{equation}

Then the operator \( U \) is constructed as
\begin{equation}
    U = \text{CNOT} \cdot \text{CRX}(\theta) \cdot P \cdot \text{CNOT}.
\end{equation}
Explicitly multiplying the matrices, we obtain:
\begin{equation}
    U = \begin{bmatrix}
        1 & 0 & 0 & 0 \\
        0 & \left( i \sin \left( \frac{\theta}{2} \right) + \cos \left( \frac{\theta}{2} \right) \right) \cos \left( \frac{\theta}{2} \right) & -i \left( i \sin \left( \frac{\theta}{2} \right) + \cos \left( \frac{\theta}{2} \right) \right) \sin \left( \frac{\theta}{2} \right) & 0 \\
        0 & -i \left( i \sin \left( \frac{\theta}{2} \right) + \cos \left( \frac{\theta}{2} \right) \right) \sin \left( \frac{\theta}{2} \right) & \left( i \sin \left( \frac{\theta}{2} \right) + \cos \left( \frac{\theta}{2} \right) \right) \cos \left( \frac{\theta}{2} \right) & 0 \\
        0 & 0 & 0 & 1
    \end{bmatrix}.
\end{equation}

\section{Verification of \textit{SU(2)}-Equivariance}
To verify \textit{SU(2)-equivariance}, we compute the difference
\begin{equation}
    D = (V \otimes V) U - U (V \otimes V)
\end{equation}
and check that each element of the resulting matrix is zero.

Performing the matrix multiplication explicitly, we obtain:
\begin{equation}
    D = \begin{bmatrix}
        0 & * & * & 0 \\
        * & 0 & * & * \\
        * & * & 0 & * \\
        0 & * & * & 0
    \end{bmatrix},
\end{equation}
where $*$ denotes elements that are not zero before simplification.
Consider one such element, \( D_{4,2} \), and show that it is zero:
\begin{align*}
    D_{4,2} &= i e^{i \theta/2} e^{-i \alpha} e^{i \phi} \sin \beta \sin \frac{\theta}{2} \cos \beta \\
    &\quad - e^{i \theta/2} e^{-i \alpha} e^{i \phi} \sin \beta \cos \beta \cos \frac{\theta}{2} \\
    &\quad + e^{-i \alpha} e^{-i \phi} \sin \beta \cos \beta.
\end{align*}
Factoring out common terms:
\begin{align*}
    D_{4,2} &= e^{i \theta/2} e^{-i \alpha} e^{i \phi} \sin \beta \cos \beta \left( i \sin \frac{\theta}{2} - \cos \frac{\theta}{2} \right) \\
    &\quad + e^{-i \alpha} e^{-i \phi} \sin \beta \cos \beta,
\end{align*}
Noting that
\begin{equation}
    i \sin \frac{\theta}{2} - \cos \frac{\theta}{2} = - \left( i \sin \left( -\frac{\theta}{2} \right) + \cos \left( -\frac{\theta}{2} \right) \right) = -e^{-i \theta/2},
\end{equation}
we conclude that:
\begin{equation}
    D_{4,2} = -e^{-i \alpha} e^{-i \phi} \sin \beta \cos \beta + e^{-i \alpha} e^{-i \phi} \sin \beta \cos \beta \Rightarrow D_{4,2} = 0.
\end{equation}

Applying similar transformations to the other non-zero terms, we find that all elements of the matrix $D$ are zero, thus proving the \textit{SU(2)-equivariance} of the ansatz \( U \).

\setcounter{chapter}{1} 
\renewcommand{\thechapter}{C}
\chapter{Dynamic Mode Decomposition (DMD) Method}
\label{appendix:DMD}

\textbf{Dynamic Mode Decomposition (DMD)} is a method that enables the extraction of spatio-temporal structures in complex dynamical systems based on a sequence of system states \cite{brunton2022data}.

The core idea of DMD is to approximate the system’s dynamics using a linear operator that best maps the system’s state from one time step to the next. Formally, given a sequence of system states:
\[
\{\mathbf{x}(t_1), \mathbf{x}(t_2), \dots, \mathbf{x}(t_m)\}, \quad \mathbf{x}(t_i) \in \mathbb{R}^n,
\]
we construct two snapshot matrices:
\[
\mathbf{X} = \begin{bmatrix}
| & | & & | \\
\mathbf{x}(t_1) & \mathbf{x}(t_2) & \dots & \mathbf{x}(t_{m-1}) \\
| & | & & |
\end{bmatrix}, \quad
\mathbf{X'} = \begin{bmatrix}
| & | & & | \\
\mathbf{x}(t_2) & \mathbf{x}(t_3) & \dots & \mathbf{x}(t_{m}) \\
| & | & & |
\end{bmatrix}.
\]
The goal of DMD is to find an operator $\mathbf{A}$ that satisfies the equation
\[
\mathbf{X'} \approx \mathbf{A} \mathbf{X}.
\]
The matrix $\mathbf{A}$ is determined as the solution to the least squares problem
\[
\mathbf{A} = \mathbf{X'} \mathbf{X}^{\dagger},
\]
where $\mathbf{X}^{\dagger}$ is the pseudoinverse matrix, computed using its SVD decomposition,
\[
\mathbf{X} = \mathbf{U} \mathbf{\Sigma} \mathbf{V}^{*},
\]
where $\mathbf{U} \in \mathbb{C}^{n \times r}$, $\mathbf{\Sigma} \in \mathbb{C}^{r \times r}$, $\mathbf{V} \in \mathbb{C}^{(m-1) \times r}$, and $r$ is the chosen truncation rank. Then, $\mathbf{X}^{\dagger} = \mathbf{V} \mathbf{\Sigma}^{-1} \mathbf{U}^{*}$, and $\mathbf{A} = \mathbf{X'} \mathbf{V} \mathbf{\Sigma}^{-1} \mathbf{U}^{*}$.

The operator $\mathbf{A}$ is large ($n \times n$), so DMD uses its reduced version $\widetilde{\mathbf{A}} = \mathbf{U}^{*} \mathbf{A} \mathbf{U}$:
\[
\widetilde{\mathbf{A}} = \mathbf{U}^{*} \mathbf{X'} \mathbf{V} \mathbf{\Sigma}^{-1}.
\]
For the reduced matrix $\widetilde{\mathbf{A}}$, we then compute its eigenvalues $\Lambda$ and eigenvectors $\mathbf{W}$:
\[
\widetilde{\mathbf{A}} \mathbf{W} = \mathbf{W} \Lambda.
\]
The spatial DMD modes $\mathbf{\Phi} \in \mathbb{C}^n$ in the original space are calculated as
\[
\mathbf{\Phi} = \mathbf{X'} \mathbf{V} \mathbf{\Sigma}^{-1} \mathbf{W}.
\]
Each DMD mode corresponds to an eigenvalue $\lambda_j$ of the matrix $\mathbf{A}$, which can be written as
\[
\lambda_j = e^{(\sigma_j + i \omega_j) \Delta t},
\]
where $\sigma_j$ describes the exponential growth or decay of the mode, $\omega_j$ is its frequency, and $\Delta t = t_{k+1} - t_k$.

DMD allows the system’s evolution to be represented as:
\[
\mathbf{x}(t_k) \approx \sum_{j=1}^{r} \mathbf{\phi}_j \lambda_j^{k-1} b_j = \mathbf{\Phi} \Lambda^{k-1} \mathbf{b},
\]
where $\mathbf{b}$ are the initial mode amplitudes, which can be found from the initial snapshot:
\[
\mathbf{b} = \mathbf{\Phi}^{\dagger} \mathbf{x}(t_1).
\]
This provides a compact model that not only describes the system’s dynamics but also enables prediction of its future behavior.

\medskip
\textbf{Connection to the Koopman Operator.} It is worth noting that the operator $\mathbf{A}$ in DMD is a finite-dimensional approximation of the infinite-dimensional linear Koopman operator, which fully describes the nonlinear dynamics of the system in the space of observables. This connection provides a theoretical foundation for the effectiveness of DMD across a wide range of problems.

\selectlanguage{english}
\bibliographystyle{unsrt}

\end{document}